\documentclass[prd,preprint,eqsecnum,nofootinbib,amsmath,amssymb,tightenlines,preprintnumbers]{revtex4-2}

\usepackage[colorlinks=true, linkcolor=blue, citecolor=blue, urlcolor=blue]{hyperref}

\usepackage{bm}
\usepackage{slashed}

\usepackage{xcolor}

\usepackage{tikz}
\newcommand{\Plus}{{\mathord{\tikz\draw[line width=0.2ex, x=1ex, y=1ex] (0.5,0) -- (0.5,1)(0,0.5) -- (1,0.5);}}}

\usepackage{pifont}
\def\cnum#1{\ding{\numexpr171+#1\relax}}

\def\Nf{N_{\rm f}}

\def\eps{\epsilon}
\def\vareps{\varepsilon}
\def\qhat{\hat q}

\def\csch{\operatorname{csch}}

\def\min{{\rm min}}
\def\max{{\rm max}}

\def\el{{\rm el}}
\def\coll{{\rm coll}}

\def\BH{{\rm BH}}
\def\BK{{\rm BK}}

\def\LPM{{\rm LPM}}
\def\LPMplus{{{\rm LPM}\Plus}}
\def\Elpma{E_{\rm LPM}}
\def\Elpm{\hat E_{\rm LPM}}
\def\kgamma{k_\gamma}
\def\bkgamma{{\bm k}_\gamma}
\def\form{{\rm form}}
\def\tform{t_\form}
\def\Sxtra{S_{\rm extra}}
\def\brem{{\rm brem}}
\def\pair{{\rm pair}}
\def\me{m_{\rm e}}

\def\Re{\operatorname{Re}}
\def\Im{\operatorname{Im}}
\def\tr{\operatorname{tr}}
\def\pr{{\rm pr}}
\def\Mo{{\bar M}_0}
\def\Mpr{M_\pr}
\def\hpr{h^\pr}
\def\Omegapr{\Omega_\pr}
\def\gammaE{\gamma_{\rm\scriptscriptstyle E}}
\def\yfrak{\eta}

\def\mp{m_{\rm p}}

\def\dlangle{\langle\!\langle}
\def\drangle{\rangle\!\rangle}

\def\b{{\bm b}}
\def\p{{\bm p}}
\def\v{{\bm v}}
\def\k{{\bm k}}
\def\kc{k_{\rm c}}
\def\P{{\bm P}}
\def\omegapl{\omega_{\rm pl}}
\def\grad{{\bm\nabla}}

\def\kc{k_{\rm c}}
\def\PiT{\Pi_{\rm T}}

\def\bpT{\p_{\rm T}}
\def\qmax{q_{\rm max}}
\def\qmin{q_{\rm min}}

\def\PSI{\bar\psi}
\def\lnGamma{\operatorname{ln\Gamma}}
\def\LNGAMMA{\operatorname{\overline{ln\Gamma}}}

\begin {document}

\title{Calculating extremely high energy bremsstrahlung in matter}

\author{Peter Arnold}
\author{Joshua Bautista}
\affiliation{%
    Department of Physics,
    University of Virginia,
    Charlottesville, Virginia 22904-4714, USA
}%
\author{Omar Elgedawy}
\affiliation{%
  State Key Laboratory of Nuclear Physics and
  Technology, Institute of Quantum Matter, South China Normal
  University, Guangzhou 510006, China
}
\affiliation{%
  Guangdong Basic Research Center of Excellence for
  Structure and Fundamental Interactions of Matter, Guangdong
  Provincial Key Laboratory of Nuclear Science, Guangzhou
  510006, China
}
\affiliation{%
  CPHT, CNRS, \'{E}cole polytechnique, Institut Polytechnique de Paris,
  91120 Palaiseau, France
}
\author{Shahin Iqbal}
\affiliation{%
National Centre for Physics,
  Shahdra Valley Road,
  Islamabad, 45320 Pakistan
}
\affiliation{%
  Theoretical Physics Department,
  CERN,
  CH-1211 Geneva 23, Switzerland
}%

\preprint{CPHT-RR014.042026}
\preprint{CERN-TH-2026-090}

\date {\today}

\begin {abstract}%
{%
  Ultra-relativistic electrons initiate electromagnetic showers in ordinary
  matter that evolve through bremsstrahlung and pair production.
  At very high energy, the quantum mechanical duration of bremsstrahlung
  becomes longer than the mean free time to elastically scatter from the
  medium, leading to a significant suppression known as the
  Landau-Pomeranchuk-Migdal (LPM) effect.
  For some ranges of bremsstrahlung photon and initial electron energies
  $(\kgamma,E)$, the duration becomes so long that
  it will also overlap
  with subsequent pair production by the bremsstrahlung photon,
  disrupting LPM suppression and drastically changing
  LPM predictions.
  We have previously calculated this change for extremely high energies
  ($\kgamma \gg 2$ TeV or more, depending on the medium),
  for which the electron mass and medium-induced photon mass could be ignored.
  In this paper, we extend that analysis to lower
  (but still ultra-relativistic)
  energy by accounting for those masses, leading to a rich map of
  behavior in different regions of $(\kgamma,E)$.
}%
\end {abstract}

\maketitle
\newpage

\thispagestyle {empty}

{\def\boldmath{}\tableofcontents}
\newpage


\section{Introduction}

Very high energy electrons or positrons passing through ordinary matter
lose energy predominantly through showering via repeated
bremsstrahlung ($e \to e\gamma$) and pair production ($\gamma \to e\bar e$)
mediated by the electric fields inside atoms.
At sufficiently high energy, the quantum duration of
bremsstrahlung (known as the formation or coherence time $\tform$)
becomes larger than the mean free time
between elastic scatterings from the medium,
leading to a drastic suppression of the naive bremsstrahlung rate,
known as the Landau-Pomeranchuk-Migdal (LPM) effect
\cite{LP1,LP2,Migdal}.%
\footnote{
  English translations of the papers \cite{LP1,LP2}
  are available in \cite{LPenglish}.
}
At extremely high energies, depending on the initial electron energy $E$ and the
bremsstrahlung photon energy $\kgamma$,
the formation time can become
long enough
to allow for quantum overlap of the bremsstrahlung process
with subsequent pair production $\gamma \to e\bar e$ by the
bremsstrahlung photon.
For the regions of $(\kgamma,E)$ where this occurs,
it was long ago believed that the effect would be
to further suppress the already-suppressed LPM bremsstrahlung rate
\cite{Galitsky}.  We have recently revisited this issue
\cite{softqed1,softqed2}, concluding instead that overlapping
pair production \textit{disrupts} LPM suppression,
leading to bremsstrahlung rates much \textit{larger} than the
original LPM rate.
We refer to this pair-production disruption of the LPM effect as the
$\LPMplus$ effect.

Ref.\ \cite{softqed1} provides a qualitative argument for
this $\LPMplus$ disruption of the LPM effect, which is then
confirmed by a
derivation in the case of extremely high energy bremsstrahlung with
photon energy $\kgamma \gg \Elpma$, where $\Elpma$ is a medium-dependent
energy scale that is, for example, $\Elpma \simeq 2.5$ TeV for Gold
and $234$ PeV for air \cite{SpencerReview}.
Ref.\ \cite{softqed2} introduces and summarizes our new results from
generalizing that calculation to lower energies, and the
purpose of the current paper is to provide the derivation of
those results.
Specifically, our earlier calculations for $\kgamma \gg \Elpma$ were
able to ignore both the mass $m$ of the high-energy electrons and
the medium-induced photon mass $m_\gamma$.  This paper generalizes that
analysis to also cover $\kgamma \lesssim \Elpma$ by including
the previously-ignored masses.

\bigskip

In the next section, we briefly outline our calculation's
assumptions and approximations along with some of our notation.
In section \ref{sec:LPM}, we warm up by reproducing Migdal's results
for the original LPM rate for both high-energy bremsstrahlung and
high-energy pair production, including the effect of masses.
Section \ref{sec:massless} then partially generalizes the earlier
calculation \cite{softqed1} of
the $\LPMplus$ effect
to now include the
mass $m$ of the high-energy electrons while still ignoring the
medium-induced photon mass (also known as the dielectric effect).
We complete the full calculation in
section \ref{sec:dielectric} by incorporating non-zero $m_\gamma$.
Throughout sections \ref{sec:LPM}--\ref{sec:dielectric}, we
also provide simpler results for limiting cases.
Section \ref{sec:logs} offers insight into the physical origin of
various logarithms that appear in those limits.
Section \ref{sec:photonuclear} warns about the
applicability of our results at extreme energies $k_\gamma \gtrsim 10^{20}$ eV
--- a warning that could be resolved by including into our
calculation the additional possibility of disrupting
LPM bremsstrahlung via
direct collision of the bremsstrahlung photon
with a nucleus ($\gamma A \to$ hadrons).

As in refs.\ \cite{softqed1,softqed2}, it is simplest to make our first
quantitative calculations of the $\LPMplus$ effect by (i) assuming that
the size $L$ of the medium is large compared to the formation time
for bremsstrahlung and (ii) focusing on computing the
differential rate $d\Gamma/d\kgamma$ for the electron to
lose energy $\kgamma$ through radiative processes.
Section \ref{sec:finiteL} explains our qualitative expectation for
how our result would change for smaller values of $L$,
and, for those interested,
section \ref{sec:collEloss} clarifies for what region of $(\kgamma,E)$
radiative processes for energy loss dominate over non-radiative ones.


\section {Brief Review of Assumptions and Notation}
\label {sec:assumptions}

Our notation and general assumptions are mostly covered in
the companion paper \cite{softqed2} and the first section
of ref.\ \cite{softqed1}, but we briefly summarize
them here for convenience.
We work throughout in natural units.

For simplicity,
we only consider the case where the medium is large compared to the
bremsstrahlung formation length.

We will simplify our derivation by assuming
that the medium-induced photon mass $m_\gamma$ is
small compared to the electron mass $m$, and in particular we will
eventually make the parametric assumption that $m_\gamma \ll \alpha^{1/2} m$.
This assumption is very well satisfied for ordinary matter
(but, in contrast, not for an ultra-relativistic QED plasma%
\footnote{
  For an ultra-relativistic QED plasma ($eT \gg$ the vacuum electron
  mass),
  the medium-induced photon mass and medium-induced electron mass
  would both be of order $e T$.
}%
).

In much of the QED literature on the LPM effect, results are presented in
terms of a scale $\Elpma$ that characterizes the energy scale at which
the LPM effect becomes important for democratic bremsstrahlung
(``democratic'' meaning
that neither daughter of $e{\to}e\gamma$ has energy very small
compared to the other).  This scale is usually defined
in terms of the radiation length $X_0$ for the medium.
In our derivations, we instead characterize the medium by
the diffusion constant $\qhat$ for the transverse momentum of high-energy
electrons, and we define the scale
\begin {equation}
  \Elpm \equiv \frac{m^4}{\qhat} .
\label {eq:Elpmhat}
\end {equation}
$\Elpm$ agrees with the conventional $\Elpma$ to within a factor of 2.
In the context of the LPM effect, the relevant
value of $\qhat$ (and so also $\Elpm$)
depends logarithmically on the bremsstrahlung formation time and
so logarithmically  on $(\kgamma,E)$, and $\qhat$ can vary by as much as a
factor of 2 in our context.
The relation of our $\Elpm$ to
the conventional definition of $\Elpma$ varies between roughly
\begin {equation}
  \Elpm \simeq
  \Elpma \times
  \begin {cases}
    2 , & \mbox{for negligible or mild LPM suppression}; \\
    1 , & \mbox{for \textit{extreme} LPM suppression}.
  \end {cases}
\label {eq:Elpmcases}
\end {equation}
We will not need the explicit formula for $\qhat$ in our derivation, but,
for the sake of concreteness, it is
\begin {equation}
   \qhat \simeq 8\pi (Z\alpha)^2 n \ln(\cdots) ,
\label {eq:qhat}
\end {equation}
where (for a single-element medium) $Z$ is atomic number,
$n$ is the number density of atoms,
and the argument of the logarithm varies roughly from
$Z^{-1/3} a_{\rm Bohr} m$ to $Z^{-1/3} a_{\rm Bohr}/R_{\rm nucleus}$
depending on $(\kgamma,E)$.
See ref.\ \cite{softqed1} for details.%
\footnote{
  In particular, see section 1.2 and appendix A of ref.\ \cite{softqed1}.
  As explained there, the factor of 2 difference between the two limits in
  (\ref{eq:Elpmcases}) above is accidental and only approximate.
}

The description of bremsstrahlung and pair production in terms of $\qhat$
is a leading-log approximation that assumes that there are many elastic
collisions with the medium during the formation time.  It is equivalent
to the leading-log approximation originally used by Migdal \cite{Migdal}.
Following Migdal, it is conventional to slightly adjust the overall
scale of the argument of the logarithm in (\ref{eq:qhat}) to reproduce
as well as possible the Bethe-Heitler formula \cite{BH} for
the high-energy bremsstrahlung when there is only a single scattering
(i.e.\ in the limit of no LPM suppression).
Described in terms of $\qhat$, the
Bethe-Heitler result for the bremsstrahlung rate is
approximately
\begin {equation}
  \left[ \frac{d\Gamma}{dx_\gamma} \right]_\BH \simeq 
  \frac{\alpha \qhat}{6\pi m^2} \,
  \bigl[ 2 P_{e\to\gamma}(x_\gamma) + x_\gamma \bigr] ,
\label{eq:BHrate}
\end {equation}
where $P_{e\to\gamma}(x_\gamma)$ is the unregulated
Dokshitzer-Gribov-Lipatov-Altarelli-Parisi (DGLAP) splitting function
\begin {equation}
  P_{e\to\gamma}(x_\gamma) = \frac{1 + (1{-}x_\gamma)^2}{x_\gamma}
\label {eq:Peg}
\end {equation}
and the logarithm in (\ref{eq:qhat}) is, for this limit,%
\footnote{
  Migdal \cite{Migdal} used 190 instead of 184 inside the logarithm on
  the far right-hand side of (\ref{eq:BHlog}).
  The refined coefficient 184 can be found, for example, in
  Klein's LPM review \cite{SpencerReview} and in the $Z > 4$ entry for
  $L_{\rm rad}$
  in Table 34.2 of the 2024 Review of Particle Physics \cite{RPP2024}.
}
\begin {equation}
  \ln(\cdots)
  \simeq \ln\bigl(1.34\times Z^{-1/3} a_{\rm Bohr} m\bigr)
  \simeq \ln\bigl(184\,Z^{-1/3}\bigr) .
\label {eq:BHlog}
\end {equation}

Since our purpose in this paper is to explore the consequences of the
$\LPMplus$ effect, and not to get bogged down in discussing
further improvement of these approximations, we simply use the
multiple-scattering ($\qhat$) approximation throughout and
then present our final results in terms of $\Elpm$, with the
understanding that $\Elpm$ is related by (\ref{eq:Elpmcases})
to the values of $\Elpma$ traditionally
quoted for media.%
\footnote{
  Beware that some authors'
  conventions for defining $\Elpma$ differ by a factor of 2.
  For $\Elpma$, we use the convention of ref.\ \cite{SpencerReview} and of
  eq.\ (34.33) of the 2024 Review of Particle Physics \cite{RPP2024}.
}

\begin {figure}[t]
\begin {center}
  \includegraphics[scale=0.7]{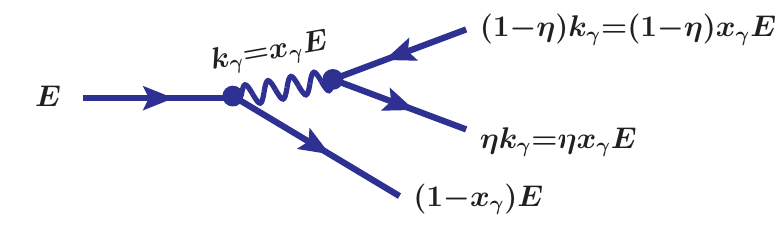}
  \caption{
    \label{fig:notation}
    Our notation for energies and energy fractions $(x_\gamma,\eta)$ for
    $e \to e\gamma$ and $\gamma \to e\bar e$.
    The overlap of these two processes (the $\protect\LPMplus$ effect) is only
    important when $x_\gamma \ll 1$, in which case there is no
    confusion differentiating the two otherwise-identical final-state
    electrons because one is much softer than the other.
    (See the discussion in section 6 of ref.\ \cite{softqed1}.)
    The energy fraction $\eta$ in this paper was called
    $\mathfrak{y}_{\scriptscriptstyle{\rm E}}$
    in ref.\ \cite{softqed1}.
  }
\end {center}
\end {figure}

The notation used in this paper for various particle energies is summarized in
fig.\ \ref{fig:notation}.
More notation will be introduced in the next section, for which a summary
will be given in table \ref{tab:notation}.

\pagebreak[4] 

\section{Review: Calculating the ordinary LPM effect}
\label {sec:LPM}

\begin{figure}
  \includegraphics[scale=1.0]{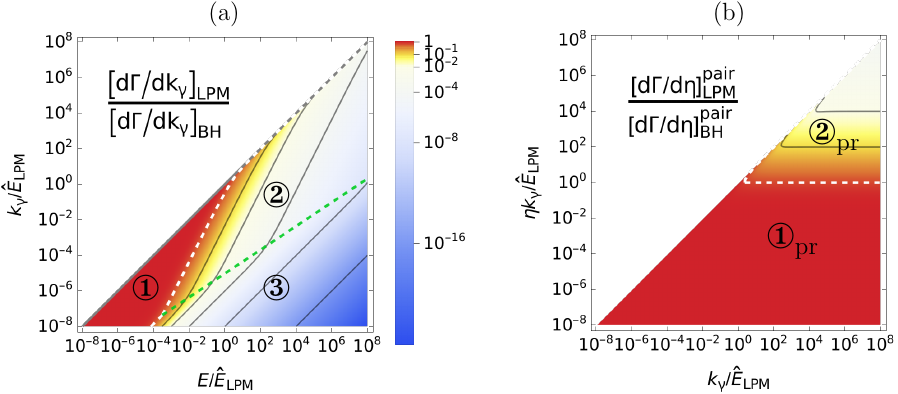}
  \caption{
     \label{fig:LPMoverBH}
     (a) Log-log-log contour plot \cite{softqed2} of the ratio $\LPM/\BH$
     of the original
     LPM bremsstrahlung rate (\ref{eq:LPMrate})
     to the Bethe-Heitler rate (\ref{eq:BHrate}) vs.\ $E/\Elpm$ and
     $\kgamma/\Elpm$ for $m_\gamma/m = 10^{-4}$.  Note the non-uniform
     spacing $(10^{-1},10^{-2},10^{-4},10^{-8},10^{-16})$ of the contour lines.
     The diagonal $\kgamma{=}E$ corresponds to $\LPM/\BH = 1$.
     The labeling of the different regions is
     \cnum{1} Bethe-Heitler (BH),
     \cnum{2} deep LPM, and
     \cnum{3} dielectric (like Bethe-Heitler but
       dominated by the effect of a medium-induced photon mass).
     (b) The same for the ratio of the LPM (\ref{eq:LPMpair})
     and Bethe-Heitler (\ref{eq:BHpair}) rates for
     pair production $\gamma{\to}e\bar e$ vs.\
     $\kgamma/\Elpm$ and $E_e/\Elpm = \yfrak\kgamma/\Elpm$.
  }
\end{figure}

Before we proceed to calculate the effect of pair production overlapping
bremsstrahlung, it is useful to first review the calculation of the
ordinary LPM effect (separately) for both bremsstrahlung and
pair production and
reproduce the results of Migdal \cite{Migdal}.
Rather than following Migdal in detail, we will use the formalism and
notation that we later adapt to the $\LPMplus$ case.
We have previously reviewed \cite{softqed1} the ordinary LPM calculation
in the specific case of the deep-LPM regime where all particle masses
can be ignored.  For bremsstrahlung, that corresponded
to deep inside region 2 of fig.\ \ref{fig:LPMoverBH}a, which shows the
LPM suppression factor
LPM/BH relative to the ordinary (Bethe-Heitler) bremsstrahlung rate
\cite{BH}.
Here, we now include the effects of the
electron mass $m$ and the medium-induced
photon mass $m_\gamma$ into our review of Migdal's result, which
describes the smooth transitions to regions 1 (the Bethe-Heitler region) and
3 (the dielectric region) of fig.\ \ref{fig:LPMoverBH}a.

\subsection{LPM bremsstrahlung}
\label {sec:LPMbrem}

\subsubsection{General Formula}

\begin {figure}[t]
\begin {center}
  \includegraphics[scale=0.6]{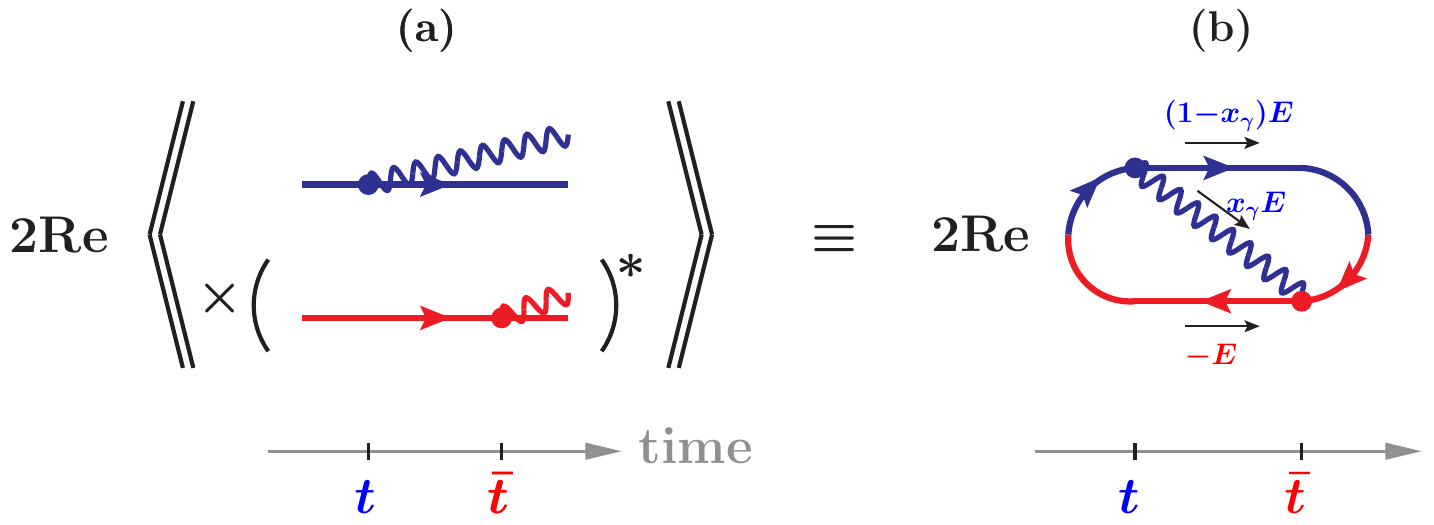}
  \caption{
    \label{fig:lpm}
    (a) A graphical representation of the LPM bremsstrahlung rate,
    consisting of (i) the amplitude (blue) times the conjugate amplitude
    (red) for time ordering $t < \bar t$, implicitly integrated over $t$
    and $\bar t$, (ii) $2\Re(\cdots)$ to add in
    the other time ordering $\bar t < t$, and (iii) averaging
    $\dlangle \cdots \drangle$ the rate
    over the randomness of the amorphous medium.
    All electron lines are implicitly summed over arbitrary numbers of
    elastic collisions with the medium.
    (b) depicts the short-hand graphical notation that we use for (a). 
  }
\end {center}
\end {figure}

As in ref.\ \cite{softqed1}, we use the formalism of
Zakharov \cite{Zakharov3} to organize the calculation.%
\footnote{
  See also his earlier papers \cite{Zakharov1,Zakharov2}.
  In parallel with
  Baier, Dokshitzer, Mueller, Peigne, and Schiff (BDMPS)
  \cite{BDMPS1,BDMPS2,BDMPS3,BDMS},
  Zakharov was working out how to generalize the LPM effect to QCD.
}
Fig.\ \ref{fig:lpm}a depicts the leading-order contribution to
medium-induced bremsstrahlung in time-ordered perturbation theory.
The top (blue) diagram represents
a contribution to the amplitude, the bottom (red) diagram represents
a similar contribution to the complex amplitude.
Though not explicitly shown, each electron line in the diagrams should
be understood to undergo arbitrarily many elastic scatterings with
the medium.  Fig.\ \ref{fig:lpm}b represents a short-hand graphical
notation for fig.\ \ref{fig:lpm}a that will be used throughout this
paper.  Zakharov conceptually re-interpreted the diagram of
fig.\ \ref{fig:lpm}b as three high-energy particles $e^+\gamma e^-$
propagating forward in time, and he packaged that evolution
into the form of
a two-dimensional Schr\"odinger-like equation that described the
transverse dynamics of the particles by an effective Hamiltonian
\begin {equation}
  {\cal H} =
  \frac{p_{\perp e^-}^2{+}m^2}{2(1{-}x_\gamma)E}
     + \frac{k_{\perp \gamma}^2{+}m_\gamma^2}{2x_\gamma E} 
     - \frac{p_{\perp e^+}^2{+}m^2}{2E}
  + V(\b_{e^-} - \b_{e^+}) .
\label {eq:Heff}
\end {equation}
The $\b_i$ are transverse positions and
the ``potential'' $V$ accounts for the medium-averaging of the
interactions of the three high-energy particles with the medium.
In the multiple-scattering ($\qhat$)
approximation (equivalent to the large-logarithm
approximation that Migdal used for making analytic calculations of
the LPM effect),%
\footnote{
  See sections 3.1--3.2 of our earlier paper \cite{softqed1} for a brief
  review of the source of (\ref{eq:Vqhat}) in the present notation and context.
  However, $\qhat$ has logarithmic dependence on the scale $\b$ itself
  (arising from the Rutherford tail of the probability distribution for
  individual $\p_\perp$ kicks from the medium), and our treatment here
  of $\qhat(b)$ as \textit{constant} is equivalent to the leading-log,
  multiple-scattering approximation mentioned in section \ref{sec:assumptions}
  (which is also, in completely different language, the large-logarithm
  approximation made by Migdal).  If desired, one technique that could be used
  to systematically improve
  this approximation is the Improved Opacity Expansion (IOE)
  \cite{IOE1,IOE2,IOE3}.
}
\begin {equation}
   V(\b) \simeq -\tfrac{i}{4} \qhat b^2 .
\label {eq:Vqhat}
\end {equation}
In (\ref{eq:Heff}), $m$ is the usual electron mass.
The medium-induced photon mass $m_\gamma$ appearing in (\ref{eq:Heff})
is more often referred to in the literature as the plasma frequency
$\omegapl$.  It is dominated by forward scattering from
atomic electrons and given by%
\footnote{
  See, for example, Klein's LPM review \cite{SpencerReview}
  (which uses units where $\alpha = e^2$).
}
\begin {equation}
  m_\gamma \simeq \sqrt{ \frac{4\pi\alpha Z n}{m} }
\label {eq:mgamma}
\end {equation}
in the simple case of a
medium composed of atoms of atomic number $Z$ and density $n$.
Since this plasma frequency is typically defined and calculated classically in
a non-relativistic context ($\kgamma \ll m$),
we provide a derivation in terms of
relativistic forward Compton scattering in appendix \ref{app:mgamma}
for the sake of completeness.%
\footnote{
  The appendix also elucidates how the calculation differs
  between the cases of ordinary matter vs.\ an ultra-relativistic
  plasma.
  In the latter case,
  the effective mass $m_\gamma$ of very high energy
  photons is not the same as the plasma frequency $\omegapl$.
}

Following ref.\ \cite{softqed1} (but here including the medium-induced
photon mass $m_\gamma$), we use momentum conservation
$\p_{e^-}{+}\bkgamma{+}\p_{e^+}=0$ (after medium averaging) and
choose the $z$ axis to be in the direction of the photon
($\k_{\perp\gamma}{=}0$) to reduce (\ref{eq:Heff}) to
an effective 1-particle Hamiltonian
\begin {subequations}
\label {eq:HO}
\begin {equation}
  {\cal H} = \biggl( \frac{p_\perp^2}{2\Mo} + \tfrac12 \Mo\Omega_0^2 b^2 \biggr)
           + \biggl( \frac{m^2}{2\Mo} + \frac{m_\gamma^2}{2\kgamma} \biggr)
\label {eq:HeffHO}
\end {equation}
with $\p_\perp$ representing $\p_{\perp,e^-} = -\p_{\perp,e^+}$, and
\begin {equation}
  \Mo \equiv \frac{(1{-}x_\gamma)E}{x_\gamma} \,,
  \qquad
  \Omega_0 \equiv \sqrt{ - \frac{i\qhat}{2\Mo} }
  = \sqrt{ - \frac{i x_\gamma \qhat}{2(1{-}x_\gamma)E} } \,.
\label {eq:MOmega0}
\end {equation}
\end {subequations}

For the massless case ($m=m_\gamma=0$), ref.\ \cite{softqed1} reviews,
in the language used here, that Zakharov's version of the
rate formula corresponding to our fig.\ \ref{fig:lpm} is%
\footnote{
  See in particular section 3.1 and appendix C of ref.\ \cite{softqed1}.
}
\begin {equation}
  \left[ \frac{d\Gamma}{dx_\gamma} \right]_\LPM
  =
  \frac{ \alpha\,P_{e\to\gamma}(x_\gamma) }{ \Mo^2 }
  \Re \int_0^\infty d(\Delta t) \>
  \grad_{\b'} \cdot \grad_{\b} \,
  G(\b',\Delta t;\b,0) \Bigl|_{\b'=\b=0} ,
\label {eq:Zrate0}
\end {equation}
where $G(\b',\Delta t;\b,0)$ is the
propagator $\langle\b'|e^{-i{\cal H}\,\Delta t}|\b\rangle$ associated with
harmonic oscillator Hamiltonian (\ref{eq:HO}).
The derivatives $\grad_{\b}$ and $\grad_{\b'}$ are position-space
versions of factors of transverse momentum associated with each
photon vertex.  The factor of
$\alpha\,P_{e\to\gamma}(x_\gamma)$ is also associated with the vertices,
with $P_{e\to\gamma}(x_\gamma)$ given by (\ref{eq:Peg}).

When the electron mass $m$ is included in the calculation, the
factors associated with the splitting vertices become more complicated.
The electron-helicity conserving amplitudes give the same factors of
$\alpha\,P_{e\to\gamma}(x_\gamma) \grad_{\b'}\cdot\grad_{\b}$ as before,
but there is also a helicity-flip amplitude that is proportional
to $m^2$.  In total, this corresponds to replacing
\begin {equation}
  \alpha\,P_{e\to\gamma}(x_\gamma) \grad_{\b'} \cdot \grad_{\b} \longrightarrow
  \alpha \left[
    P_{e\to\gamma}(x_\gamma) \grad_{\b'} \cdot \grad_{\b}
    + x_\gamma m^2
  \right]
\label{eq:PwithFlip}
\end {equation}
in (\ref{eq:Zrate0}).
See appendix \ref{app:BremWithFlip} for more detail.

Since the effective Hamiltonian (\ref{eq:HO}) is just a
harmonic oscillator Hamiltonian plus a constant, the corresponding
propagator is simply
\begin {multline}
  G(\b',\Delta t;\b,0) =
  \frac{\Mo\Omega_0 \csc(\Omega_0\,\Delta t)}{2\pi i}
  \exp\Bigl( i\Mo\Omega_0 \bigl[
    \tfrac12(b^2 + b'^2) \cot(\Omega_0\,\Delta t)
    - \b\cdot\b' \csc(\Omega_0\,\Delta t)
  \bigr] \Bigr)
\\
  \times
  e^{-i(h_m+h_\gamma) \Delta t} ,
\label {eq:Gprop}
\end {multline}
where
\begin {equation}
  h_m \equiv \frac{m^2}{2\Mo}
  \qquad \mbox{and} \qquad
  h_\gamma \equiv \frac{m_\gamma^2}{2\kgamma}
\label {eq:hmgamma}
\end {equation}
are the two constant terms in ${\cal H}$,
the $e^{-i(h_m+h_\gamma) \Delta t}$ in (\ref{eq:Gprop})
is the contribution of those constants to
$e^{-i{\cal H}\,\Delta t}$, and the rest is the propagator for a 2-dimensional
harmonic oscillator.
Using (\ref{eq:Gprop}) in (\ref{eq:Zrate0}), and making the replacement
(\ref{eq:PwithFlip}), yields
\begin {multline}
  \left[ \frac{d\Gamma}{dx_\gamma} \right]_\LPM
  =
  - \frac{\alpha}{\pi} \Re \int_0^\infty d(\Delta t) \>
  \biggl\{
    P_{e\to\gamma}(x_\gamma) \,
       \Omega_0^2 \csc^2(\Omega_0\,\Delta t)
\\
    + i x_\gamma h_m 
       \Omega_0 \csc(\Omega_0\,\Delta t)
  \biggr\}
  \, e^{-i(h_m+h_\gamma) \Delta t} .
\label {eq:ZrateHO1}
\end {multline}
As in the massless case ($h_m{=}h_\gamma{=}0$) reviewed in ref.\ \cite{softqed1},
this integral has a UV divergence from $\Delta t \to 0$, but one can
sidestep dealing with
that divergence by subtracting the (on-shell) bremsstrahlung rate in
vacuum ($\qhat \to 0$ and so $\Omega_0 \to 0$),
which must vanish by energy-momentum conservation.  This yields
\begin {multline}
  \left[ \frac{d\Gamma}{dx_\gamma} \right]_\LPM
  =
  - \frac{\alpha}{\pi} \Re \int_0^\infty d(\Delta t) \>
  \biggl\{
    P_{e\to\gamma}(x_\gamma) \,
      \left[ \Omega_0^2 \csc^2(\Omega_0\,\Delta t)
                - \frac{1}{(\Delta t)^2} \right]
\\
    + i x_\gamma h_m 
      \left[ \Omega_0 \csc(\Omega_0\,\Delta t)
                - \frac{1}{\Delta t} \right]
  \biggr\}
  \, e^{-i(h_m+h_\gamma) \Delta t} .
\label {eq:ZrateHO2}
\end {multline}
The integrals needed are (see appendix \ref{app:intLObrem})
\begin {subequations}
\label {eq:intLObrem}
\begin {align}
  \int_0^\infty dt \>
    \left[ \Omega^2 \csc^2(\Omega t) - \frac{1}{t^2} \right] e^{-i h t}
  &= -i\Omega  + i h\,\PSI(1;\tfrac{h}{2\Omega}) ,
\label {eq:bremint1}
\\
  \int_0^\infty dt \>
    \left[ \Omega \csc(\Omega t) - \frac{1}{t} \right] e^{-i h t}
  &= -\PSI(\tfrac12;\tfrac{h}{2\Omega}) ,
\label {eq:bremint2}
\end {align}
\end {subequations}
where we introduce the (unconventional) notation
\begin {equation}
  \PSI(r\,;z) \equiv \psi(r{+}z) - \ln z ,
\label {eq:Psidef}
\end {equation}
and $\psi(z) = \Gamma'(z)/\Gamma(z)$ is the digamma function.
The definition (\ref{eq:Psidef}) can be thought of as
$\psi(r{+}z)$ minus its large-$z$ behavior and gives
$\PSI(r;z) \to 0$ as $|z|\to\infty$.

Eq.\ (\ref{eq:ZrateHO2}) then evaluates to
\begin {subequations}
\label {eq:LPMrate}
\begin {equation}
  \left[ \frac{d\Gamma}{dx_\gamma} \right]_\LPM
  \simeq
  \frac{\alpha}{\pi}
  \Re\Bigl\{
    i P_{e\to\gamma}(x_\gamma)
    \Bigl[
      \Omega_0
      - ( h_m{+}h_\gamma )
        \,\PSI\bigl(
          1 ;
          \tfrac{h_m{+}h_\gamma}{2\Omega_0}
        \bigr)
    \Bigr]
    + i x_\gamma h_m
        \,\PSI\bigl(
          \tfrac12 ;
          \tfrac{h_m{+}h_\gamma}{2\Omega_0}
        \bigr)
  \Bigr\} .
\label {eq:LPMrate0}
\end {equation}
This formula can be slightly simplified in the limit $m_\gamma \ll m$
(applicable to ordinary matter) that we use in the rest of
this paper.  Were we interested only in $x_\gamma \sim 1$, then
$h_\gamma \ll h_m$ and we could remove
all the $h_\gamma$'s from (\ref{eq:LPMrate0}).  If, on the other hand,
we focus on $x_\gamma \ll 1$, then 
the $x_\gamma h_m \,\PSI\bigl( \tfrac12; \cdots)$ term above
becomes negligible compared to the other terms in (\ref{eq:LPMrate0}).
Combining those two observations, dropping $h_\gamma$ from just
the $x_\gamma h_m \,\PSI\bigl( \tfrac12; \cdots)$ term will make
a negligible change for any value of $x_\gamma$:
\begin {equation}
  \left[ \frac{d\Gamma}{dx_\gamma} \right]_\LPM
  \simeq
  \frac{\alpha}{\pi}
  \Re\Bigl\{
    i P_{e\to\gamma}(x_\gamma)
    \Bigl[
      \Omega_0
      - ( h_m{+}h_\gamma )
        \,\PSI\bigl(
          1 ;
          \tfrac{h_m{+}h_\gamma}{2\Omega_0}
        \bigr)
    \Bigr]
    + i x_\gamma h_m
        \,\PSI\bigl(
          \tfrac12 ;
          \tfrac{h_m}{2\Omega_0}
        \bigr)
  \Bigr\} .
\label {eq:LPMrate1}
\end {equation}
\end {subequations}
This is equivalent to the result given by Migdal%
\footnote{
  Since Migdal sometimes sets $m{=}1$, we find it easier to compare
  the above result to eqs.\ (72--76) of Klein's review
  \cite{SpencerReview}, including the discussion surrounding Klein eq.\ (84)
  in order to include the dielectric effect ($m_\gamma$),
  together with the digamma versions of
  eqs.\ (8A) in the appendix of Migdal \cite{Migdal}.
  Remember that both Migdal and
  Klein work in electromagnetic units where $e^2 = \alpha$.
  Use appendix A.1 of ref.\ \cite{softqed1} (including footnotes)
  to translate $\qhat$ into the type of expressions used by Migdal and
  Klein; use the digamma identity $\psi(z) = \psi(1+z) - z^{-1}$
  to rewrite Migdal's $\psi(s-is)$ in terms of $\psi(1+s-is)$;
  and then note that
  $\Re[i\PSI(r;z)] = \Im\psi\bigl((r{+}z)^*\bigr) + \arg z$
  and that Migdal's $s-i s = (\sqrt{2i}\,s)^*$.
}
with one caveat.
In Migdal's presentation, our $h_\gamma$ above is
replaced by $h_\gamma/(1{-}x_\gamma)$, which is another change to
the more general result (\ref{eq:LPMrate0}) that is negligible
because the dielectric effect is only important when $x_\gamma \ll 1$.


\subsubsection{Limits}

The limiting cases of the LPM rate (\ref{eq:LPMrate}) depend on which
of $|\Omega_0|$, $h_m$, and $h_\gamma$ is the largest.

Deep inside region 1 of fig.\ \ref{fig:LPMoverBH}a,
$h_m$ is the largest,
and this limit of (\ref{eq:LPMrate}) reproduces the approximation
(\ref{eq:BHrate}) to the Bethe-Heitler rate.
In taking the limit, it is worth noting that the large-argument expansion%
\footnote{
  for $|\arg w| < \pi$, which will always be the case in our application.
}
\begin {equation}
   \psi(w) = \ln w - \tfrac12 w^{-1} - \tfrac1{12} w^{-2} + O(w^{-3})
\end {equation}
of the digamma function grows logarithmically with $w$, but the subtraction in
our definition of $\PSI(r;z)$ cancels this log divergence, leaving
\begin {equation}
   \PSI(r;z) =
   (r-\tfrac12) z^{-1} - \bigl[\tfrac12 r(r{-}1) + \tfrac1{12}\bigr] z^{-2}
   + O(z^{-3})
\end {equation}
and so
\begin {equation}
   \Re[i\PSI(1;z)] \simeq \tfrac12 \Re(i z^{-1}) - \tfrac1{12} \Re(i z^{-2})
   \qquad \mbox{and} \qquad
   \Re[i\PSI(\tfrac12;z)] \simeq \tfrac1{24} \Re(i z^{-2}) .
\label {eq:PSIsmall}
\end {equation}

Deep inside region 2 of fig.\ \ref{fig:LPMoverBH}a, $\Omega_0$ is the largest,
and the limit of (\ref{eq:LPMrate}) is
simply what one gets by setting $m$ and $m_\gamma$ to zero
in (\ref{eq:LPMrate}):
\begin {equation}
  \left[ \frac{d\Gamma}{dx_\gamma} \right]_\LPM
  \simeq
  \frac{\alpha}{\pi} \, P_{e\to\gamma}(x_\gamma)
  \Re(i\Omega_0)
  =
  \frac{\alpha}{2\pi} \, P_{e\to\gamma}(x_\gamma)
  \sqrt{ \frac{x_\gamma \qhat}{(1{-}x_\gamma)E} }
  \qquad
  \mbox{(deep LPM)} .
\label {eq:deepLPM}
\end {equation}
We note for later reference
that (as reviewed in ref.\ \cite{softqed1}) in this case
($h_m$ and $h_\gamma$ ignorable)
$1/|\Omega_0|$ sets the time scale for $\Delta t$ in
the integrals (\ref{eq:ZrateHO2}), and so the LPM bremsstrahlung
formation time is
\begin{equation}
  \tform^\LPM \sim \frac{1}{|\Omega_0|}
  \qquad
  \mbox{(deep LPM)} .
\label {eq:tformLPM}
\end {equation}

Deep inside region 3, $h_\gamma$ is the largest, giving
\begin {equation}
  \left[ \frac{d\Gamma}{dx_\gamma} \right]_\LPM
  \simeq
  \frac{\alpha\qhat}{3\pi m_\gamma^2} \, x_\gamma^2 P_{e\to\gamma}(x_\gamma)
  \qquad
  \mbox{(dielectric region)} .
\label {eq:dielectric}
\end {equation}
In (\ref{eq:dielectric}), one may replace $P_{e\to\gamma}(x_\gamma)$ by
its soft-photon limit $2/x_\gamma$ because the dielectric effect is only
ever important in the soft-photon limit.

For later reference, the energy (inverse time) scales which control
ordinary LPM brem\-sstrah\-lung in the different regions 1--3
are summarized in the top section of table \ref{tab:scales}.
Parametrically, the boundaries between these regions are summarized in
the top section of table \ref{tab:boundaries}.
As an example, the boundary $2|3$ between regions 2 and 3 represents
the switchover between (i) $|\Omega_0|$ being large compared to
$h_m$ and $h_\gamma$ [the condition for the limiting formula
(\ref{eq:deepLPM})] and (ii) $h_\gamma$ being large compared to
$|\Omega_0|$ and $h_m$ [the condition for the limiting formula
(\ref{eq:dielectric})].

The location of the ``boundary'' between two regions is only
meaningful parametrically because our boundaries represent
smooth transitions between limiting behaviors, not
abrupt transitions.  Nonetheless, our \textit{convention} is
to draw the dashed
boundary lines on fig.\ \ref{fig:LPMoverBH} (and similar figures
later) where the
limiting rate formulas for the two regions
are exactly equal.  For example,
the $2|3$ boundary line in fig.\ \ref{fig:LPMoverBH}a is drawn
at the location where (\ref{eq:deepLPM}) equals (\ref{eq:dielectric}).


\begin {table}

\setlength{\tabcolsep}{2pt}
\begin {tabular}{|cl|ccccccccc|}
\toprule
  \multicolumn{2}{|c|}{region}
  & $|\Omega_0|$ & $h_m$ & $h_\gamma$
  & $|\Omegapr|$ & $\hpr_m$ & $\Gamma_\pair$
  & \multicolumn{2}{c}{$e$ mass for}
  &
\\
\cline{6-7}\cline{9-10} &&&&&&&&& \\[-3em]
  & & & &
  & \multicolumn{2}{c}{(democratic)}
  &
  & $e{\to}e\gamma$ & $\gamma{\to}e\bar e$
  & 
\\
\hline
  1. & BH
  & $\downarrow$ & $\frac{x_\gamma m^2}{2(1{-}x_\gamma)E}$ & $\downarrow$
  &&&
  & \underline{include} & &
\\
  2. & deep LPM ~~(general)
  & $\sqrt{\!\frac{x_\gamma\qhat}{2(1{-}x_\gamma)E}}$ & $\uparrow$ & $\downarrow$
  &&&
  & ignorable & &
\\
  & \phantom{deep LPM {}}($\kgamma{\ll}\Elpma$)
  & $\simeq \sqrt{ \!\frac{x_\gamma\qhat}{2E} }$ & $\uparrow$ & $\downarrow$
  &&&
  & ignorable & &
\\
  3. & dielectric
  & $\uparrow$ & $\uparrow$ & $\frac{m_\gamma^2}{2k_\gamma}$
  &&&
  & ignorable & &
\\[4pt]
\hline
  4$_{\rm a}$. & deep $\LPMplus$
  & $\uparrow$ & $\uparrow$ & $\updownarrow$
  & $\sim \! \sqrt{\!\frac{\qhat}{\kgamma}}$ & $\downarrow$
    & $\sim \! \alpha\sqrt{\!\frac{\qhat}{\kgamma}}$
  & ignorable & ignorable &
\\[6pt]
  4$_{\rm b}$. & deep $\LPMplus$
  & $\uparrow$ & $\uparrow$ & $\updownarrow$
  & $\uparrow$ & $\frac{m^2}{2\kgamma}$
    & $\sim \! \alpha \frac{\qhat}{m^2}$
  & ignorable & \underline{include} &
\\[6pt]
\hline
  5. & dielectric$\Plus$
  & $\uparrow$ & $\uparrow$ & $\frac{m_\gamma^2}{2\kgamma}$
  & $\uparrow$ & $\frac{m^2}{2\kgamma}$ & $\uparrow$
  & ignorable & \underline{include} &
\\[6pt]
\botrule
\end {tabular}
\caption{%
\label{tab:scales}%
  Behavior of different energy scales deep inside the various regions of figs.\
  \ref{fig:LPMoverBH}a, \ref{fig:overBH_nodie}, and \ref{fig:overBH}.
  A $\downarrow$ or $\uparrow$ means that the entry is the same as the line
  below or above.  Explicit formulas are only written in cases
  where the scale determines the bremsstrahlung formation time.
  In regions 1--3, the explicit formula listed in each
  row is the largest of $(|\Omega_0|,h_m,h_\gamma)$ in that region,
  and its inverse is $\sim \tform$.
  In regions 4$_{\rm a}$--5, two scales are written explicitly in each row, and
  their inverses represent the parametric extremes of a range of
  bremsstrahlung formation times that contribute equally to the
  energy loss rate $d\Gamma/dx_\gamma$, generating a logarithmic
  enhancement to the rate.  The ratio of those two scales
  is (parametrically) the scale of the argument of the logarithms in
  the limiting formulas for $d\Gamma/dx_\gamma$ given for each region
  in the main text.  Pair production scales $(\Omegapr,\hpr_m)$ are
  specialized to democratic splittings $\yfrak \sim 1{-}\yfrak \sim 1$,
  which dominate the total pair production rate $\Gamma_\pair$.
  The last two columns of the table indicate whether
  or not calculations may ignore, or must include, the mass of
  (i) the initial electron and/or (ii) the masses of the pair-produced
  $e^-e^+$.  We have only shown pair production scales in
  rows where pair production has an effect on bremsstrahlung.
}
\end {table}


\def\vbar{\,$|$\,}
\def\ph{\phantom{${}_{\rm b}$}}
\begin {table}

\setlength{\tabcolsep}{7pt}
\begin {tabular}{llll}
\toprule
\\[-14pt]
  \ph 1\vbar2 &
     $h_m \sim |\Omega_0|$ & \kern-0.8em
     $\begin{cases}
           x_\gamma \sim \frac{\qhat E}{m^4}& (\kgamma{\ll}\Elpma) \\
       1{-}x_\gamma \sim \frac{m^4}{\qhat E}& (\kgamma{\gg}\Elpma)
     \end{cases}$ & \kern-0.4em
     $\begin{array}{ll}
           \kgamma \sim \frac{E^2}{\Elpma}\\
       E{-}\kgamma \sim \Elpma           
     \end{array}$
\\
  \ph 1\vbar3 &
     $h_m \sim h_\gamma$ &
     $x_\gamma \sim \frac{m_\gamma}{m}$ &
     $\kgamma \sim \frac{m_\gamma}{m} E$
\\
  \ph 2\vbar3 &
     $|\Omega_0| \sim h_\gamma$ &
     $x_\gamma \sim \left(\frac{m_\gamma^4}{\qhat E}\right)^{1/3}$ &
     $\kgamma \sim \left(\frac{m_\gamma}{m}\right)^{4/3} \Elpma^{1/3} E^{2/3}$
\\[4pt]
\hline
\\[-14pt]
  \ph 2\vbar4$_{\rm a}$ &
     $|\Omega_0| \sim \Gamma_\pair
      \ln\bigl( \frac{|\Omega_\pr|}{\Gamma_\pair} \bigr)$ &
     $x_\gamma \sim \alpha \ln\bigl( \frac{1}{\alpha} \bigr)$ &
     $\kgamma \sim \alpha E \ln\bigl( \frac{1}{\alpha} \bigr)$
\\
  \ph 2\vbar4$_{\rm b}$ &
     $|\Omega_0| \sim \Gamma_\pair
      \ln\bigl( \frac{h_m^\pr}{\Gamma_\pair} \bigr)$ &
     $x_\gamma \sim \frac{\alpha^2 \qhat E}{m^4}
      \ln^2\bigl( \frac{m^4}{\alpha^{3/2} \qhat E} \bigr)$ &
     $\kgamma \sim \frac{(\alpha E)^2}{\Elpma}
      \ln^2\bigl( \frac{\Elpma}{\alpha^{3/2} E} \bigr)$
\\[4pt]
\hline
\\[-14pt]
  \ph 2\vbar5 &
     $|\Omega_0| \sim \Gamma_\pair \ln\bigl( \frac{m}{m_\gamma} \bigr)$ &
     $x_\gamma \sim \frac{\alpha^2\qhat E}{m^4}
      \ln^2\bigl( \frac{m}{m_\gamma} \bigr)$ &
     $\kgamma \sim \frac{(\alpha E)^2}{\Elpma}
      \ln^2\bigl( \frac{m}{m_\gamma} \bigr)$
\\
  \ph 3\vbar5 &
     $\frac{\qhat x_\gamma^2}{m_\gamma^2}
      \sim \Gamma_\pair\ln\bigl( \frac{m}{m_\gamma} \bigr)$ &
     $x_\gamma \sim \alpha^{1/2} \, \frac{m_\gamma}{m}
      \ln^{1/2}\bigl( \frac{m}{m_\gamma} \bigr)$ &
     $\kgamma \sim \alpha^{1/2} \, \frac{m_\gamma}{m} E
      \ln^{1/2}\bigl( \frac{m}{m_\gamma} \bigr)$
\\
  4$_{\rm b}$\vbar5 &
     $\Gamma_\pair \sim h_\gamma$ &
     $x_\gamma \sim \left(\frac{m_\gamma}{m}\right)^{\kern-0.5pt 2}
      \frac{m^4}{\alpha\qhat E}$ &
     $\kgamma \sim \left(\frac{m_\gamma}{m}\right)^{\kern-0.5pt 2}
      \frac{\Elpma}{\alpha}$
\\[6pt]
\botrule
\end {tabular}
\caption{%
\label{tab:boundaries}%
  Boundaries between the different regions of figs.\ \ref{fig:LPMoverBH}a,
  \ref{fig:overBH_nodie}, and \ref{fig:overBH}.
  Here, for convenience,
  we have expressed each boundary in three (parametrically equivalent) ways.
  Above, $\Gamma_\pair$ implicitly refers to the $k_\gamma \gg \Elpma$
  limit (\ref{eq:pairmassless}) for boundaries involving region $4_{\rm a}$
  and to the $k_\gamma \ll \Elpma$
  limit (\ref{eq:pairBH}) for boundaries involving regions $4_{\rm b}$ or $5$.
  The entry for the $1|2$ boundary is split into two lines, corresponding
  to the two cases $\kgamma{\ll}\Elpma$ and $\kgamma{\gg}\Elpma$.
  In the entry for the $2|4_{\rm b}$ case, $\hpr_m$ refers to the size
  $\hpr_m({\rm democratic}) \sim m^2/\kgamma$ of (\ref{eq:hprm})
  for democratic pair production
  (i.e.\ neither $\yfrak$ nor $1{-}\yfrak$ small).
}
\end {table}


\subsection{LPM pair production}
\label {sec:LPMpair}

\begin {figure}[t]
\begin {center}
  \includegraphics[scale=0.6]{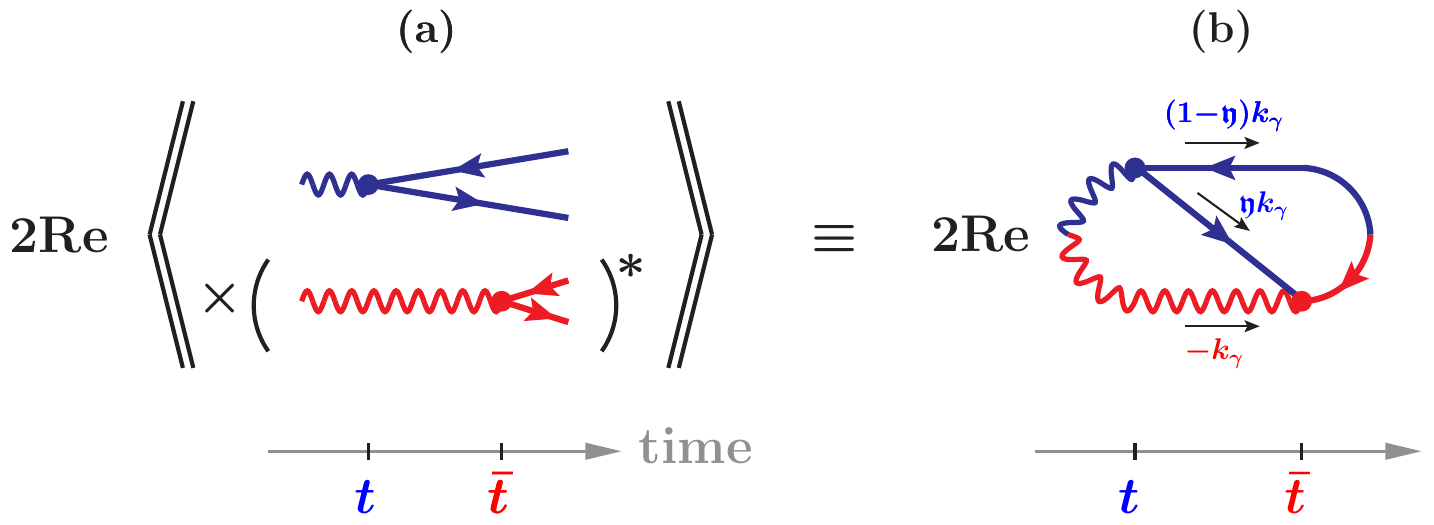}
  \caption{
    \label{fig:pair}
    Like fig.\ \ref{fig:lpm} but for pair production
    $\gamma \to e^- e^+$ from a photon with energy $k_\gamma$ to
    daughters with energy fractions $\yfrak$ and $1{-}\yfrak$ of $k_\gamma$.
  }
\end {center}
\end {figure}

We now similarly review
LPM pair production, corresponding to fig.\ \ref{fig:pair}.

\subsubsection{General Formula}
\label {sec:PairGeneral}

Ref.\ \cite{softqed1} also reviewed the LPM calculation of
pair production in the massless
case, which was a good approximation in the deep-LPM
regime corresponding to deep inside
region $2_\pr$ of our fig.\ \ref{fig:LPMoverBH}b.
The effective Hamiltonian analogous to (\ref{eq:Heff}) was
\begin {equation}
  {\cal H}_\pair =
  \frac{p_{\perp e^-}^2{+}m^2}{2(1{-}\yfrak) k_\gamma}
     + \frac{p_{\perp e^+}^2{+}m^2}{2\yfrak k_\gamma}
  -\frac{k_{\perp \gamma}^2 + m_\gamma^2}{2k_\gamma}
  + V(\b_{e^-} - \b_{e^+}) ,
\label {eq:Hpair}
\end {equation}
which in the multiple-scattering approximation (and again choosing
the $z$ axis to be in the direction of the photon) becomes
\begin {equation}
  {\cal H}_\pair =
     \biggl( \frac{p_\perp^2}{2\Mpr} + \tfrac12 \Mpr\Omegapr^2 b^2 \biggr)
     + \biggl( \frac{m^2}{2\Mpr} - \frac{m_\gamma^2}{2\kgamma} \biggr)
\label {eq:HpairHO}
\end {equation}
with
\begin {equation}
  \Mpr \equiv \yfrak(1{-}\yfrak)k_\gamma \,,
  \qquad
  \Omegapr \equiv \sqrt{ - \frac{i\qhat}{2\Mpr} }
  = \sqrt{ - \frac{i \qhat}{2\yfrak(1{-}\yfrak)\kgamma} } \,.
\label {eq:MOmegapr}
\end {equation}
For the massless case, the analog of (\ref{eq:Zrate0}) is%
\footnote{
  \label {foot:Nf=1}%
  We implicitly assume that the only relevant fermions that can
  be pair produced are electrons.
  The effects of muon pair production are suppressed
  unless one considers immensely larger photon energies $\kgamma$ of order
  $\Elpm^{(\mu)} = m_\mu^4/\qhat \sim (m_\mu/\me)^4 \Elpma$.
}
\begin {equation}
  \left[ \frac{d\Gamma}{d\yfrak} \right]^\LPM_\pair
  =
  \frac{ \alpha\,P_{\gamma\to e}(\yfrak) }{ \Mpr^2 }
  \Re \int_0^\infty d(\Delta t) \>
  \grad_{\b'} \cdot \grad_{\b} \,
  G_\pr(\b',\Delta t;\b,0) \Bigl|_{\b'=\b=0} ,
\label {eq:Zpair0}
\end {equation}
with DGLAP splitting function
\begin {equation}
  P_{\gamma\to e}(\yfrak) = \yfrak^2 + (1{-}\yfrak)^2 .
\end {equation}
For the massive case, we must again include the vertex contributions
that do not conserve fermion helicity, for which the analog of
(\ref{eq:PwithFlip}) is to replace
\begin {equation}
  \alpha\,P_{\gamma\to e}(\yfrak) \grad_{\b'} \cdot \grad_{\b} \longrightarrow
  \alpha \left[
    P_{\gamma\to e}(\yfrak) \grad_{\b'} \cdot \grad_{\b}
    + m^2
  \right]
\label{eq:pairPwithFlip}
\end {equation}
in (\ref{eq:Zpair0}).
See appendix \ref{app:PairWithFlip} for more detail.

Because $m_\gamma \ll m$ and $\Mpr < \kgamma$,
the $m_\gamma^2$ term in (\ref{eq:HpairHO}) will \textit{always}
be negligible compared to the $m^2$ term, and so we may ignore the
dielectric effect altogether in what follows.
The propagator associated with (\ref{eq:HpairHO}) is then
\begin {multline}
  G_\pr(\b',\Delta t;\b,0) =
  \frac{\Mpr\Omegapr \csc(\Omegapr\,\Delta t)}{2\pi i}
\\ \times
  \exp\Bigl( i\Mpr\Omegapr \bigl[
    \tfrac12(b^2 + b'^2) \cot(\Omegapr\,\Delta t)
    - \b\cdot\b' \csc(\Omegapr\,\Delta t)
  \bigr] \Bigr)
  \times
  e^{-i\hpr_m \Delta t} ,
\label {eq:GpairProp}
\end {multline}
analogous to (\ref{eq:Gprop}).  Above,
\begin {equation}
  \hpr_m \equiv \frac{m^2}{2\Mpr} \,.
\label {eq:hprm}
\end {equation}
The analog to the bremsstrahlung rate (\ref{eq:ZrateHO1}) is then
\begin {equation}
  \left[ \frac{d\Gamma}{d\yfrak} \right]_\pair^\LPM
  =
  - \frac{\alpha}{\pi} \Re \int_0^\infty d(\Delta t) \>
  \biggl\{
    P_{\gamma\to e}(\yfrak) \, \Omegapr^2 \csc^2(\Omegapr\,\Delta t)
    + i \hpr_m 
      \Omegapr \csc(\Omegapr\,\Delta t)
  \biggr\}
  \, e^{-i\hpr_m \Delta t} .
\label {eq:ZpairHO1}
\end {equation}
Regulating the integral by subtracting the (vanishing) vacuum contribution gives
\begin {multline}
  \left[ \frac{d\Gamma}{d\yfrak} \right]_\pair^\LPM
  =
  - \frac{\alpha}{\pi} \Re \int_0^\infty d(\Delta t) \>
  \biggl\{
    P_{\gamma\to e}(\yfrak) \,
      \left[ \Omegapr^2 \csc^2(\Omegapr\,\Delta t)
                - \frac{1}{(\Delta t)^2} \right]
\\
    + i \hpr_m 
      \left[ \Omegapr \csc(\Omegapr\,\Delta t)
                - \frac{1}{\Delta t} \right]
  \biggr\}
  \, e^{-i\hpr_m \Delta t} ,
\label {eq:ZpairHO2}
\end {multline}
and thence
\begin {equation}
  \left[ \frac{d\Gamma}{d\yfrak} \right]_\pair^\LPM
  \simeq
  \frac{\alpha}{\pi}
  \Re\Bigl\{
    i P_{\gamma\to e}(\yfrak)
    \Bigl[
      \Omegapr
      - \hpr_m
        \,\PSI\bigl(
          1 \,;
          \tfrac{\hpr_m}{2\Omegapr}
        \bigr)
    \Bigr]
    + i \hpr_m
        \,\PSI\bigl(
          \tfrac12 \,;
          \tfrac{\hpr_m}{2\Omegapr}
        \bigr)
  \Bigr\} .
\label {eq:LPMpair}
\end {equation}
This is equivalent to Migdal's result for pair production.


\subsubsection{Limits}
\label {sec:LPMpairLimits}

For pair production, the Bethe-Heitler limit
(region $1_\pr$ of fig.\ \ref{fig:LPMoverBH}b) corresponds to
the limit $\hpr_m \gg |\Omegapr|$ of (\ref{eq:LPMpair}).
Using (\ref{eq:PSIsmall}), that is
\begin {equation}
  \left[ \frac{d\Gamma}{d\yfrak} \right]_\pair^\LPM \simeq 
  \frac{\alpha \qhat}{6\pi m^2} \,
  \bigl[ 2 P_{\gamma\to e}(\yfrak) + 1 \bigr]
  \qquad \mbox{($\gamma{\to}e\bar e$ BH)} .
\label{eq:BHpair}
\end {equation}
Integrating over $\yfrak$ the total rate in this limit is
\begin {equation}
  \Gamma_\pair^\LPM \simeq 
  \frac{7\alpha\qhat}{18\pi m^2}
  \qquad \mbox{($\kgamma\ll\Elpma$)} .
\label {eq:pairBH}
\end {equation}
We've characterized the parametric condition for this formula as
$\kgamma\ll\Elpma$ because the integral over $\yfrak$ is dominated by
``democratic'' splittings where neither $\yfrak$ nor $1{-}\yfrak$ is
parametrically small. In that case, using (\ref{eq:MOmegapr})
and (\ref{eq:hprm}),
the limit $\hpr_m \gg |\Omegapr|$ that gave (\ref{eq:BHpair})
is equivalent to $\kgamma \ll m^4/\qhat \equiv \Elpm \sim \Elpma$.

The deep-LPM limit of pair production
(region $2_\pr$ of fig.\ \ref{fig:LPMoverBH}b) corresponds to
the opposite limit $\hpr_m \ll |\Omegapr|$ of (\ref{eq:LPMpair}),
\begin {equation}
  \left[ \frac{d\Gamma}{d\yfrak} \right]_\pair^\LPM \simeq 
  \frac{\alpha}{\pi} \, P_{\gamma\to e}(\yfrak) \, \Re(i\Omega_\pr)
  =
  \frac{\alpha}{2\pi} \, P_{\gamma\to e}(\yfrak)
     \sqrt{ \frac{\qhat}{\yfrak(1{-}\yfrak)\kgamma} }
  \qquad \mbox{($\gamma{\to}e\bar e$ deep LPM)} .
\end {equation}
Integrating over $\yfrak$,
\begin {equation}
  \Gamma_\pair^\LPM \simeq 
  \frac{3\alpha}{8} \sqrt{ \frac{\qhat}{\kgamma} }
  \qquad \mbox{($\kgamma\gg\Elpma$)} .
\label {eq:pairmassless}
\end {equation}


\subsection{Summary of notation}

For easy reference, a summary of selected notation introduced so far
is provided in table \ref{tab:notation}.

\begin {table}
\setlength{\tabcolsep}{2pt}
\begin {tabular}{ll}
\toprule
  $m$
  & electron mass
\\
  $m_\gamma$
  & medium-induced photon mass
\\
  $\kgamma$
  & bremsstrahlung photon energy
\\
  $x_\gamma \equiv \kgamma/E$
  & bremsstrahlung photon energy fraction
\\
  $\eta$
  & energy of pair-produced electron divided by $k_\gamma$
\\
  $\qhat$
  & $\p_\perp$ diffusion constant normalized so
    $\langle (\Delta p_\perp)^2 \rangle_{\rm typical} = \qhat\,\Delta t$
\\
  $\Elpm \equiv m^4/\qhat$
  & energy scale where the LPM effect becomes important even for
    $\kgamma \sim E$
\\
  ${\cal H}$
  & non-unitary ``Hamiltonian'' for evolution of ``$e^-\gamma e^+$''
    for bremsstrahlung
\\
  ${\cal H}_\pair$
  & similar for pair production
\\
  $V(\b)$
  & imaginary ``potential'' in ${\cal H}$
    from averaged interactions with medium
\\
  $\Mo$, $\Omega_0$
  & ``mass'' and complex frequency (\ref{eq:MOmega0}) in ``harm.\ osc.''\
    interpretation of ${\cal H}$ 
\\
  $\Mpr$, $\Omegapr$
  & similar for pair production (\ref{eq:MOmegapr})
\\
  $h_m \equiv m^2/2\Mo$
  & constant term in $\cal H$ accounting for electron mass
\\
  $\hpr_m \equiv m^2/2\Mpr$
  & similar for pair production
\\
  $h_\gamma \equiv m_\gamma^2/2\kgamma$
  & constant term in $\cal H$ accounting for medium-induced photon mass
\\
  $\Gamma_\pair$
  & medium-induced pair production rate
\\
  ${\cal G}_\brem$, ${\cal G}_\pair$
  & related to LPM bremsstrahlung or pair production rates by
  $2\Re{\cal G} = \Gamma$
\\
  $\bar\psi(r;z)$
  & subtracted digamma function (\ref{eq:Psidef})
\\[4pt]
\botrule
\end {tabular}
\caption{%
\label{tab:notation}%
  Selected notation.  (${\cal G}$ will be introduced later.)
}
\end {table}


\section{LPM+ ignoring the dielectric effect \boldmath$(m_\gamma = 0)$}
\label {sec:massless}

\begin{figure}
  \includegraphics[scale=1.0]{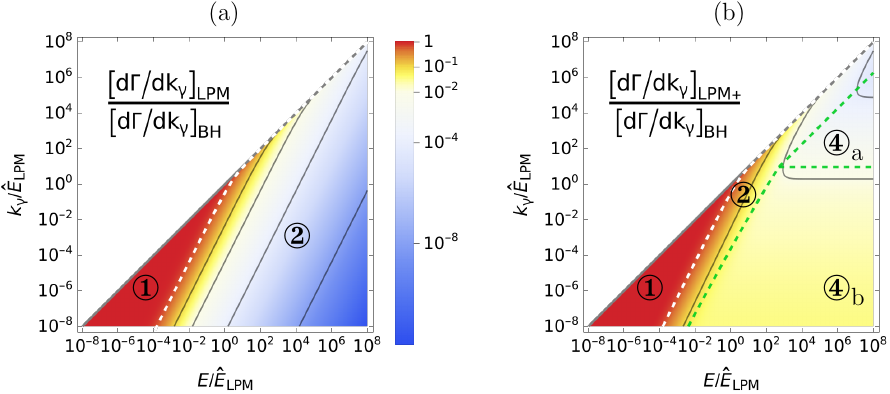}
  \caption{
     \label{fig:overBH_nodie}
     (a) A version of the LPM/BH plot from fig.\ \ref{fig:LPMoverBH} for the
     case $m_\gamma = 0$ (i.e.\ neglecting the dielectric effect).
     (b) Log-log-log contour plot of the ratio $\protect\LPMplus$/BH
     of the
     $\protect\LPMplus$ bremsstrahlung rate (\ref{eq:LPM+rate_nodie})
     to the Bethe-Heitler rate (\ref{eq:BHrate}) vs.\ $E/\Elpm$ and
     $\kgamma/\Elpm$ for $m_\gamma = 0$.  Note the non-uniform
     spacing $(10^{-1},10^{-2},10^{-4},10^{-8})$ of the contour lines.
     The labeling of the different regions is
     \cnum{1} Bethe-Heitler (BH),
     \cnum{2} deep LPM, and
     \cnum{4} deep $\protect\LPMplus$.
     For later reference, region 4 is further divided into
     sub-regions \cnum{4}$_{\rm a}$ for $k_\gamma \gg \Elpma$
     [the first case of (\ref{eq:deepLPM+})] and
     \cnum{4}$_{\rm b}$ for $k_\gamma \ll \Elpma$
     [the second case of (\ref{eq:deepLPM+})].
  }
\end{figure}

We now turn to calculating the $\LPMplus$ correction to the ordinary
LPM effect.  In this section, we ignore the dielectric effect
(medium-induced photon mass) by taking $m_\gamma{=}0$.
We will return to
the dielectric effect in section \ref{sec:dielectric}.
As a starting point, the ordinary LPM/BH ratio of fig.\ \ref{fig:LPMoverBH}a
(which does not account for overlapping pair production) simplifies
to fig.\ \ref{fig:overBH_nodie}a for $m_\gamma{=}0$.

\subsection{When LPM+ corrections will be important}

Before we dive into the calculation, it is useful to first understand
parametrically when the $\LPMplus$ correction will be important.
Pair production will have a negligible effect
on the LPM bremsstrahlung rate when the LPM bremsstrahlung formation
time is small compared to the typical delay $1/\Gamma_\pair$
before the bremsstrahlung photon pair produces.
In the deep LPM regime (region 2 of fig.\ \ref{fig:overBH_nodie}a),
that means that the $\LPMplus$ correction will only be significant
when
\begin {equation}
   \tform^\LPM \gtrsim \frac{1}{\Gamma_\pair} \,,
\end {equation}
where $\tform^\LPM \sim 1/|\Omega_0|$.
From (\ref{eq:MOmega0}) for $\Omega_0$ and the limits
(\ref{eq:pairmassless}) and (\ref{eq:pairBH})
for $\Gamma_\pair$, this is parametrically
\begin {equation}
  \sqrt{ \frac{(1{-}x_\gamma)E}{x_\gamma\qhat} }
  \gtrsim
  \frac{1}{\alpha}
  \begin{cases}
     \sqrt{\kgamma/\qhat} \,, & \mbox{for $\kgamma \gg \Elpma$}; \\
     m^2/\qhat              , & \mbox{for $\kgamma \ll \Elpma$}.
  \end{cases}
\label {eq:deepLPM+}
\end {equation}
Remembering that $k_\gamma = x_\gamma E$ and that
$\Elpma \sim \Elpm \equiv m^4/\qhat$, this condition can only be satisfied for
$x_\gamma \ll 1$ and can then be rewritten as
\begin {equation}
  x_\gamma \lesssim
  \begin{cases}
     \alpha             , & \mbox{for $\kgamma \gg \Elpma$}; \\
     \alpha^2 E/\Elpma  , & \mbox{for $\kgamma \ll \Elpma$}
  \end{cases}
\label {eq:24boundary}
\end {equation}
(up to logarithms that will be discussed later).
Replacing $\lesssim$ above by $\sim$ then gives the boundary where
the ordinary LPM effect transitions into the $\LPMplus$ effect that we
will calculate.  Looking ahead to the final results
for this section, that boundary is indicated by the green dashed line in
fig.\ \ref{fig:overBH_nodie}b between region 2 and region $4_{\rm a}{+}4_{\rm b}$.
The most important qualitative point for our calculation is that
$x_\gamma \ll 1$ whenever the $\LPMplus$ effect is significant,
and so, as in our previous work
\cite{softqed1} that ignored masses (corresponding to $k_\gamma \gg \Elpma$
here), we may use soft photon approximations to calculate the
$\LPMplus$ effect more generally.

The condition (\ref{eq:24boundary}) was originally discussed by Galitsky and
Gurevich \cite{Galitsky} in 1964 for the case of $\kgamma \ll \Elpma$.
However, for the reasons discussed in ref.\ \cite{softqed1}, we will
find qualitatively different results for how pair production affects
the bremsstrahlung rate in the deep $\LPMplus$ region.


\subsection{Diagrams relevant for LPM+ corrections to the energy loss rate}
\label {sec:diags}

\begin {figure}[t]
\begin {center}
  \includegraphics[scale=0.4]{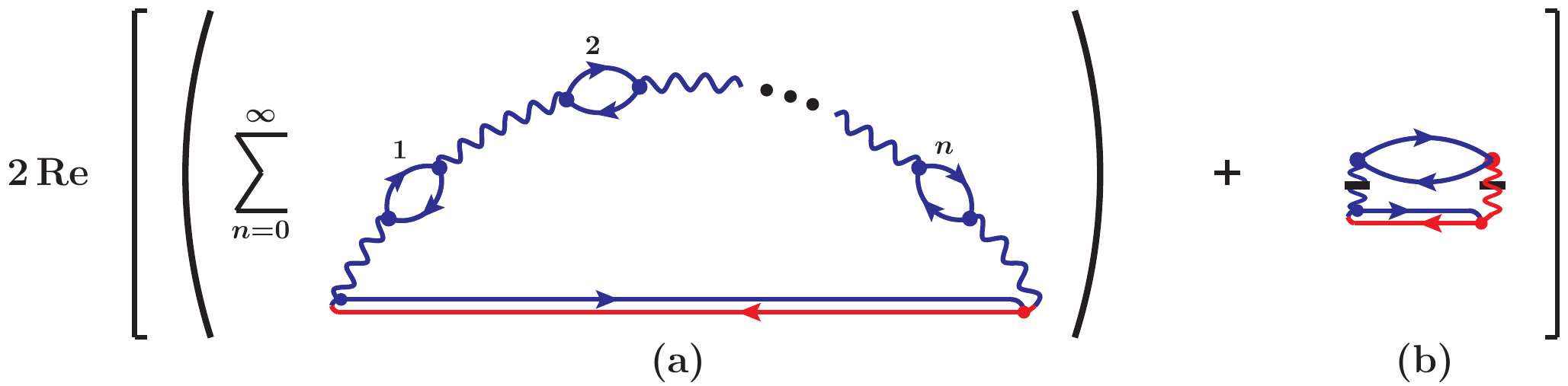}
  \caption{
    \label{fig:LPM+diags}
    Lightcone-time ordered rate diagrams contributing to
    the energy loss rate $d\Gamma/dx_\gamma$.
    Photon lines
    crossed by black bars represent longitudinally polarized photons,
    which in lightcone gauge act instantaneously in lightcone time.
    Uncrossed photon lines represent transversely polarized photons.
    See ref.\ \cite{softqed1} for details.
    Following that reference, we
    have drawn the bottom two electron lines close together
    to visually indicate that the transverse spatial separation
    of those two particles is relatively
    small in the soft-photon limit where
    the $\protect\LPMplus$ effect is significant.
  }
\end {center}
\end {figure}

As in ref.\ \cite{softqed1}, this paper focuses on
calculating the differential rate $d\Gamma/dx_\gamma$ for an initial
high-energy electron of
energy $E$ to lose energy $x_\gamma E$.  The usual LPM result would
be that this energy loss rate
is just the LPM bremsstrahlung rate reviewed in section
\ref{sec:LPMbrem}.
$\LPMplus$ corrections to the energy loss rate include the sum of
corrections from the initial bremsstrahlung $e{\to}e\gamma$
overlapping (i) subsequent real pair production
$\gamma{\to}e\bar e$ of the bremsstrahlung photon or (ii)
virtual pair production $\gamma{\to}e\bar e{\to}\gamma$.
Ref.\ \cite{softqed1} argued that many time-ordered diagrams cancel in
this sum, and that the relevant diagrams are those of
fig.\ \ref{fig:LPM+diags}.
The $n{=}0$ term of the
sum is equivalent to fig.\ \ref{fig:lpm}b, representing
the ordinary LPM bremsstrahlung result.

We should mention that our previous work \cite{softqed1}
at first justified fig.\ \ref{fig:LPM+diags} by imagining a large-$\Nf$ limit
of QED with $\Nf$ flavors of high-energy electrons.
Later in that paper,%
\footnote{
  section 6 of ref.\ \cite{softqed1}.
}
we outlined arguments that, because of soft-photon enhancements, the same
diagrams that dominate in the large-$\Nf$ limit are
\textit{also} the diagrams that dominate in the soft-photon limit
(the only case where the $\LPMplus$ effect is important)
for any value of $\Nf$, including $\Nf{=}1$.
We believe
this to be true, but we did not
claim to give a fully systematic and rigorous diagrammatic analysis.
For the purpose of the present paper, we assume that
this conclusion is correct, and we present our results for the
case $\Nf{=}1$.%
\footnote{
  See footnote \ref{foot:Nf=1}.
}

Following ref.\ \cite{softqed2}, it will
be convenient to assemble the result of
fig.\ \ref{fig:LPM+diags} by separately evaluating (i) the
one-bubble ($n{=}1$) term of the sum, (ii) the rest $(n{\ge}2)$ of the sum
over $n{>}0$, and (iii) the very last diagram of the figure.
We will correspondingly write the $\LPMplus$ energy loss rate
as
\begin {subequations}
\label {eq:LPM+rate_nodie}
\begin {equation}
  \left[ \frac{d\Gamma}{dx_\gamma} \right]_\LPMplus
  =
  \left[ \frac{d\Gamma}{dx_\gamma} \right]_\LPM
  + \delta \! \left[\frac{d\Gamma}{dx_\gamma}\right]
\end {equation}
with the correction to the ordinary
LPM rate (\ref{eq:LPMrate}) split up as
\begin {equation}
  \delta \! \left[\frac{d\Gamma}{dx_\gamma}\right] =
  \left[\frac{d\Gamma}{dx_\gamma}\right]_\perp
  +
  \left[\frac{d\Gamma}{dx_\gamma}\right]_{\rm L}
  =
  \left[\frac{d\Gamma}{dx_\gamma}\right]_{(n=1)}
  +
  \left[\frac{d\Gamma}{dx_\gamma}\right]_{(n\ge2)}
  +
  \left[\frac{d\Gamma}{dx_\gamma}\right]_{\rm L} .
\label {eq:deltaLPM+}
\end {equation}
\end {subequations}
The subscripts $\perp$ and L respectively represent the contributions
from processes involving intermediate transversely (fig.\ \ref{fig:LPM+diags}a)
or longitudinally (fig.\ \ref{fig:LPM+diags}b) polarized photons.


\subsection{The \boldmath$n{=}1$ bubble diagram}
\label{sec:neq1}

\begin {figure}[t]
\begin {center}
  \includegraphics[scale=0.4]{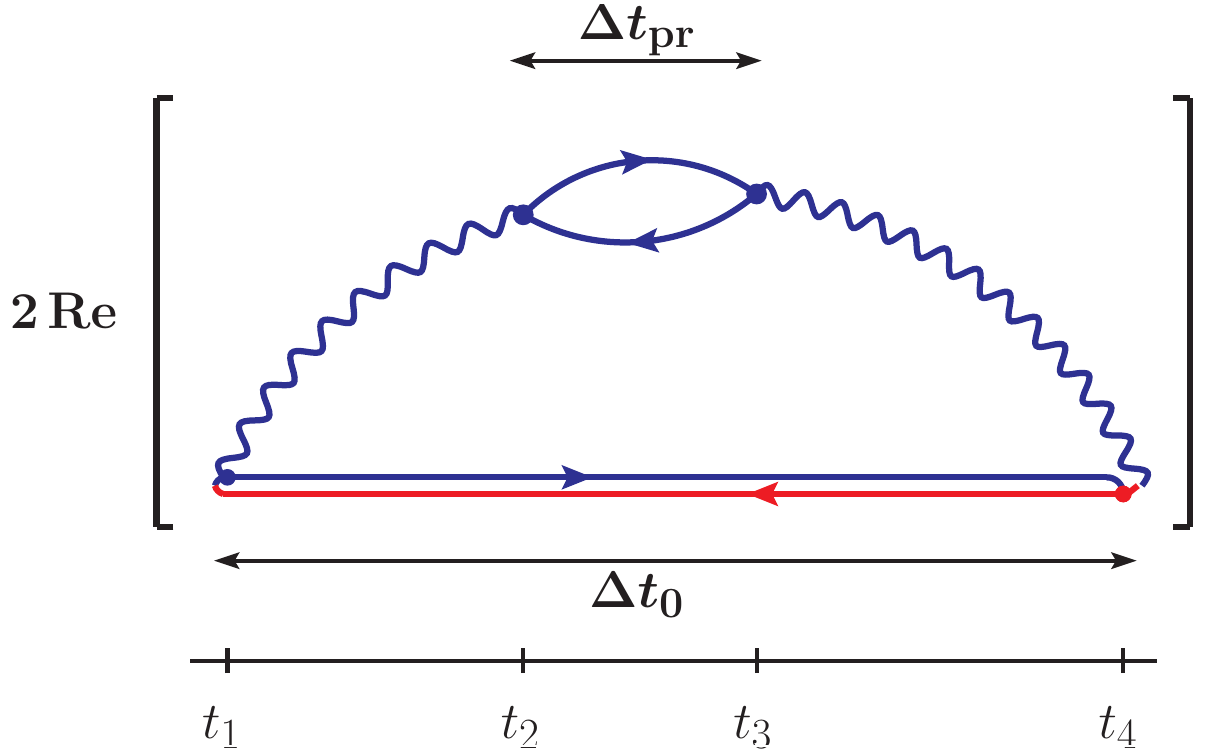}
  \caption{
    \label{fig:neq1}
    Single-bubble ($n{=}1$) contribution to fig.\ \ref{fig:LPM+diags}a.
  }
\end {center}
\end {figure}

Consider first the one-bubble diagram shown in fig.\ \ref{fig:neq1},
where we label the duration of the bremsstrahlung
process $\Delta t_0$ and the duration of the virtual pair production
bubble $\Delta t_\pr$.  Both times must be integrated over.

\subsubsection{Review of the massless case $m=0$}
\label{sec:neq1massless}

Consider fig.\ \ref{fig:neq1} \textit{before} integrating over the
times associated with the splitting vertices and before taking
$2\Re(\cdots)$.
Ref.\ \cite{softqed1} showed that, in the soft photon ($x_\gamma{\ll}1$)
limit, the calculation of fig.\ \ref{fig:neq1} then factorizes into
(i) a standard LPM calculation (sec.\ \ref{sec:LPMbrem}) of the bremsstrahlung
rate, multiplied by
(ii) a standard LPM calculation (sec.\ \ref{sec:LPMpair}) of
the pair production rate.%
\footnote{
  See sections 4.2--4.3 of ref.\ \cite{softqed1}.
}
Specifically,
\begin {equation}
  \left[ \frac{d\Gamma}{dx_\gamma} \right]_{(n=1)}
  = \int_0^1 d\yfrak \>
    \left[ \frac{d\Gamma}{dx_\gamma \, d\yfrak} \right]_{(n=1)}
\label {eq:Atotal}
\end {equation}
where
\begin {equation}
  \left[ \frac{d\Gamma}{dx_\gamma \, d\yfrak} \right]_{(n=1)}
  \simeq
  - 2\Re
  \int d({\rm times}) \>
  \left[ \frac{d{\cal G}}{dx_\gamma\,d(\Delta t_0)} \right]_\brem
  \left[ \frac{d{\cal G}}{d\yfrak\,d(\Delta t_\pr)} \right]_\pair
\label {eq:Afactorize}
\end {equation}
and where the $d{\cal G}/{dx\,d(\Delta t)}$ represent the ordinary
LPM bremsstrahlung or pair rate formula
(\ref{eq:ZrateHO1}) or (\ref{eq:ZpairHO1})
before integrating over the time separation $\Delta t$ or
taking $2\Re(\cdots)$.  For the massless approximation made in
ref.\ \cite{softqed1}, those were simply
\begin {subequations}
\begin {align}
   \left[ \frac{d{\cal G}}{dx_\gamma\,d(\Delta t_0)} \right]_\brem
   &=
   - \frac{\alpha}{2\pi} \,P_{e\to\gamma}(x_\gamma)
   \, \Omega_0^2 \csc^2(\Omega_0\,\Delta t_0) ,
\label {eq:dGbrem}
\\
   \left[ \frac{d{\cal G}}{d\yfrak\,d(\Delta t_\pr)} \right]_\pair ~
   &=
   - \frac{\alpha}{2\pi} \,P_{\gamma\to e}(\yfrak)
   \, \Omegapr^2 \csc^2(\Omegapr\,\Delta t_\pr) .
\label {eq:dGpair}
\end {align}
\end {subequations}
The overall minus sign in (\ref{eq:Afactorize}) arose because the
pair production in fig.\ \ref{fig:neq1} is virtual instead of real.
The time integrals in (\ref{eq:Afactorize}) are over the relative
time separations of the vertices in fig.\ \ref{fig:neq1} that maintain
the time ordering $t_1 < t_2 < t_3 < t_4$ of that time-ordered
diagram.  Since the integrand only depends on the separations
$\Delta t_0$ and $\Delta t_\pr$, those time integrations could be
rewritten as
\begin {equation}
  \int d({\rm times}) \> \cdots
  ~=~
  \int_0^\infty d(\Delta t_\pr) \, \int_{\Delta t_\pr}^\infty d(\Delta t_0  ) \>
  (\Delta t_0   {-} \Delta t_\pr) \> \cdots .
\label {eq:inttimes2}
\end {equation}
The integral (\ref{eq:Afactorize}) is then
UV divergent from $\Delta t_\pr \to 0$, but
ref.\ \cite{softqed1} identified this divergence as coming from the
vacuum contribution to the (in-medium) photon self-energy loop in
fig.\ \ref{fig:neq1} and argued that that particular contribution to
fig.\ \ref{fig:neq1} is parametrically suppressed in the soft photon
limit $x_\gamma \ll 1$ (the only case where
the $\LPMplus$ effect, and so this diagram, is significant).
Leaving out the unimportant vacuum contribution to the
self-energy corresponded to
replacing (\ref{eq:dGpair}) by%
\footnote{
  See section 4.4.2 of ref.\ \cite{softqed1}.
}
\begin {equation}
   \left[ \frac{d{\cal G}}{d\yfrak\,d(\Delta t_\pr)} \right]_\pair
   \longrightarrow ~
   - \frac{\alpha}{2\pi} \,P_{\gamma\to e}(\yfrak)
   \, \left[
        \Omegapr^2 \csc^2(\Omegapr\,\Delta t_\pr) - \frac{1}{(\Delta t_\pr)^2}
      \right]
\label {eq:dGpairSub}
\end {equation}
since $\Omegapr \to 0$ in vacuum ($\qhat\to 0$).
Combining (\ref{eq:Afactorize}), (\ref{eq:dGbrem}), and (\ref{eq:dGpairSub}),
we were able to do the time integrals (\ref{eq:inttimes2}) analytically to
find
\begin {equation}
  \left[ \frac{d\Gamma}{dx_\gamma \, d\yfrak} \right]_{(n=1)}
  \simeq
  \frac{\alpha^2}{2\pi^2} \, P_{e\to\gamma}(x_\gamma) \,P_{\gamma\to e}(\yfrak) \,
  \Re(i\Omega_\pr)
    \left[
      \ln\bigl(\tfrac{|\Omegapr|}{2\pi|\Omega_0|}\bigr) + \gammaE
    \right] ,
\label {eq:old1}
\end {equation}
where $\gammaE = 0.57721\cdots$ is the Euler-Mascheroni constant.
Finally, we were able to do the $\yfrak$ integral
(\ref{eq:Atotal}) analytically as well.


\subsubsection{Including the electron mass}
\label{sec:neq1mass}

The arguments made in ref.\ \cite{softqed1}
for the factorization (\ref{eq:Afactorize})
did not rely upon ignoring the electron mass, and so
all we need do now is no longer set $m{=}0$ when extracting
the formulas (\ref{eq:dGbrem}) and (\ref{eq:dGpair}) from 
(\ref{eq:ZrateHO1}) and (\ref{eq:ZpairHO1}).
Remembering that we are still treating $m_\gamma{=}0$ for now,
then
\begin {subequations}
\begin {align}
   \left[ \frac{d{\cal G}}{dx_\gamma\,d(\Delta t_0)} \right]_\brem
   &=
   - \frac{\alpha}{2\pi}
   \biggl\{
     P_{e\to\gamma}(x_\gamma) \,
       \Omega_0^2 \csc^2(\Omega_0\,\Delta t_0)
     + i x_\gamma h_m 
       \Omega_0 \csc(\Omega_0\,\Delta t_0)
   \biggr\}
   \, e^{-i h_m \Delta t_0} ,
\label {eq:dGbremMass}
\\
   \left[ \frac{d{\cal G}}{d\yfrak\,d(\Delta t_\pr)} \right]_\pair ~
   &=
   - \frac{\alpha}{2\pi}
   \biggl\{
     P_{\gamma\to e}(\yfrak) \,
       \Omegapr^2 \csc^2(\Omegapr\,\Delta t_\pr)
     + i \hpr_m 
       \Omegapr \csc(\Omegapr\,\Delta t_\pr)
   \biggr\}
   \, e^{-i\hpr_m \Delta t_\pr} .
\label {eq:dGpairMass}
\end {align}
\end {subequations}

Fortunately, there is a very useful simplification that can be made.
The region (\ref{eq:24boundary}) where the $\LPMplus$ effect is
significant (region 4 of fig.\ \ref{fig:overBH_nodie}b and its boundary)
is deeply inside the original LPM region
(region 2 of fig.\ \ref{fig:overBH_nodie}a) where
$|\Omega_0| \gg h_m$, as discussed earlier for the deep-LPM rate
(\ref{eq:deepLPM}).  That deep-LPM rate was insensitive to the electron mass.
Because of the factorization (\ref{eq:Afactorize}), we may therefore
similarly ignore the electron mass also in the factor
(\ref{eq:dGbremMass}) extracted from the original
LPM bremsstrahlung rate, replacing (\ref{eq:dGbremMass}) by
the massless case (\ref{eq:dGbrem}):%
\footnote{
  In more detail, remembering that $\Omega_0$ (\ref{eq:MOmega0}) has
  an imaginary part, the exponential fall-off of
  $\csc^2(\Omega_0\,\Delta t_0)$ and $\csc(\Omega_0\,\Delta t_0)$
  in (\ref{eq:dGbremMass}) for $\Delta t_0 \gg 1/|\Omega_0|$ means
  that, when integrating over $\Delta t_0$, we can approximate
   $e^{-i h_m \Delta t_0} \simeq 1$ in 
  (\ref{eq:dGbremMass}) since $h_m \ll |\Omega_0|$.
  Similarly, the $i x_\gamma h_m \Omega_0$ term
  in (\ref{eq:dGbremMass}) is ignorable compared to the $\Omega_0^2$ term.
}
\begin {subequations}
\label {eq:dMass}
\begin {equation}
   \left[ \frac{d{\cal G}}{dx_\gamma\,d(\Delta t_0)} \right]_\brem
   \simeq
   - \frac{\alpha}{2\pi} \,
     P_{e\to\gamma}(x_\gamma) \,
       \Omega_0^2 \csc^2(\Omega_0\,\Delta t_0) .
\label {eq:dGbremMass2}
\end {equation}
For the pair production factor (\ref{eq:dGpairMass}), however, no such
simplification is available because $\hpr_m \ll |\Omegapr|$
only when $k_\gamma \gg \Elpma$.  To get a full description of all
of region 4 of fig.\ \ref{fig:overBH_nodie}b (including the transitions
at its boundaries), we need to keep the mass in (\ref{eq:dGpairMass}).

This discussion of when and where the electron mass $m$ may be ignored
is summarized by the last two columns of table \ref{tab:scales}.
Those particular column entries also apply to the smooth transitions between
two regions if either region is sensitive to $m$ in that column.

As in the massless case, we may ignore the vacuum contribution
to the photon self-energy, corresponding now to replacing (\ref{eq:dGpairMass})
by
\begin {multline}
   \left[ \frac{d{\cal G}}{d\yfrak\,d(\Delta t_\pr)} \right]_\pair
   \longrightarrow~
   - \frac{\alpha}{2\pi}
   \biggl\{
     P_{\gamma\to e}(\yfrak)
       \biggl[
         \Omegapr^2 \csc^2(\Omegapr\,\Delta t_\pr) - \frac{1}{(\Delta t_\pr)^2}
       \biggr]
\\
     + i \hpr_m
       \biggl[
         \Omegapr \csc(\Omegapr\,\Delta t_\pr) - \frac{1}{\Delta t_\pr}
       \biggr]
   \biggr\}
   \, e^{-i\hpr_m \Delta t_\pr} .
\label {eq:dPairMass2}
\end {multline}
\end {subequations}

The result for $[d\Gamma/dx_\gamma\,d\yfrak]_{(n=1)}$ may now be obtained
by using eqs.\ (\ref{eq:dMass}) in (\ref{eq:Afactorize}) and
performing the integrations over $\Delta t_0$ and
$\Delta t_\pr$ using (\ref{eq:inttimes2}).
The $\Delta t_0$ integration is the same as the massless case,
\begin {equation}
  \int_{\Delta t_\pr}^\infty d(\Delta t_0) \>
    (\Delta t_0 {-} \Delta t_\pr) \,
    \Omega_0^2 \csc^2(\Omega_0\,\Delta t_0)
  = -\ln\bigl( 1 - e^{-2i\Omega_0\,\Delta t_\pr} \bigr) ,
\label {eq:intpr}
\end {equation}
leaving us with
\begin {equation}
  \left[ \frac{d\Gamma}{dx_\gamma \, d\yfrak} \right]_{(n=1)}
  \simeq
  - \frac{\alpha}{\pi} \, P_{e\to\gamma}(x_\gamma) \,
  \Re \int_0^\infty d(\Delta t_\pr) \>
  \left[ \frac{d{\cal G}}{d\eta\,d(\Delta t_\pr)} \right]_\pair
  \ln\bigl( 1 - e^{-2i\Omega_0\,\Delta t_\pr} \bigr) .
\label {eq:int0}
\end {equation}

Generalizing the argument of ref.\ \cite{softqed1} to the massive
case, we now argue that the last integral is dominated by
$|\Omega_0\,\Delta t_\pr| \ll 1$, which allows further simplification
of the integral.
Remembering that $\Omegapr$ (\ref{eq:MOmegapr}) has an imaginary part,
the $[d{\cal G}/d\eta\,d(\Delta t_\pr)]_\pair$ factor (\ref{eq:dGpairMass})
in the integral is dominated by
(i) $\Delta t_\pr \lesssim 1/[\max(|\Omega_\pr|,\hpr_m)]$.
Next note that the formulas (\ref{eq:MOmega0}) and (\ref{eq:MOmegapr})
for $\Omega_0$ and $\Omegapr$ imply that
(ii) $|\Omega_0| \ll |\Omega_\pr|$ in the soft-photon limit ($x_\gamma\ll 1$)
relevant to calculating the $\LPMplus$ effect.  The combination of these two
inequalities gives $|\Omega_0\,\Delta t_\pr| \ll 1$, which allows the
approximation
\begin {equation}
  \ln\bigl( 1 - e^{-2i\Omega_0\,\Delta t_\pr} \bigr)
  \simeq
  \ln(2i\Omega_0\,\Delta t_\pr)
\label {eq:logApprox}
\end {equation}
in (\ref{eq:int0}).

Changing integration variable from $\Delta t_\pr$ to
$\tau \equiv i\Omegapr \, \Delta t_\pr$ in the integral
(\ref{eq:int0}) and using (\ref{eq:dGpairMass}) and (\ref{eq:logApprox}) gives
\begin {equation}
  \left[ \frac{d\Gamma}{dx_\gamma \, d\yfrak} \right]_{(n=1)}
  \simeq
  \frac{\alpha^2}{2\pi^2} \, P_{e\to\gamma}(x_\gamma)
  \Re\left\{
      i P_{\gamma\to e}(\yfrak) \, \Omegapr \,
        I_1\bigl(
          \tfrac{2|\Omega_0|}{|\Omegapr|} \,,\, \tfrac{\hpr_m}{\Omegapr}
        \bigr)
      + i\hpr_m \,
        I_2\bigl(
          \tfrac{2|\Omega_0|}{|\Omegapr|} \,,\, \tfrac{\hpr_m}{\Omegapr}
        \bigr)
  \right\} ,
\label {eq:int1}
\end {equation}
where
\begin {subequations}
\label {eq:I12def}
\begin{align}
  I_1(c,a) &\equiv
  \int_0^\infty d\tau\>
  \Bigl( \frac{1}{\sinh^2\tau} - \frac{1}{\tau^2} \Bigr)
  \ln(c\tau) \, e^{-a\tau} ,
\label {eq:I1def}
\\
  I_2(c,a) &\equiv
  \int_0^\infty d\tau\>
  \Bigl( \frac{1}{\sinh\tau} - \frac{1}{\tau} \Bigr)
  \ln(c\tau) \, e^{-a\tau} ,
\label {eq:I2def}
\end {align}
\end {subequations}
and we have used the fact that $\Omega_0$ and $\Omegapr$ both have the
same complex phase $\sqrt{-i}$ to rewrite
$\Omega_0/\Omegapr = |\Omega_0|/|\Omegapr|$.
The integrals are carried out in appendix \ref{app:integrals}
with result
\begin {subequations}
\label {eq:I12}
\begin{align}
  I_1(c,a) &=
  \left[
     a\, \PSI\bigl(1;\tfrac{a}{2}\bigr)
     - 1
  \right]
  \left[ \ln\bigl(\tfrac{c}{2}\bigr) - \gammaE \right]
  + a
    \left[
       \gamma_1\bigl(\tfrac{a}{2}\bigr)
       + \tfrac12 \ln^2\bigl(\tfrac{a}{2}\bigr)
       - \ln\bigl(\tfrac{a}{2}\bigr) + 1
       \right]
\nonumber\\ & \hspace{22em}
  + 2\lnGamma\bigl(\tfrac{a}{2}\bigr) - \ln(2\pi) ,
\label {eq:I1}
\\
  I_2(c,a) &=
  -
     \PSI\bigl(\tfrac12;\tfrac{a}{2}\bigr)
  \left[ \ln\bigl(\tfrac{c}{2}\bigr) - \gammaE \right]
  - \gamma_1\bigl(\tfrac12{+}\tfrac{a}{2}\bigr)
  - \tfrac12 \ln^2\bigl(\tfrac{a}{2}\bigr) ,
\label {eq:I2}
\end {align}
\end {subequations}
where $\PSI(r;z)$ is defined by (\ref{eq:Psidef}),
and $\lnGamma(z)$ is the log-Gamma function.%
\footnote{
  The log-Gamma function $\lnGamma(z)$ in (\ref{eq:I12}) is
  defined with the convention that all cuts run along the negative
  real $z$ axis.  For complex $z$, as in our application, this is a
  different choice of Riemann surface than $\ln\bigl(\Gamma(z)\bigr)$
  with the conventional Riemann surface choice for $\ln w$
  (which would be to cut $\ln w$ along the negative real $w{=}\Gamma(z)$ axis
  instead of along the negative $z$ axis).
  As a practical matter, most numerical libraries have a direct
  implementation of $\lnGamma(z)$, such as LogGamma[$z$] in
  Mathematica \cite{Mathematica}.
}
The more obscure special function $\gamma_1(z)$ is a generalized
Stieltjes coefficient, defined in terms of the Laurent
expansion about the $s{=}1$ pole of the Hurwitz $\zeta$ function as%
\footnote{
  Useful properties of $\zeta(s,q)$ may be found in, for
  example, section 25.11 of ref.\ \cite{DLMF}
  and section 9.5 of ref.\ \cite{GR}.
  Besides online sources like Wolfram MathWorld and
  Wikipedia, one possible entry into the literature on
  generalized Stieltjes constants $\gamma_n(q)$ is
  a brief historical review in section 1 of ref.\ \cite{Stieltjes}.
  But in our paper
  (specifically appendices \ref{app:integrals} and \ref{app:limits})
  we will only need the definition (\ref{eq:Hurwitz}) and a few
  properties of $\zeta(s,q)$.
}
\begin {equation}
  \zeta(s,q) \equiv \sum_{k=0}^\infty \frac{1}{(k+q)^s}
  = \frac{1}{s-1} + \sum_{n=0}^\infty \, \frac{(-)^n}{n!}\,\gamma_n(q) \, (s{-}1)^n
  = \frac{1}{s-1} - \psi(q) - \gamma_1(q)\,(s{-}1) + \cdots .
\label {eq:Hurwitz}
\end {equation}
Mathematica \cite{Mathematica}
has a numerical implementation StieltjesGamma[$n$,$q$] of $\gamma_n(q)$.

We have been unable to analytically integrate (\ref{eq:int1})
over $\yfrak$ as in (\ref{eq:Atotal}); so our
final answer for fig.\ \ref{fig:neq1} is
\begin {equation}
  \left[ \frac{d\Gamma}{dx_\gamma} \right]_{(n=1)}
  \simeq
  \frac{\alpha^2}{2\pi^2} \, P_{e\to\gamma}(x_\gamma) \int_0^1 d\yfrak \>
  \Re\left\{
      i P_{\gamma\to e}(\yfrak) \, \Omegapr \,
        I_1\bigl(
          \tfrac{2|\Omega_0|}{|\Omegapr|} \,,\, \tfrac{\hpr_m}{\Omegapr}
        \bigr)
      + i\hpr_m \,
        I_2\bigl(
          \tfrac{2|\Omega_0|}{|\Omegapr|} \,,\, \tfrac{\hpr_m}{\Omegapr}
        \bigr)
  \right\} ,
\label {eq:dGammaneq1}
\end {equation}  
with the $\yfrak$ integral left to be done numerically.
Appendix \ref{app:numerics} discusses
subtleties to controlling numerical error in that integration.
In particular, it is useful to know the large-$q$ asymptotic expansion
of $\gamma_1(q)$, which we present in appendix \ref{app:asymptotic}.


\subsection{The \boldmath$n{\ge}2$ bubble diagrams}
\label {sec:nge2}

\subsubsection{Review and generalization}

In the massless case, the $n\ge 2$ contribution to fig.\ \ref{fig:LPM+diags}a
was found to be%
\footnote{
  See section 5.2 of ref.\ \cite{softqed1}.
}
\begin {equation}
  \left[ \frac{d\Gamma}{dx_\gamma} \right]_{n\ge2}
  \simeq
  2\Re \int_0^\infty d(\Delta t_0) \>
    \left[ \frac{d{\cal G}}{dx_\gamma\,d(\Delta t_0)} \right]_\brem
    \left[ e^{-{\cal G}_\pair \Delta t_0} - 1 + {\cal G}_\pair \Delta t_0 \right]
\label{eq:nge2a}
\end {equation}
with
\begin {equation}
  {\cal G}_\pair \equiv
  \int_0^1 d\yfrak \int_0^\infty d(\Delta t_\pr) \> 
  \left[ \frac{d{\cal G}}{d\yfrak\,d(\Delta t_\pr)} \right]_\pair ,
\label{eq:calGrate0}
\end {equation}
which is related to the total pair production rate by
\begin {equation}
   \Gamma_\pair = 2\Re({\cal G}_\pair) .
\label{eq:GammavcalG}
\end {equation}
Taking the massless form (\ref{eq:dGbremMass2})
of $[d{\cal G}/dx_\gamma\,d(\Delta t_0)]_\brem$
and using the integral \cite{softqed1}
\begin {equation}
  {\cal I}(\Omega,{\cal G}) \equiv
  \int_0^\infty dt \> \Omega^2 \csc^2(\Omega t) \,
       \left[ e^{-{\cal G} t} - 1 + {\cal G} t \right]
  = \left[
       \psi\bigl(1{+}\tfrac{\cal G}{2i\Omega}\bigr) + \gammaE
    \right] {\cal G} ,
\label {eq:integral}
\end {equation}
the rate (\ref{eq:nge2a}) gives
\begin {equation}
  \left[ \frac{d\Gamma}{dx_\gamma} \right]_{n\ge2}
  \simeq
  - \frac{\alpha}{\pi} \,P_{e\to\gamma}(x_\gamma) \,
  \Re \left\{
    \left[
        \psi\bigl(1{+}\tfrac{{\cal G}_\pair}{2i\Omega_0}\bigr) + \gammaE
    \right]
    {\cal G}_\pair
  \right\} .
\label{eq:nge2b}
\end {equation}

We now need to generalize those results to non-zero electron mass $m$.
For the same reasons discussed in section \ref{sec:neq1mass},
the electron mass can be ignored in $[d{\cal G}/dx_\gamma\,d(\Delta t_0)]_\brem$
whenever the $\LPMplus$ effect is important, and so the derivation just
discussed is unchanged except that we need to use the massive
version (\ref{eq:dPairMass2}) of $[d{\cal G}/d\yfrak\,d(\Delta t_\pr)]_\pair$
in (\ref{eq:calGrate0}) to get ${\cal G}_\pair$.
Using our result for the $\Delta t_\pr$ integral from
section \ref{sec:PairGeneral},%
\footnote{
  Eq.\ (\ref{eq:LPMpair}) without the $2\Re(\cdots)$ of (\ref{eq:GammavcalG}).
}
our final result for $n\ge2$ is then simply (\ref{eq:nge2b}) with
\begin {equation}
  {\cal G}_\pair =
  \frac{\alpha}{2\pi} \int_0^1 d\yfrak \>
  \Bigl\{
    i P_{\gamma\to e}(\yfrak)
    \Bigl[
      \Omegapr
      - \hpr_m \,
        \PSI\bigl( 1 ; \tfrac{\hpr_m}{2\Omegapr} \bigr)
    \Bigr]
    + i \hpr_m \,
        \PSI\bigl( \tfrac12 ; \tfrac{\hpr_m}{2\Omegapr} \bigr)
  \Bigl\} .
\label{eq:calGpair}
\end {equation}
We do not know an analytic result for this $\yfrak$ integral; so we
leave it also to be done numerically.


\subsubsection{Validity of (\ref{eq:nge2a})}
\label{sec:nge2validity}

For later reference in section \ref{sec:dielectric},
we should clarify that
the relatively simple
form of (\ref{eq:nge2a}) for $n{\ge}2$ bubbles in fig.\ \ref{fig:LPM+diags}a
relied \cite{softqed1} on the approximation that
the bremsstrahlung time scale $\Delta t_0$ dominating the integral
(\ref{eq:nge2a}) is parametrically large compared to the
pair-production time scale $\Delta t_\pr$ dominating the integral
(\ref{eq:calGrate0}).
For the massless case, those
two scales were $\Delta t_0 \sim 1/\Gamma_\pair$ and
$\Delta t_\pr \sim 1/|\Omegapr|$, which indeed have
$\Delta t_0 \gg \Delta t_\pr$ everywhere in the soft photon limit
$x_\gamma \ll 1$ relevant to the $\LPMplus$ effect.
In contrast, the massless $n{=}1$ diagram reviewed in section
\ref{sec:neq1massless} gets important contributions from a range
$1/|\Omegapr| \lesssim \Delta t_0 \lesssim 1/|\Omega_0|$ of $\Delta t_0$
that includes $\Delta t_0 \sim 1/|\Omegapr|$.
The $e^{-{\cal G}_\pair \Delta t_0}$ term in (\ref{eq:nge2a}) represents
the sum, in the $\Delta t_0 \gg 1/|\Omegapr|$ and soft-photon approximations,
over all $n \ge 0$ in fig.\ \ref{fig:LPM+diags}a, with the $n$th term of the
Taylor expansion in ${\cal G}_\pair$ corresponding to the diagram with $n$
bubbles.  The $-1$ in (\ref{eq:nge2a}) then subtracts the
ordinary LPM bremsstrahlung result ($n{=}0$), and the
$+{\cal G}_\pair \Delta t_0$ subtracts the $n{=}1$ term, for which the
approximation made here is invalid (beyond determining the coefficient
of the logarithm).  The failure for $n{=}1$ of the approximations
that give (\ref{eq:nge2a}) can also be seen \textit{a posteriori}
from (\ref{eq:nge2a}) itself; were one to remove the very last
term $+{\cal G}_\pair \Delta t_0$ that subtracts away the $n{=}1$
contribution, then [referring back to (\ref{eq:dGbremMass2})]
the $\Delta t_0$ integral in (\ref{eq:nge2a}) would become
logarithmically divergent due to the $\Delta t_0 \to 0$ behavior
of the integrand.

To justify having now used (\ref{eq:nge2a}) also in the massive case, we should
re-check the assumption underlying that formula.
Again, the $\Delta t_0$ integral in (\ref{eq:nge2a}) is dominated
by $\Delta t_0 \sim 1/\Gamma_\pair$, but now consider the case
$\kgamma \lesssim \Elpma$, where the mass significantly
modifies the pair production amplitude ${\cal G}_\pair$.
The $\Delta t_\pr$ integral of (\ref{eq:dPairMass2}) is then
dominated by $\Delta t_\pr \sim 1/\hpr_m$, and so the condition
$\Delta t_0 \gg \Delta t_\pr$ for the validity of (\ref{eq:nge2a}) becomes
$\Gamma_\pair \ll \hpr_m$.  From (\ref{eq:pairBH}), (\ref{eq:hprm}),
and $\Elpma \sim \Elpm \equiv m^4/\qhat$, that condition is
$\kgamma \ll \Elpma/\alpha$, which is indeed satisfied parametrically for
$\kgamma \lesssim \Elpma$.


\subsection{The longitudinally-polarized photon contribution}

\begin {figure}[t]
\begin {center}
  \includegraphics[scale=0.7]{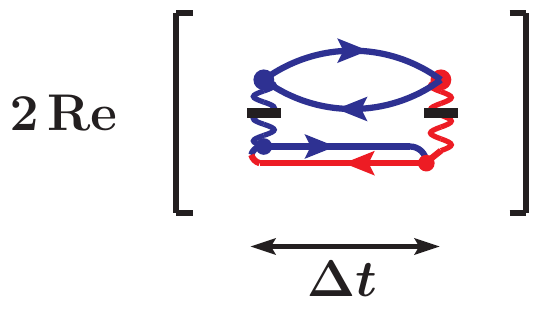}
  \caption{
    \label{fig:L}
    $e \to ee\bar e$ mediated by longitudinally-polarized photons.
  }
\end {center}
\end {figure}

Now turn to the contribution of fig.\ \ref{fig:LPM+diags}b, which involves
longitudinally-polarized photons.
In the massless case, ref.\ \cite{softqed1} found that this contribution
was%
\footnote{
  See specifically eq.\ (D.14) of appendix D of ref.\ \cite{softqed1}.
}
\begin {equation}
  \left[ \frac{d\Gamma}{dx_\gamma} \right]_{\rm L}
  = \int_0^1 d\yfrak \>
    \left[ \frac{d\Gamma}{dx_\gamma \, d\yfrak} \right]_{\rm L}
\label {eq:Lrate0}
\end {equation}
with
\begin {equation}
  \left[ \frac{d\Gamma}{dx_\gamma \, d\yfrak} \right]_{\rm L} =
  -\frac{\alpha^2}{\pi^2} \, \frac{4\yfrak(1{-}\yfrak)}{x_\gamma} \,
  \Re \int_0^\infty \frac{d(\Delta t)}{\Delta t} \>
    \Omega_\pr \csc(\Omega_\pr\,\Delta t)
  \qquad \mbox{(for $m{=}0$)}.
\label {eq:II3}
\end {equation}
Above, $\Delta t$ is the time between the two instantaneous longitudinal
photon exchanges, as depicted in fig.\ \ref{fig:L}.%
\footnote{
  We work in lightcone gauge, and technically the longitudinal photon
  exchanges are instantaneous in lightcone time $x^+$.
  However, for the high-energy, nearly-collinear splitting processes
  relevant to the LPM effect, that also means instantaneous in ordinary
  time $t$ to excellent approximation.
}
The $\Omega_\pr \csc(\Omega_\pr\,\Delta t)$ factor describes
(up to a proportionality constant)
the propagation of the soft $e^-e^+$ pair in the medium,
and this factor limits $\Delta t$ to $\Delta t \lesssim 1/|\Omega_\pr|$.
There is in principle a similar factor $\Omega_0 \csc(\Omega_0\,\Delta t)$
associated with the propagation of the
other two (hard) ``particles'' in fig.\ \ref{fig:L}, but,
since $|\Omega_0| \ll |\Omega_\pr|$ in the soft-photon limit
relevant to the $\LPMplus$ effect, that factor can be approximated
by the $1/\Delta t$ factor in (\ref{eq:II3}).
The remaining factors in (\ref{eq:II3}) include the factors associated
with the longitudinal photon exchanges and vertices.

In sections \ref{sec:neq1} and \ref{sec:nge2}, we discussed
that we need only include the effects of the electron mass $m$
on the production and evolution of the soft $e^-e^+$ pair.
The same is true here for the same reasons.
So, we must replace
\begin {equation}
  \Omega_\pr \csc(\Omega_\pr\,\Delta t)
  \longrightarrow
  \Omega_\pr \csc(\Omega_\pr\,\Delta t) \, e^{-i\hpr_m \Delta t}
\label {eq:Lchange}
\end {equation}
in (\ref{eq:II3}) to account for the effect of electron mass
in (\ref{eq:Hpair}) on the soft pair evolution $e^{-i{\cal H}_\pair\,\Delta t}$.

When we reviewed the LPM pair production rate in section \ref{sec:LPMpair},
the mass {\it also} had an effect on the vertices for emitting a
transversely-polarized photon.  This introduced the second term
in (\ref{eq:ZpairHO1}) --- the pair-production analog of
the helicity-flip term for bremsstrahlung $e \to e\gamma$\
in (\ref{eq:ZrateHO1}).
However, there is
no such additional term for interactions involving longitudinal
photons,%
\footnote{
  Specifically, one may check that the relativistic
  matrix element given by eq.\ (D.11) of
  ref.\ \cite{softqed1} is independent of the electron mass
  in the limit (relevant here) that all electrons/positrons have
  ultra-relativistic energies ($E_i \gg m$).
  In light-cone gauge, longitudinally polarized photons couple to
  $\bar u_{\sigma'}(p') \gamma^+ u_\sigma(p)
    = 2 \delta_{\sigma',\sigma} \sqrt{p^+ p^{'+}}$
  for bremsstrahlung and
  $\bar v_{\sigma'}(p') \gamma^+ u_\sigma(p)
    = 2 \delta_{\sigma',-\sigma} \sqrt{p^+ p^{'+}}$
  for pair production, where $\sigma$ and $\sigma'$ are helicities.
  There are no terms that do not conserve electron helicity.
  (Readers may find useful tables II and III of ref.\ \cite{LB}
  or the version presented in tables A.1 and A.2 of ref.\ \cite{KL}.)
}
and so (\ref{eq:Lchange}) is the
only modification necessary.

All told then,
\begin {equation}
  \left[ \frac{d\Gamma}{dx_\gamma \, d\yfrak} \right]_{\rm L} =
  -\frac{\alpha^2}{\pi^2} \, \frac{4\yfrak(1{-}\yfrak)}{x_\gamma} \,
  \Re \int_0^\infty \frac{d(\Delta t)}{\Delta t} \>
    \Omega_\pr \csc(\Omega_\pr\,\Delta t) \, e^{-i\hpr_m \Delta t} .
\label {eq:II3mass}
\end {equation}
Like the LPM bremsstrahlung rate (\ref{eq:ZrateHO1}), this integral
has a UV divergence from $\Delta t \to 0$, but once again we may
sidestep that divergence by subtracting the (vanishing) vacuum
contribution to the rate, giving
\begin {equation}
  \left[ \frac{d\Gamma}{dx_\gamma \, d\yfrak} \right]_{\rm L} =
  -\frac{\alpha^2}{\pi^2} \, \frac{4\yfrak(1{-}\yfrak)}{x_\gamma} \,
  \Re \int_0^\infty \frac{d(\Delta t)}{\Delta t} \>
  \left[ \Omega_\pr \csc(\Omega_\pr\,\Delta t) - \frac{1}{\Delta t} \right]
  \, e^{-i\hpr_m \Delta t} .
\label {eq:IImass}
\end {equation}
Changing integration variable to $\tau \equiv i\Omega_\pr \, \Delta t$,
\begin {equation}
  \left[ \frac{d\Gamma}{dx_\gamma \, d\yfrak} \right]_{\rm L} =
  -\frac{\alpha^2}{\pi^2} \, \frac{4\yfrak(1{-}\yfrak)}{x_\gamma} \,
  \Re\left\{ i \Omegapr \, I_3\bigl( \tfrac{\hpr_m}{\Omegapr} \bigr) \right\}
\label {eq:dGL}
\end {equation}
with
\begin {equation}
  I_3(a) \equiv
  \int_0^\infty \frac{d\tau}{\tau} \>
  \left( \frac{1}{\sinh\tau} - \frac{1}{\tau} \right) e^{-a\tau} .
\label {eq:I3def}
\end {equation}
This integral is evaluated in appendix \ref{app:integrals} with result
\begin {equation}
  I_3(a)
  =
  2 \Bigl[
      \lnGamma\bigl(\tfrac12{+}\tfrac{a}{2}\bigr)
      - \tfrac{a}{2} \ln\bigl(\tfrac{a}{2}\bigr)
      + \tfrac{a}{2}
      - \tfrac12 \ln(2\pi)
    \Bigr] .
\label {eq:I3}
\end {equation}
Once again, we leave the final integral (\ref{eq:Lrate0}) over $\yfrak$
to numerics, and so the final form of our result is
\begin {equation}
  \left[ \frac{d\Gamma}{dx_\gamma} \right]_{\rm L} =
  -\frac{4\alpha^2}{\pi^2x_\gamma} \,
  \int_0^1 d\yfrak \> \yfrak(1{-}\yfrak) \,
  \Re\left\{ i \Omegapr \, I_3\bigl( \tfrac{\hpr_m}{\Omegapr} \bigr) \right\}
  .
\label {eq:Lrate}
\end {equation}


\subsection {Limits}

\subsubsection {$k_\gamma \gg \Elpma$}

In the limit $\kgamma \gg \Elpma$ of extremely high-energy photons,
we have $\hpr_m \ll |\Omegapr|$, which corresponds to $a{\to}0$ in
the application of (\ref{eq:I12}) to (\ref{eq:int1}).
In this limit, we find that our $n{=}1$ bubble result
(\ref{eq:dGammaneq1}) matches (the $\eta$ integral of)
the differential rate (\ref{eq:old1}), originally found
in ref.\ \cite{softqed1}.%
\footnote{
  For the relevant limit of $\gamma_1(q)$, we use the
  small-$q$ expansion
  $\gamma_1(q) \simeq q^{-1}\ln q$ from (\ref{eq:smallq}).
  Combined with the more well known small-argument
  expansions of $\psi(z)$ and $\Gamma(z)$, this gives
  $I_1(c;a) \simeq \gammaE - \ln(\pi c)$
  and
  $a\,I_2(c;a) \to 0$ from (\ref{eq:I12}).
}
Integrating over $\eta$ then gives \cite{softqed1}
\begin {equation}
  \left[ \frac{d\Gamma}{dx_\gamma} \right]_{(n=1)}
  \simeq
  \frac{\alpha}{2\pi} \, P_{e\to\gamma}(x_\gamma) \,
  \Gamma_\pair \,
  \left[
    - \ln\left( \tfrac{\pi x_\gamma}{2} \right)
    + \gammaE + \tfrac16
  \right] ,
\label {eq:old1b}
\end {equation}
with $\Gamma_\pair$ given by the deep-LPM limiting formula
(\ref{eq:pairmassless}).

$\hpr_m \ll |\Omegapr|$ also simplifies
the pair production amplitude (\ref{eq:calGpair}) to
\begin {equation}
  {\cal G}_\pair \simeq
  \frac{\alpha}{2\pi} \int_0^1 d\yfrak \>
  \Bigl\{
    i P_{\gamma\to e}(\yfrak) \, \Omegapr
  \Bigl\}
  =
  \frac{3\alpha}{8} \sqrt{ \frac{i\qhat}{2\kgamma} } \,.
\end {equation}
In this limit, (\ref{eq:GammavcalG}) then gives
\begin {equation}
  {\cal G}_\pair \simeq \bigl(\tfrac{i}{2}\bigr)^{\!1/2} \, \Gamma_\pair
\label{eq:GpairDeepLPM}
\end {equation}
with $\Gamma_\pair$ again given by the deep-LPM limiting formula
(\ref{eq:pairmassless}).  Then (\ref{eq:nge2b}) for $n{\ge}2$
may be rewritten in the form
\begin {equation}
  \left[ \frac{d\Gamma}{dx_\gamma} \right]_{(n\ge2)}
  \simeq
  \frac{\alpha}{2\pi} \,P_{e\to\gamma}(x_\gamma) \,
    \Gamma_\pair
    \left[
       -\psi\bigl(1{+}\tfrac{\Gamma_\pair}{2\sqrt2 \,|\Omega_0|}\bigr) - \gammaE
    \right] .
\end {equation}

Finally, taking the $a{\to}0$ limit of (\ref{eq:I3}) in (\ref{eq:Lrate})
and then performing the integration over $\yfrak$ gives%
\footnote{
  The $a{\to}0$ limit of (\ref{eq:I3}) is $I_3(0)=-\ln2$.
}
\begin {equation}
  \left[ \frac{d\Gamma}{dx_\gamma} \right]_{\rm L}
  \simeq
  \frac{\alpha}{\pi x_\gamma}
    \Gamma_\pair
    \left[ \tfrac23 \ln 2 \right].
\end {equation}

In total, the $k_\gamma \gg \Elpma$ limit of the differential energy loss
rate (\ref{eq:LPM+rate_nodie}) is then 
\begin {subequations}
\label {eq:largekregion}
\begin {equation}
  \left[ \frac{d\Gamma}{dx_\gamma} \right]_\LPMplus
  =
  \left[ \frac{d\Gamma}{dx_\gamma} \right]_\LPM
  +
  \delta \! \left[\frac{d\Gamma}{dx_\gamma}\right]
\label {eq:thisone}
\end {equation}
with
\begin {equation}
  \delta \! \left[\frac{d\Gamma}{dx_\gamma}\right]
  \simeq
  \frac{\alpha}{\pi x_\gamma} \,
  \Gamma_\pair \,
  \left[
    -\psi\bigl(1{+}\tfrac{\Gamma_\pair}{2\sqrt2 \,|\Omega_0|}\bigr)
    - \ln\left( \tfrac{\pi x_\gamma}{2} \right)
    + \tfrac16 + \tfrac23\ln2
  \right] , \qquad (\kgamma\gg\Elpma)
\label {eq:dGamlargek}
\end {equation}
\end {subequations}
where we've made the small-$x_\gamma$ approximation
$P_{e\to\gamma}(x_\gamma) \simeq 2/x_\gamma$ in the $\LPMplus$ corrections.

Above, $[d\Gamma/dx_\gamma]_\LPM$ is the ordinary LPM bremsstrahlung rate
(\ref{eq:LPMrate}) \textit{including} effects of electron mass.
We were able to ignore the mass of the initial electron
for $\LPMplus$ corrections
because $\LPMplus$ corrections are only important in the
bremsstrahlung deep-LPM regime.
But, if we want a formula (\ref{eq:thisone}) that also covers regions
where the $\LPMplus$ correction is unimportant, we must keep the mass in
$[d\Gamma/dx_\gamma]_\LPM$ to correctly handle the transition from the
LPM region (region 2 of fig.\ \ref{fig:overBH_nodie}) to the
Bethe-Heitler region (region 1).  For $k_\gamma \gg \Elpma$, that
transition to Bethe-Heitler becomes relevant as $k_\gamma{\to}E$ ---
a transition that was excluded, for the sake of simplicity,
from the analysis of ref.\ \cite{softqed1}.

The distinction between regions 1+2 versus 4$_{\rm a}$
in fig.\ \ref{fig:overBH_nodie}b
simply corresponds to whether the
$[d\Gamma/dx_\gamma]_\LPM$ or $\delta[d\Gamma/dx_\gamma]$ term
dominates the total rate (\ref{eq:largekregion}).
Deep inside region 4$_{\rm a}$ corresponds (ignoring logarithms for now)
to the $\Gamma_\pair \gg |\Omega_0|$
limit of (\ref{eq:largekregion}), which gives
\begin {multline}
  \left[\frac{d\Gamma}{dx_\gamma}\right]_{\LPMplus}
  \simeq
  \delta \left[\frac{d\Gamma}{dx_\gamma}\right]
  \simeq
  \frac{\alpha}{\pi x_\gamma} \,
  \Gamma_\pair \,
  \left[
    -\ln\bigl(\tfrac{\pi x_\gamma \Gamma_\pair}{4\sqrt2 \,|\Omega_0|}\bigr)
    + \tfrac16 + \tfrac23\ln2
  \right]
\\
  \simeq
  \frac{\alpha}{2\pi} \, P_{e\to\gamma}(x_\gamma)
  \, \Gamma_\pair \,
  \left[
    \ln\bigl(\tfrac{32}{3\pi\alpha}\bigr)
    + \tfrac16 + \tfrac23\ln2
  \right] .
  \qquad \mbox{(deep region 4$_{\rm a}$)}
\label {eq:deep4a}
\end {multline}
We've reverted to using $P_{e\to\gamma}(x_\gamma)$ because the logarithm in
this result has a physical interpretation in terms of the
vacuum Weizs\"acker-Williams (DGLAP) probability of finding a photon inside
the original electron times the probability for a nearly-onshell photon
to pair produce inside the medium \cite{softqed1}.
(See section \ref{sec:logs} for a brief discussion.)

We define the transition region between regions 2 and $4_{\rm a}$ as the
place where the limiting rate formulas for the two regions --- 
in this case (\ref{eq:deepLPM}) and (\ref{eq:deep4a}) ---
are parametrically equal to each other, which gives
the $2|4_{\rm a}$ entry of table \ref{tab:boundaries}.
If one ignores logarithms, this is the same as $\Gamma_\pair \sim |\Omega_0|$.
As earlier, our convention in plots like fig.\ \ref{fig:overBH_nodie}b
is to draw the dashed boundary lines where the limiting formulas for the
two regions are exactly equal to each other.


\subsubsection {$k_\gamma \ll \Elpma$}

The limit $k_\gamma \ll \Elpma$ corresponds to the opposite limit,
$\hpr_m \gg |\Omegapr|$.  For the $n{=}1$ contribution (\ref{eq:int1}),
we then need the large-$a$ limit of eqs.\ (\ref{eq:I12}) for
$I_1(c,a)$ and $I_2(c,a)$.  Using the asymptotic expansion of
$\gamma_1(q)$ from appendix \ref{app:asymptotic},
\begin {equation}
  I_1(c,a) \simeq \frac{ \ln\bigl(\frac{a}{c}\bigr) + \gammaE}{3a} \,,
  \qquad
  I_2(c,a) \simeq \frac{ \ln\bigl(\frac{a}{c}\bigr) + \gammaE - 1}{6a^2} \,,
\end {equation}
and then
\begin {equation}
  \left[ \frac{d\Gamma}{dx_\gamma \, d\yfrak} \right]_{(n=1)}
  \simeq
  \frac{\alpha^2}{2\pi^2} \, P_{e\to\gamma}(x_\gamma) \,
  \Re\left\{
    \frac{i\Omegapr^2}{3\hpr_m}
    \left[
      \Bigl( P_{\gamma\to e}(\yfrak) + \tfrac12 \Bigr)
        \Bigl(
          \ln\bigl(\tfrac{\hpr_m}{2\Omega_0}\bigr) + \gammaE
        \Bigr)
      - \tfrac12
    \right]
  \right\} .
\label {eq:old1asymp}
\end {equation}
Integrating over $\yfrak$, and
remembering that both $\Omegapr$ (\ref{eq:MOmegapr}) and
$\hpr_m$ (\ref{eq:hprm}) depend on $\yfrak$,
gives
\begin {equation}
  \left[ \frac{d\Gamma}{dx_\gamma} \right]_{(n=1)}
  \simeq
  \frac{\alpha}{2\pi} \, P_{e\to\gamma}(x_\gamma) \, \Gamma_\pair
  \Re
  \left[
    \ln \bigl( \tfrac{m^2}{4\kgamma\Omega_0} \bigr)
    + \gammaE
    + \tfrac53
  \right] ,
\label{eq:neq1low}
\end {equation}
where here $\Gamma_\pair$ refers to the Bethe-Heitler limit
(\ref{eq:pairBH}) of the pair production rate.

For the $n{\ge}2$ bubble contribution, we first need the $\hpr_m\gg|\Omega_0|$
limit of (\ref{eq:calGpair}), which evaluates to
\begin {equation}
   {\cal G}_\pair
   \simeq
   \frac{\alpha}{2\pi} \int_0^1 d\yfrak \>
     \left[ \tfrac13 \, P_{\gamma\to e}(\yfrak) + \tfrac16 \right]
     \frac{i\Omegapr^2}{\hpr_m}
   = \frac{7\alpha\qhat}{36\pi m^2}
   = \frac{\Gamma_\pair}{2} \,.
\label {eq:Gpairlow}
\end {equation}
Note that ${\cal G}_\pair$ is real-valued in this limit, whereas
in the limit (\ref{eq:GpairDeepLPM}) it has a complex phase.
The $n{\ge}2$ rate (\ref{eq:nge2b}) becomes
\begin {equation}
  \left[ \frac{d\Gamma}{dx_\gamma} \right]_{(n\ge2)}
  \simeq
  \frac{\alpha}{2\pi} \, P_{e\to\gamma}(x_\gamma) \, \Gamma_\pair
  \Re
  \left[
    -\psi\bigl( 1 {+} \tfrac{\Gamma_\pair}{4i\Omega_0} \bigr)
    - \gammaE
  \right] .
\end {equation}
Finally, taking the large-$a$ limit
$I_3(a) \simeq -\frac{1}{6a}$ of (\ref{eq:I3})
in (\ref{eq:Lrate})
and integrating over $\yfrak$ gives
\begin {equation}
  \left[ \frac{d\Gamma}{dx_\gamma} \right]_{\rm L}
  \simeq
  \frac{\alpha}{\pi x_\gamma} \, \Gamma_\pair
  \times
    \tfrac27
  .
\label {eq:LrateLow}
\end {equation}
In total, the $\kgamma{\ll}\Elpma$ analog of (\ref{eq:dGamlargek}) is then
\begin {equation}
  \delta \! \left[\frac{d\Gamma}{dx_\gamma}\right]
  \simeq
  \frac{\alpha}{\pi x_\gamma} \,
  \Gamma_\pair \,
  \Re
  \left[
    -\psi\bigl(1{+}\tfrac{\Gamma_\pair}{4i\Omega_0}\bigr)
    + \ln \bigl( \tfrac{m^2}{4\kgamma\Omega_0} \bigr)
    + \tfrac{41}{21}
  \right] . \qquad (\kgamma\ll\Elpma)
\label {eq:dGamsmallk}
\end {equation}

Similar to what happened in the $\kgamma\gg\Elpma$ limit, the
distinction between regions 1+2 versus 4$_{\rm b}$ in
fig.\ \ref{fig:overBH_nodie}b is determined by whether
the $[d\Gamma/dx_\gamma]_\LPM$ or $\delta[d\Gamma/dx_\gamma]$ term
dominates the total rate (\ref{eq:thisone}), now using
(\ref{eq:dGamsmallk}).  Deep inside region 4$_{\rm b}$ corresponds
to the $\Gamma_\pair \gg |\Omega_0|$ limit of (\ref{eq:dGamsmallk}),
yielding
\begin {equation}
  \left[\frac{d\Gamma}{dx_\gamma}\right]_{\LPMplus}
  \simeq
  \delta \left[\frac{d\Gamma}{dx_\gamma}\right]
  \simeq
  \frac{\alpha}{2\pi} \, P_{e\to\gamma}(x_\gamma) \,
  \Gamma_\pair
  \left[
    \ln \bigl( \tfrac{m^2/\kgamma}{\Gamma_\pair} \bigr)
    + \tfrac{41}{21}
  \right]
  \qquad \mbox{(deep region 4$_{\rm b}$)}
\label {eq:deep4b}
\end {equation}
as the analog of (\ref{eq:deep4a}).


\section{LPM+ including the dielectric effect
         \boldmath$(0 < m_\gamma \ll m)$}
\label{sec:dielectric}

\subsection{Incorporating \boldmath$m_\gamma$}

The dielectric effect modifies the $\LPMplus$ analysis
of section \ref{sec:massless} by including a factor of
$e^{-i h_\gamma\,\Delta t_0}$ in the bremsstrahlung amplitude
(\ref{eq:dGbremMass2}):
\begin {equation}
   \left[ \frac{d{\cal G}}{dx_\gamma\,d(\Delta t_0)} \right]_\brem
   \longrightarrow~
   - \frac{\alpha}{2\pi} \,
     P_{e\to\gamma}(x_\gamma) \,
       \Omega_0^2 \csc^2(\Omega_0\,\Delta t_0) \, e^{-i h_\gamma \Delta t_0} ,
\label {eq:dGbremDie}
\end {equation}
where $h_\gamma = m_\gamma^2/2\kgamma$ as in (\ref{eq:hmgamma}).
In principle, we should also include $m_\gamma$ in the calculation
of the pair production amplitude (\ref{eq:dPairMass2}) and
(\ref{eq:calGpair}), but, as discussed in section \ref{sec:LPMpair},
that effect is ignorable for ordinary matter ($m_\gamma \ll m$).
We also do not have to worry about the dielectric effect for the
instantaneous exchanges of longitudinally polarized photons in
fig.\ \ref{fig:LPM+diags}b.%
\footnote{
  As explained in section 4.1 of
  ref.\ \cite{softqed1}, the $\LPMplus$ diagrams of
  fig.\ \ref{fig:LPM+diags} are evaluated using Light Cone Perturbation
  Theory, which uses light-cone gauge.
  In light-cone gauge, the longitudinally polarized photon propagator
  is
  $G_{\rm L}(k) = i n^\mu n^\nu/(n\cdot k)^2
   \propto \delta^{\mu}_{-} \delta^{\nu}_{-} / (k^+)^2$,
  where $(n^+,n^-,{\bm n}^\perp)=(0,1,{\bm 0})$.
  (See in particular appendix D.1.2 of ref.\ \cite{softqed1}.)
  The medium-induced photon mass $m_\gamma$ is generated by the contribution
  to the photon self-energy $\Pi^{\mu\nu}$ from Compton scattering of
  photons from atomic electrons, as reviewed in our appendix
  \ref{app:mgamma}.  That self-energy is of order $m_\gamma^2$, which is
  parametrically small compared to the inverse longitudinal photon propagator
  $[G_{\rm L}(k)]^{-1} \sim (k^+)^2$ and so may be ignored.
}

Without further approximation,
doing the $\Delta t_0$ and $\Delta t_\pr$
integrals for the $n{=}1$ bubble diagram as in section \ref{sec:neq1mass}
becomes more complicated using (\ref{eq:dGbremDie}),
but there is a simple workaround.
We find that computing the \text{difference}
\begin {subequations}
\label {eq:dGperp}
\begin {equation}
  \delta_{m_\gamma}\! \left[\frac{d\Gamma}{dx_\gamma}\right]_\perp
  \equiv
  \left[ \frac{d\Gamma}{dx_\gamma} \right]_\perp^{m_\gamma\not=0}
  -
  \left[ \frac{d\Gamma}{dx_\gamma} \right]_\perp^{m_\gamma=0}
\label {eq:dgammadef}
\end {equation}
that $m_\gamma$ makes to
\begin {equation}
  \left[\frac{d\Gamma}{dx_\gamma}\right]_\perp
  \equiv
  \left[\frac{d\Gamma}{dx_\gamma}\right]_{(n=1)}
  +
  \left[\frac{d\Gamma}{dx_\gamma}\right]_{(n\ge2)}
\end {equation}
\end {subequations}
is easier than directly computing the $m_\gamma{\not=}0$ result itself.
In this language, the total energy-loss rate is
\begin {equation}
  \left[ \frac{d\Gamma}{dx_\gamma} \right]_\LPMplus
  =
  \left[ \frac{d\Gamma}{dx_\gamma} \right]_\LPM
  + \delta\!\left[ \frac{d\Gamma}{dx_\gamma} \right]_{m_\gamma=0}
  + \delta_{m_\gamma} \! \left[\frac{d\Gamma}{dx_\gamma}\right]_\perp ,
\end {equation}
where $\delta[d\Gamma/dx_\gamma]_{m_\gamma=0}$
represents the result for (\ref{eq:deltaLPM+}) found
in section \ref{sec:massless}.

The trick to simplifying the calculation
is to return to the formula (\ref{eq:nge2a}) for $n{\ge}2$ bubble diagrams
but \textit{remove}
the $+{\cal G}_\pair \Delta t_0$ term that subtracted the
$n{=}1$ diagram:
\begin {align}
  \left[ \frac{d\Gamma}{dx_\gamma} \right]_\perp
  &\simeq
  2\Re \int_0^\infty d(\Delta t_0) \>
    \left[ \frac{d{\cal G}}{dx_\gamma\,d(\Delta t_0)} \right]_\brem
    \left[ e^{-{\cal G}_\pair \Delta t_0} - 1 \right]
\nonumber\\
  &\simeq
  - \frac{\alpha}{\pi} \,
     P_{e\to\gamma}(x_\gamma) \,
    \Re \int_0^\infty d(\Delta t_0) \>
       \Omega_0^2 \csc^2(\Omega_0\,\Delta t_0)
    \left[ e^{-{\cal G}_\pair \Delta t_0} - 1 \right]
    e^{-i h_\gamma \Delta t_0} .
\label {eq:nge1}
\end {align}
As discussed in section \ref{sec:nge2validity}, that $+{\cal G}_\pair \Delta t_0$
term was necessary for the small-$\Delta t_0$
convergence of the integral, and
the lack of convergence for (\ref{eq:nge1}) reflects the failure
for the $n{=}1$ diagram of the $\Delta t_0 \gg \Delta t_\pr$
approximation used for (\ref{eq:nge2a}).
But now consider instead using (\ref{eq:nge1}) for the $m_\gamma{\not=}0$
\textit{difference} (\ref{eq:dgammadef}):
\begin {equation}
  \delta_{m_\gamma}\! \left[\frac{d\Gamma}{dx_\gamma}\right]_\perp
  = 
  - \frac{\alpha}{\pi} \,
     P_{e\to\gamma}(x_\gamma) \,
    \Re \int_0^\infty d(\Delta t_0) \>
       \Omega_0^2 \csc^2(\Omega_0\,\Delta t_0)
    \left[ e^{-{\cal G}_\pair \Delta t_0} - 1 \right]
    \left[ e^{-i h_\gamma \Delta t_0} - 1 \right] .
\label {eq:dgamma1}
\end {equation}
Unlike (\ref{eq:nge1}), this integral is convergent.
Moreover, when the $\LPMplus$ effect is important
($\Gamma_\pair \gtrsim |\Omega_0|$), the integral is dominated by
\begin {equation}
  \Delta t_0 \sim \min( \Gamma_\pair^{-1} , h_\gamma^{-1} ) ,
\end {equation}
which is indeed parametrically large ($\Delta t_0 \gg \Delta t_\pr$) compared to
\begin {equation}
  \Delta t_\pr \sim \min\bigl( |\Omega_\pr|^{-1} , (\hpr_m)^{-1} \bigr) .
\end {equation}

To evaluate (\ref{eq:dgamma1}), the last two factors of the integrand can
be algebraically rearranged as
\begin {multline}
  \left[ e^{-{\cal G}_\pair \Delta t_0} - 1 \right]
  \left[ e^{-i h_\gamma \Delta t_0} - 1 \right]
  =
  \left[
    e^{-({\cal G}_\pair{+}i h_\gamma) \Delta t_0} - 1
    + ({\cal G}_\pair{+}i h_\gamma) \Delta t_0
  \right]
\\
  -
  \left[
    e^{-{\cal G}_\pair \Delta t_0} - 1 + {\cal G}_\pair \Delta t_0
  \right]
  -
  \left[
    e^{-i h_\gamma \Delta t_0} - 1 + i h_\gamma \Delta t_0
  \right] .
\end {multline}
Then the integral (\ref{eq:dgamma1}) for the effect of $m_\gamma$ on
$n\ge 1$ bubble diagrams can be rewritten in terms of the $m_\gamma{=}0$
rate $[d\Gamma/dx_\gamma]_{(n\ge2)}$ for $n\ge 2$ diagrams as
\begin {equation}
  \delta_{m_\gamma}\! \left[\frac{d\Gamma}{dx_\gamma}\right]_\perp
  = 
  - \frac{\alpha}{\pi} \,
     P_{e\to\gamma}(x_\gamma) \,
    \Re\bigl[
       {\cal I}(\Omega_0,{\cal G}_\pair{+}i h_\gamma)
        - {\cal I}(\Omega_0,{\cal G}_\pair) - {\cal I}(\Omega_0,i h_\gamma)
    \bigr]
\end {equation}
where ${\cal I}(\Omega,{\cal G})$ is the $n{\ge}2$ bubble integral
(\ref{eq:integral}), with result
\begin {multline}
  \delta_{m_\gamma}\! \left[\frac{d\Gamma}{dx_\gamma}\right]_\perp
  =
\\
  - \frac{\alpha}{\pi} \,
     P_{e\to\gamma}(x_\gamma) \,
    \Re\Bigl[
       ({\cal G}_\pair{+}i h_\gamma) \,
       \psi\bigl(1{+}\tfrac{{\cal G}_\pair{+}i h_\gamma}{2i\Omega_0}\bigr)
       -
       {\cal G}_\pair \, \psi\bigl(1{+}\tfrac{{\cal G}_\pair}{2i\Omega_0}\bigr)
       -
       i h_\gamma \, \psi\bigl(1{+}\tfrac{h_\gamma}{2\Omega_0}\bigr)
    \Bigr] .
\label {eq:delGperp}
\end {multline}
When combined with the $m_\gamma{=}0$ result, the
${\cal G}_\pair \, \psi\bigl(1{+}\tfrac{{\cal G}_\pair}{2i\Omega_0}\bigr)$
term above cancels the similar term in (\ref{eq:nge2b}).
In total, combining (\ref{eq:dGammaneq1}), (\ref{eq:nge2b}),
(\ref{eq:dGperp}), and (\ref{eq:delGperp}),
\begin {multline}
  \left[\frac{d\Gamma}{dx_\gamma}\right]_\perp
  \simeq
    \frac{\alpha^2}{2\pi^2} \, P_{e\to\gamma}(x_\gamma) \int_0^1 d\yfrak \>
  \Re\left\{
      i P_{\gamma\to e}(\yfrak) \, \Omegapr \,
        I_1\bigl(
          \tfrac{2|\Omega_0|}{|\Omegapr|} \,,\, \tfrac{\hpr_m}{\Omegapr}
        \bigr)
      + i\hpr_m \,
        I_2\bigl(
          \tfrac{2|\Omega_0|}{|\Omegapr|} \,,\, \tfrac{\hpr_m}{\Omegapr}
        \bigr)
  \right\}
\\
  - \frac{\alpha}{\pi} \,
     P_{e\to\gamma}(x_\gamma) \,
    \Re\Bigl\{
       ({\cal G}_\pair{+}i h_\gamma) \,
       \left[
         \psi\bigl(1{+}\tfrac{{\cal G}_\pair{+}i h_\gamma}{2i\Omega_0}\bigr)
         + \gammaE
       \right]
       -
       i h_\gamma
       \left[
          \psi\bigl(1{+}\tfrac{h_\gamma}{2\Omega_0}\bigr) + \gammaE
       \right]
    \Bigr\}
\label {eq:dGperpFinal}
\end {multline}
for $m_\gamma \not=0$.


\subsection{Final Result}

Our final result for $d\Gamma/dx_\gamma$ is
\begin {equation}
  \left[ \frac{d\Gamma}{dx_\gamma} \right]_\LPMplus
  =
  \left[ \frac{d\Gamma}{dx_\gamma} \right]_\LPM
  + \delta \! \left[\frac{d\Gamma}{dx_\gamma}\right]
  =
  \left[ \frac{d\Gamma}{dx_\gamma} \right]_\LPM
  +
  \left[\frac{d\Gamma}{dx_\gamma}\right]_\perp
  +
  \left[\frac{d\Gamma}{dx_\gamma}\right]_{\rm L}
\label {eq:TotalRate2}
\end {equation}
with (\ref{eq:LPMrate}) for the ordinary LPM bremsstrahlung rate,
(\ref{eq:dGperpFinal}) for $[d\Gamma/dx_\gamma]_\perp$, and
(\ref{eq:Lrate}) for $[d\Gamma/dx_\gamma]_{\rm L}$.
The full result is summarized in a slightly different but equivalent
way in the appendix of our companion paper \cite{softqed2}.


\subsection{Limits}
\label {sec:die+limit}

\begin{figure}
  \includegraphics[scale=0.9]{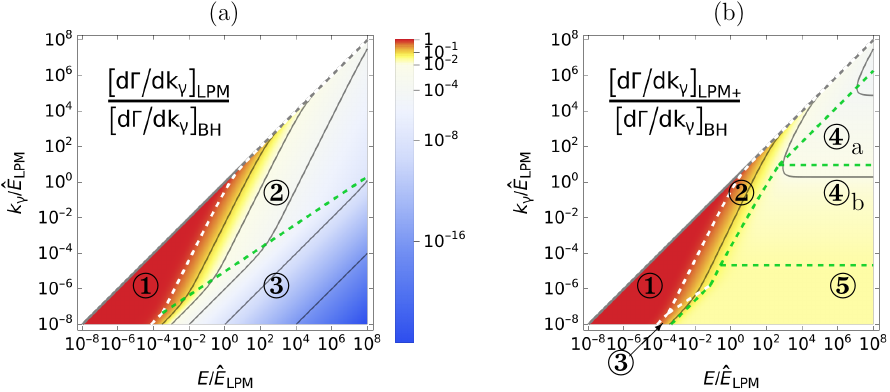}
  \caption{
     \label{fig:overBH}
     Like fig.\ \ref{fig:overBH_nodie} but here including the dielectric
     effect with $m_\gamma/m = 10^{-4}$ instead of $m_\gamma = 0$.
     Like previous figures in this paper,
     regions are labeled
     \cnum{1} Bethe-Heitler (BH),
     \cnum{2} deep LPM,
     \cnum{3} dielectric, and
     \cnum{4} deep $\protect\LPMplus$, but here we have a new region,
     \cnum{5} dielectric$\protect\Plus$.
     Note the non-uniform
     spacing $(10^{-1},10^{-2},10^{-4},10^{-8},10^{-16})$ of the contour lines.
  }
\end{figure}

Fig.\ \ref{fig:overBH} shows a plot of our final results presented
in ref.\ \cite{softqed2}, which we reproduce here
in order to discuss the new region 5, which we call the
``dielectric$\Plus$'' region, arising from having incorporated
the dielectric effect into our calculation of the $\LPMplus$ effect.

For $h_\gamma \ll |{\cal G}_\pair|$, our $m_\gamma{\not=}0$
result (\ref{eq:dGperpFinal}) for
$[d\Gamma/dx_\gamma]_\perp$ is very well approximated by the
$m_\gamma{=}0$ result
given by the sum of (\ref{eq:dGammaneq1}) and (\ref{eq:nge2b}).
Remember that the deep-$\LPMplus$ regime of our earlier
$m_\gamma{=}0$ discussion corresponded to $\Gamma_\pair \gg |\Omega_0|$
and so $|{\cal G}_\pair| \gg |\Omega_0|$.

To get an analogous limit for the case where the
dielectric effect becomes very important,
we now consider
\begin {subequations}
\label {eq:die+Conditions}
\begin {equation}
  h_\gamma \gg |{\cal G}_\pair|
\label {eq:bighgamma}
\end {equation}
with
\begin {equation}
  h_\gamma \gg |\Omega_0| .
\end {equation}
\end {subequations}
To further simplify the resulting limit, we now also strengthen our
parametric assumption that $m_\gamma \ll m$ to%
\footnote{
  Strictly speaking, the parametric assumption $m_\gamma \lesssim \alpha^{1/2} m$
  would be enough here.  However, we wrote $m_\gamma \ll \alpha^{1/2}m$
  just to emphasize that these inequalities are very well satisifed in
  the case of ordinary matter.
}
\begin {equation}
  m_\gamma \ll \alpha^{1/2} m .
\label {eq:mgammaAssumption}
\end {equation}
This continues to be an excellent approximation for
ordinary matter (for which $m_\gamma \lesssim 10^{-4} m$).%
\footnote{
  Parametrically, $m_\gamma \sim \sqrt{Z\alpha n/m}$ [see (\ref{eq:mgamma})] and
  $n \lesssim a_{\rm Bohr}^{-3} \sim (\alpha m)^3$ in ordinary matter,
  and so (parametrically)
  we always have $m_\gamma \lesssim Z^{1/2} \alpha^2 m$ in ordinary matter.
  However, the results in this paper are more general and only assume
  $m_\gamma \ll \alpha^{1/2} m$.
}
The usefulness of (\ref{eq:mgammaAssumption}) is that
combination with (\ref{eq:bighgamma}) gives $k_\gamma \ll \Elpma$,
which allows simplification of ${\cal G}_\pair$ to the limit (\ref{eq:Gpairlow}).
Also in that limit, the first term (the $\yfrak$ integral) of
(\ref{eq:dGperpFinal}) becomes (\ref{eq:neq1low}):
\begin {equation}
   \left[ \frac{d\Gamma}{dx_\gamma} \right]_{(n=1)}^{m_\gamma=0}
   \simeq
  \frac{\alpha}{2\pi} \, P_{e\to\gamma}(x_\gamma) \, \Gamma_\pair
  \Re
  \left[
    \ln \bigl( \tfrac{m^2}{4\kgamma\Omega_0} \bigr)
    + \gammaE
    + \tfrac53
  \right] .
\end {equation}
In the other term of (\ref{eq:dGperpFinal}),
the limits (\ref{eq:die+Conditions}) give
\begin {multline}
       ({\cal G}_\pair{+}i h_\gamma) \,
       \left[
         \psi\bigl(1{+}\tfrac{{\cal G}_\pair{+}i h_\gamma}{2i\Omega_0}\bigr)
         + \gammaE
       \right]
       -
       i h_\gamma
       \left[
          \psi\bigl(1{+}\tfrac{h_\gamma}{2\Omega_0}\bigr) + \gammaE
       \right]
\\
  \simeq
    {\cal G}_\pair \left[
      \ln \bigl( \tfrac{h_\gamma}{2\Omega_0} \bigr) + \gammaE + 1
    \right]
  \simeq
    \tfrac 12 \Gamma_\pair \left[
      \ln \bigl( \tfrac{h_\gamma}{2\Omega_0} \bigr) + \gammaE + 1
    \right] .
\end {multline}
Using $h_\gamma \equiv m_\gamma^2/2\kgamma$, (\ref{eq:dGperpFinal})
then simplifies to%
\footnote{
  An alternative way to derive (\ref{eq:dGperpDie+}) is
  to note that if $h_\gamma \gg |\Omega_0|$, then one may approximate
  (\ref{eq:dGbremDie}) as
  \[
   \left[ \frac{d{\cal G}}{dx_\gamma\,d(\Delta t_0)} \right]_\brem
   \longrightarrow~
   - \frac{\alpha}{2\pi} \,
     P_{e\to\gamma}(x_\gamma) \,
       \frac{ e^{-i h_\gamma \Delta t_0} }{ (\Delta t_0)^2 } \,.
  \]
  Then repeat the derivation of section \ref{sec:neq1},
  for which the integrations will now be more straightforward
  than they would be with (\ref{eq:dGbremDie}).
  A similar analysis will show that the $n\ge 2$ contributions of
  section \ref{sec:nge2} are parametrically suppressed in the
  limit (\ref{eq:die+Conditions}).
}
\begin {equation}
  \left[\frac{d\Gamma}{dx_\gamma}\right]_\perp
  \simeq
  \frac{\alpha}{2\pi} \, P_{e\to\gamma}(x_\gamma) \, \Gamma_\pair
  \left[
    \ln \bigl( \tfrac{m^2}{m_\gamma^2} \bigr)
    + \tfrac23
  \right] .
\label {eq:dGperpDie+}
\end {equation}

Since (\ref{eq:mgammaAssumption}) restricted the limit
(\ref{eq:die+Conditions}) to
$\kgamma \ll \Elpma$, we may use (\ref{eq:LrateLow}) for
$[d\Gamma/dx_\gamma]_{\rm L}$.  When the $\LPMplus$ correction
$\delta[d{\Gamma}{dx_\gamma}]$ dominates
over the ordinary LPM bremsstrahlung rate in (\ref{eq:TotalRate2}),
then
\begin {equation}
  \left[ \frac{d\Gamma}{dx_\gamma} \right]_\LPMplus \simeq
  \frac{\alpha}{2\pi} \, P_{e\to\gamma}(x_\gamma) \, \Gamma_\pair
  \left[
    \ln \bigl( \tfrac{m^2}{m_\gamma^2} \bigr)
    + \tfrac{20}{21}
  \right] .
  \qquad
  \mbox{(deep dielectric$\Plus$ region)}
\label {eq:deepDie+}
\end {equation}
This is the limit of our final result (\ref{eq:TotalRate2}) deep inside
region 5 of fig.\ \ref{fig:overBH}b.
As with other regions on this and previous plots,
our convention is that the dashed boundary lines surrounding region 5
in fig.\ \ref{fig:overBH}b are
drawn where the limiting formula (\ref{eq:deepDie+}) exactly equals
the corresponding limiting formula for the neighboring region.

\section{Physical origin of logarithms in deep-LPM+ regimes}
\label{sec:logs}

In all the regions of $(\kgamma,E)$ where our $\LPMplus$ result
is significantly different from the ordinary LPM rate
(regions 4$_{\rm a}$, 4$_{\rm b}$, and 5 of fig. \ref{fig:overBH}b), the limiting
formulas deep inside those regimes have logarithms of large
arguments.  From (\ref{eq:deep4a}), (\ref{eq:deep4b}), and
(\ref{eq:deepDie+}),
\begin {subequations}
\label {eq:deeprecap}
\begin {align}
  \left[\frac{d\Gamma}{dx_\gamma}\right]_{\LPMplus}
  &\simeq
  \frac{\alpha}{2\pi} \, P_{e\to\gamma}(x_\gamma) \,
  \Gamma_\pair \,
  \left[
    \ln\bigl(\tfrac{4\sqrt2 \,|\Omega_0|}{\pi x_\gamma \Gamma_\pair}\bigr)
    + \tfrac16 + \tfrac23\ln2
  \right]
  && \mbox{(deep $\LPMplus$ region 4$_{\rm a}$)} ,
\label {eq:deep4a2}
\\
  \left[\frac{d\Gamma}{dx_\gamma}\right]_{\LPMplus}
  &\simeq
  \frac{\alpha}{2\pi} \, P_{e\to\gamma}(x_\gamma) \,
  \Gamma_\pair
  \left[
    \ln \bigl( \tfrac{m^2/\kgamma}{\Gamma_\pair} \bigr)
    + \tfrac{41}{21}
  \right]
  && \mbox{(deep $\LPMplus$ region 4$_{\rm b}$)} ,
\label {eq:deep4b2}
\\
  \left[ \frac{d\Gamma}{dx_\gamma} \right]_\LPMplus
  &\simeq
  \frac{\alpha}{2\pi} \, P_{e\to\gamma}(x_\gamma) \, \Gamma_\pair
  \Re
  \left[
    \ln \bigl( \tfrac{m^2}{m_\gamma^2} \bigr)
    + \tfrac{20}{21}
  \right]
  && \mbox{(deep dielectric$\Plus$ region 5)} .
\label {eq:deepDie+2}
\end {align}
\end {subequations}
As discussed in ref.\ \cite{softqed1} for the first case (\ref{eq:deep4a2}),
these logarithms arise because (i) the differential energy
loss rate $d\Gamma/dx_\gamma$ gets contributions from bremsstrahlung
overlapping either real or virtual subsequent pair production and
(ii) the case of overlap with real pair production has
a collinear logarithmic enhancement associated with
in-medium LPM pair production $\gamma \to e^- e^+$ from a
photon originating from vacuum-like DGLAP splitting
$e \to e\gamma$ of the initial electron.
Specifically, ref.\ \cite{softqed1} (adapting a similar discussion
for QCD overlap effects in refs.\ \cite{seq,qcdNf}) argued that
the leading-log behavior of (\ref{eq:deep4a2}) should be
\begin {subequations}
\label {eq:LLOrate}
\begin {equation}
  \frac{d\Gamma}{dx_\gamma} \approx
    {\rm Prob}_{e\to\gamma}(x_\gamma) \, \Gamma_\pair
\end {equation}
with the probability of finding a photon in the initial electron being
\begin {multline}
  {\rm Prob}_{e\to\gamma}(x_\gamma)
  \approx
  \frac{\alpha}{2\pi} \, P_{e\to\gamma}(x_\gamma) \,
  \ln\Bigl( \frac{(p_\perp^\max)^2}{(p_\perp^\min)^2} \Bigr)
\\
  \approx
  \frac{\alpha}{2\pi} \, P_{e\to\gamma}(x_\gamma) \,
  \ln\biggl( \frac{(\Delta t)_0^\max}{(\Delta t)_0^\min} \biggr)
  \approx
  \frac{\alpha}{2\pi} \, P_{e\to\gamma}(x_\gamma) \,
  \ln\biggl( \frac{\tform^\brem}{\tform^\pair} \biggr) .
\label {eq:LLOdglap}
\end {multline}
\end {subequations}
So far, we have used the language of DGLAP splitting, analogous to parton
splitting in QCD.  But, in the context of
QED, ${\rm Prob}_{e\to\gamma}(x_\gamma)$ above is also often
referred to as the Weizs\"acker-Williams probability distribution for finding
a nearly-onshell photon in the initial electron.

Above, $p_\perp^{\rm max}$ and $p_\perp^{\rm min}$ are the limits of the
$p_\perp$ range that gives rise to a collinear logarithm, which are
related to the duration $(\Delta t)_0$ of the virtual photon by
$p_\perp^2 \propto (\Delta E)_\brem \sim 1/(\Delta t)_0$.
(See section 7 of ref.\ \cite{softqed1} for details.)
Because the electron continually receives kicks from the medium and
so cannot sustain a very small $p_\perp$ for a very long time,
the minimum $p_\perp$ corresponds to $\Delta t$ of order the
LPM bremsstrahlung formation time.  The maximum $p_\perp$ arises from
requiring $(\Delta E)_\brem \lesssim$ the off-shellness $(\Delta E)_\pair^\LPM$
of pair production
so that the photon virtuality does not suppress the LPM pair production rate.
That requirement for $\Delta E$ is equivalent to
$(\Delta t)_\brem \gtrsim \tform^\pair$.

In the deep-$\LPMplus$ region 4$_{\rm a}$, the
duration of LPM bremsstrahlung and LPM pair production are
$\tform^\brem \sim 1/\Gamma_\pair$ and
$\tform^\pair \sim 1/|\Omegapr|$ (democratic),
giving
\begin {align}
  \left[\frac{d\Gamma}{dx_\gamma}\right]_{\LPMplus}
  &\approx
  \frac{\alpha}{2\pi} \, P_{e\to\gamma}(x_\gamma) \,
  \Gamma_\pair \,
    \ln\bigl(\tfrac{|\Omegapr|~{\rm(democratic)}}{\Gamma_\pair}\bigr)
  && \mbox{(deep $\LPMplus$ region 4$_{\rm a}$)}
\end {align}
at leading log order, consistent with the more precise formula
(\ref{eq:deep4a2}) given that $\Omegapr \sim \Omega_0/x_\gamma$ for
the democratic pair production $\yfrak(1{-}\yfrak) \sim 1$ that dominates
the total LPM pair production rate $\Gamma_\pair$.

For region 4$_{\rm b}$,
it is the same except that the $k_\gamma{\ll}\Elpma$ limit of
LPM pair production is Bethe-Heitler pair production, whose time scale
is $1/\hpr_m$ (democratic) $\sim \kgamma/m^2$.
In terms of our review of the calculation of LPM pair production
in section \ref{sec:LPMpair}, this
time scale arises because the
$e^{-i\hpr_m\,\Delta t}$ factor in (\ref{eq:ZpairHO2})
cuts off the $(\Delta t)_\pr$ integration there.
Using $\tform^\brem \sim 1/\Gamma_\pair$ and $\tform^\pair \sim \kgamma/m^2$
in (\ref{eq:LLOrate}) then reproduces the leading-log behavior of the
more accurate result (\ref{eq:deep4b2}).

Region 5 is like region 4$_{\rm b}$ above except that
now the $p_\perp^{\rm min}$ of the
bremsstrahlung photon is cut off by the dielectric effect instead
of by the $\p_\perp$ kicks that the initial electron receives from the medium.
In this region, the $e^{-i h_\gamma \Delta t_0}$ factor in
(\ref{eq:dGbremDie}) is what cuts off the bremsstrahlung time
$(\Delta t)_0$.  As discussed in section \ref{sec:die+limit},
the dominance of $h_\gamma$ defines the region,
and so we take $\tform^\brem \sim 1/h_\gamma \sim \kgamma/m_\gamma^2$
in (\ref{eq:LLOrate}).
Since this region also has $k_\gamma \ll \Elpma$, thanks to
(\ref{eq:mgammaAssumption}), we also have the Bethe-Heitler limit
$\tform^\pair \sim 1/\hpr_m \sim \kgamma/m^2$ for pair production.
Eq.\ (\ref{eq:LLOrate}) then reproduces the leading-log behavior of the
more accurate result (\ref{eq:deepDie+2}).

The range of inverse time scales $(\Delta E)_\brem \sim 1/(\Delta t)_0$
that contribute above are
also summarized for regions 4$_{\rm a}$,
4$_{\rm b}$ and 5 by the entries of table \ref{tab:scales}
containing explicit formulas.

In the literature, the process where real in-medium pair production is
initiated by the vacuum-like photon content of the initial electron
is sometimes called ``\textit{direct} pair production'' $e\to ee\bar e$.%
\footnote{
  See, for example, section VI.A of ref.\ \cite{GerhardtKlein}.
}
Including the LPM effect, this rate was estimated
by Baier and Katkov (BK) \cite{BaierKatkov}.
Their estimate did not include the modification
$\tform \sim 1/\Gamma_\pair$ to the LPM bremsstrahlung formation time
that is important in our deep-$\LPMplus$ regions
$4_{\rm a}$ and $4_{\rm b}$, nor include the medium-induced photon mass
$m_\gamma$ that is important in our dielectric$\Plus$ region 5.
However, we discuss in appendix \ref{app:BK} that one may,
at leading-log order, compare their estimate to
our result for the \textit{small} $\LPMplus$ correction we find
in the deep ordinary-LPM region 2 of fig.\ \ref{fig:overBH}b,
specifically for the case $k_\gamma \ll \Elpma$.
In that case, $\tform^\brem \sim 1/|\Omega_0|$ and
$\tform^\pair \sim 1/\hpr_m$ (democratic), so that the
logarithm (\ref{eq:LLOrate}) is
\begin {multline}
  \left[\frac{d\Gamma}{dx_\gamma}\right]_{\LPMplus}
  -
  \left[\frac{d\Gamma}{dx_\gamma}\right]_ \LPM
  \approx
  \frac{\alpha}{2\pi} \, P_{e\to\gamma}(x_\gamma) \,
  \Gamma_\pair \,
    \ln\bigl(\tfrac{m^2/\kgamma}{|\Omega_0|}\bigr)
\\
  \mbox{(deep $\LPM$ region 2 for $\kgamma \ll \Elpma$)} .
\label {eq:ourBKlog}
\end {multline}
The argument of our logarithm above differs parametrically from that
of Baier and Katkov.  Appendix \ref{app:BK} discusses this difference,
whose source we believe may be a misidentification by Baier and Katkov
of the parametric
scale for the minimum virtuality of the photon contributing to
their logarithm.  We should emphasize that our own
parametric estimates of logarithms in this section are offered only
for the sake of physical understanding of our results at leading-log order
in various limits.
Our arguments for identifying the
parametric origin of the upper and lower scales of the logarithms
in the limiting cases of (\ref{eq:deeprecap}) played no role
in our full calculation
(\ref{eq:TotalRate2}) of the $\LPMplus$ rate.


\section{Influence of photo-nuclear cross-section on the LPM+ effect}
\label {sec:photonuclear}

\subsection {Overview}
\label {sec:gamAoverview}

In this paper, we have analyzed the possibility that LPM bremsstrahlung
$e{\to}e\gamma$ is disrupted by subsequent
pair production from the bremsstrahlung photon
before the bremsstrahlung is completed.  But there are other possible
ways for a photon to disappear besides electron pair production
$\gamma{\to}e\bar e$ in the Coulomb field of a nucleus.
In particular, the high-energy photon could collide
directly with the nucleus itself ($\gamma A \to$ hadrons).
Since the nucleus is a very small target, such
photo-nuclear processes are generally negligible compared to
electron pair production.  For instance, up to a logarithm, the
Bethe-Heitler cross-section for high-energy pair production is parametrically
\begin {equation}
   \sigma_\pair \sim \alpha (Z \alpha)^2 \times \pi r_m^2
   \sim \frac{Z^2 \alpha^3}{m^2} ,
\label {eq:sigmaBHpair}
\end {equation}
where $r_m \sim 1/m$ is the Compton wavelength of the electron
and ``$\pi r_m^2$'' roughly represents the corresponding cross-sectional area
(centered on the nucleus).  The
corresponding rate is $\Gamma_\pair \sim n\sigma_\pair$ which
is the behavior of (\ref{eq:pairBH}).%
\footnote{
  When comparing parametrically to (\ref{eq:pairBH}),
  use (\ref{eq:qhat}) for $\qhat$ and ignore logarithms.
}
In
contrast, up to logarithms, the high-energy photo-nuclear cross-section is
parametrically
\begin {equation}
  \sigma_{\gamma A} \sim A^\nu \sigma_{\gamma \rm p}
  \sim \alpha A^\nu \times \pi R_{\rm p}^2 ,
\label {eq:sigmagA}
\end {equation}
where $A^\nu$ is
the scaling of the photo-nuclear cross-section with
atomic weight $A$ (to be discussed),
$\sigma_{\gamma \rm p}$ is the $\gamma$-proton cross-section,
and $R_{\rm p}$ is the proton radius.
Because $R_{\rm p}$ is \textit{very} small compared to $r_m$, (\ref{eq:sigmagA})
is negligible compared to (\ref{eq:sigmaBHpair}).

The estimate (\ref{eq:sigmaBHpair}) applies only when
the LPM effect is negligible, which for pair production is when
$\kgamma \ll \Elpma$.  At higher photon energies, the pair production
cross-section and rate are LPM suppressed.
Specifically, in the $\kgamma \gg \Elpma$ limit (\ref{eq:pairmassless}),
\begin {equation}
  \Gamma_\pair^\LPM \simeq 
  \frac{3\alpha}{8} \sqrt{ \frac{\qhat}{\kgamma} }
  \qquad \mbox{($\kgamma\gg\Elpma$)} ,
\label {eq:pairmassless2}
\end {equation}
which falls with photon energy as $\kgamma^{-1/2}$.
Many years ago, Gerhardt and Klein \cite{GerhardtKlein}
pointed out that, for large enough $\kgamma$, the LPM pair
production rate will be so suppressed that the
photo-nuclear rate then becomes the dominant mechanism for disrupting
LPM bremsstrahlung.  They estimated that this takes place for
$\kgamma \gtrsim 10^{20}$ eV,%
\footnote{
  At such extreme energies, one should also consider overlapping
  $e \to e\gamma \to e\mu\bar\mu$.
  See footnote \ref{foot:Nf=1}.
}
roughly independent of the
target material (for solid targets).  They were focused on ice
($\Elpma = 303$ TeV \cite{GerhardtKlein}), for which
$\kgamma/\Elpma$ would be $\gtrsim 3\times10^5$, which would affect our
$\LPMplus$ results near the top of fig.\ \ref{fig:overBH}b.
For gold ($\Elpma = 2.5$ TeV), their estimate suggests that the
photo-nuclear rate would only be important at the extreme top of
the figure.  For context, the highest-energy cosmic ray of any type
ever observed was estimated to have energy around $3\times 10^{20}$ eV
\cite{ohmygod}.

Gerhardt and Klein went on to analyze how the photo-nuclear effect
would modify Galitsky and Gurevich's \cite{Galitsky}
original framework for estimating the effect of pair production
on the ordinary LPM bremsstrahlung rate.  That analysis will need
to be redone since our argument, in \cite{softqed1,softqed2}
and the current paper, is that Galitsky and Gurevich's conclusion
was qualitatively backward.  We have found that disruption of LPM bremsstrahlung
\textit{enhances} the bremsstrahlung rate compared to the original
LPM bremsstrahlung calculation, whereas Galitsky and Gurevich argued
that it further suppressed the original LPM rate.

We will not incorporate the photo-nuclear rate into our calculations in
this paper. Our purpose here is simply to alert readers to its importance
at extremely large values of $k_\gamma/\Elpma$, where its disruption of
LPM bremsstrahlung will make the
bremsstrahlung rate larger than the $\LPMplus$ rate we have calculated.
But we will
give an independent estimate of how large $k_\gamma$ must be for
photo-nuclear processes to be important, using a fit \cite{HalzenPhotoNuclear}
of the $\gamma$p cross-section based on
$\ln^2\!s$ (Froissart-bound saturated)
behavior at asymptotically high energy.


\subsection {Estimating importance of photo-nuclear cross-section}

Comparing (\ref{eq:pairmassless2}) with
the photo-nuclear rate
$\Gamma_{\gamma A} = n\sigma_{\gamma A}$, the photo-nuclear
rate becomes important for $k \gtrsim \kc$ with%
\footnote{
  In this section, we explicitly write $\me$ instead of simply ``$m$''
  to avoid confusion with the proton mass.
  Also, for this purpose, we roughly estimate $\Elpma \simeq \Elpm$ as in the
  ``extreme LPM suppression'' limit of (\ref{eq:Elpmcases}).
}
\begin {equation}
  \kc = \frac{9\alpha^2\qhat}{64(n\sigma_{\gamma A})^2}
      \simeq \frac{9\alpha^2\me^4}{64(n\sigma_{\gamma A})^2\Elpma} .
\label {eq:kc}
\end {equation}
For the high-energy $\gamma$p cross-section, we take a fit by Block and
Halzen \cite{HalzenPhotoNuclear} of experimental data to the form
\begin {subequations}
\label {eq:siggp}
\begin{equation}
  \sigma_{\gamma\rm p}(k_\gamma) =
  c_0
  + c_1 \ln\left( \frac{\kgamma}{\mp} \right)
  + c_2 \ln^2\left( \frac{\kgamma}{\mp} \right)
  + \beta_{{\cal P}'} \left( \frac{\kgamma}{\mp} \right)^{\mu-1} ,
\label {eq:sigpFit}
\end {equation}
where $\kgamma \simeq s/2\mp$ is the photon momentum in the proton rest
frame, $\mp$ is the proton mass, $\sqrt{s}$ is (as usual) the
center-of-mass energy, and the ``Regge intercept'' is $\mu=0.5$.
Their fit extracted from both high-energy data and matching onto the
\pagebreak[4] 
low-energy resonance region is%
\footnote{
  This is ``Fit 1'' of ref.\ \cite{HalzenPhotoNuclear}, which constrains
  both $c_0$ and $\beta_{{\cal P}'}$ from low-energy data.  They also have
  a ``Fit 2'' that constrains only $c_0$, which is not significantly
  different for our purposes.  (To test the presence of double-log energy
  growth, they also present a third fit that sets
  the coefficient $c_2$ to zero, which
  fits the high-energy data poorly.)
}
\begin {equation}
   c_0 = 105.64~\mu{\rm b} , \quad
   c_1 = -4.74 \pm 1.17~\mu{\rm b} , \quad
   c_2 = 1.17 \pm 0.16~\mu{\rm b} , \quad
   \beta_{{\cal P}'} = 64.0~\mu{\rm b} .
\end {equation}
\end {subequations}
Since we are interested only in rough estimates, we will ignore the
errors quoted on the fit of $c_1$ and $c_2$.
The $c_0$ term in (\ref{eq:sigmagA})
is the same order of magnitude as the crude
parametric estimate $\sigma_{\gamma p} \sim \alpha \pi R_{\rm p}^2$ that we
used in (\ref{eq:sigmagA}), whereas the $c_2 \ln^2(\kgamma/\mp)$ term
represents eventual double-log growth of the cross-section at very
high energy.

At the extreme energies relevant to our discussion here, the double
log term $c_2 \ln^2(\kgamma/\mp)$ will dominate $\sigma_{\gamma\rm p}$.
That limit is associated with the black-disk model of high-energy
hadronic collisions.%
\footnote{
  For readers unfamiliar with the black-disk model, we can suggest
  a few examples of introductory references that we found useful.
  For a textbook discussion of the basic model of a black disk
  with radius $R$, see section 4.12 of Perkins (2nd ed.)\ \cite{Perkins}.
  Important properties of black-disk scattering are that
  the total elastic and inelastic scattering cross-sections are equal
  and so each are half of the total cross-section, and that the
  real part of the forward scattering amplitude vanishes.
  (In our case, those properties would refer to the black
  disk model of the meson-nucleus cross-section after the photon has
  fluctuated into a meson.)
  For an extremely qualitative picture of why the
  effective radius can grow logarithmically with energy at high
  energies, see the short ``Intuitive Reasoning'' introduction of
  Froissart's Scholarpedia article \cite{Froissart} on the
  Froissart bound \cite{FroissartBound}.
  In ref.\ \cite{BlockHalzenpp}, Block and Halzen
  argue that measurements of $\rm pp$
  cross-sections at the LHC and Auger 
  confirm that
  the proton is a black disk at asymptotically high energy.
}
The photon can behave as a hadron by temporarily fluctuating
into a virtual $\rho^0$ meson (or more generally a $q\bar q$ pair)
before hitting the nucleon,
as in the picture of vector meson
dominance.  A black-disk picture of $\rho$\kern1pt p collisions
at very high energy
suggests a black-disk picture for $\rho$\kern1pt -nuclear collisions, and
one might then guess that the photo-nuclear cross-section scales
roughly as $A^{2/3}$.
For our rough estimates, we will take%
\footnote{
  There can also be direct photo-quark interactions where the photon
  does not fluctuate into a meson but instead interacts directly with a quark
  inside the nucleus.
  By themselves, one might
  expect such interactions to scale roughly proportional to $A$ but not
  be associated with the double-log growth of the cross-section with energy.
  Overall, Gerhardt and Klein
  \cite{GerhardtKlein} took $\sigma_{\gamma A}$ to scale as
  $A^{0.887}$ based on work by Engel \textit{et al.}\ \cite{Engel}.
  The latter combined contributions from both
  vector meson dominance and
  direct-quark processes but was focused on comparison to data
  at photon energies much smaller than in the application here.
}
\begin {equation}
   \sigma_{\gamma A} \simeq A^{2/3} \sigma_{\gamma\rm p}
\label {eq:Ascaling}
\end {equation}
at extremely high energy.

Let $\rho \simeq A \mp n$ be the mass density of the material.
Combining (\ref{eq:kc}) and (\ref{eq:Ascaling}) into
\begin {equation}
  \kc = \frac{9\alpha^2\qhat}{64[n A^{2/3}\,\sigma_{\gamma\rm p}(\kc)]^2}
  \simeq
  \frac{9\alpha^2\me^4\mp^2}{64[\rho A^{-1/3}\,\sigma_{\gamma\rm p}(\kc)]^2 \Elpma}
  \,,
\end {equation}
using (\ref{eq:siggp}), and solving for $\kc$,
we estimate that photo-nuclear processes become important for the
$\LPMplus$ effect at roughly the photon energies shown in table \ref{tab:kc}.
The order of magnitude estimate $\kc \sim 10^{20}$ eV for ice is very
similar to Gerhardt and Klein.

\pagebreak[4] 


\begin {table}[t]

\setlength{\tabcolsep}{4pt}
\begin {tabular}{lrrccrl}
\toprule
  & \multicolumn{1}{c}{$\Elpma$}
  & \multicolumn{1}{c}{$\rho$}
  & \multicolumn{1}{c}{$A$}
  &
  & \multicolumn{2}{c}{$k_{\rm c}$}
\\
\cline{6-7}
  ice  & 303 TeV & 0.9 gm/cm$^3$ & ${\sim}16$ &
       & $1.6 \times 10^{20}$ eV & $\simeq 5\times 10^5 \Elpma$ \\
  gold & 2.5 TeV & 19.3 g/cm$^3$ & 197 &
       & $2.2 \times 10^{20}$ eV & $\simeq 9\times 10^7 \Elpma$ \\
  air (STP)
       & 234 PeV & 0.0012 g/cm$^3$ & ${\sim}14$ &
       & $5 \times 10^{22}$ eV & $\simeq 2\times 10^5 \Elpma$ \\
\botrule
\end {tabular}
\caption{%
\label{tab:kc}%
  The photo-nuclear effect will be important for the $\protect\LPMplus$
  bremsstrahlung rate when
  $k \gtrsim \kc$, where $\kc$ above is based on the rough estimate
  of the photo-nuclear cross-section by (\ref{eq:siggp}) and
  (\ref{eq:Ascaling}).  Because this is not a precise calculation,
  we have not bothered to carefully separate out and combine the
  contributions of
  different elements in the cases of ice and air.
  Instead we have just used monatomic formulas, taking
  $A$ to be a single value roughly that of the heavier abundant
  elements (which are the ones that most influence $\qhat$, $\Elpma$,
  and the mass density $\rho$).
}
\end {table}


\section{Qualitative discussion of media with finite thickness}
\label {sec:finiteL}

Our companion paper \cite{softqed2} concludes with a very brief
qualitative statement about how we expect our results for $\LPMplus$ enhancement
of the LPM rate will be modified if the medium has a finite thickness $L$.
Here we explain our reasoning (with the caution that it is based
on purely qualitative arguments).

\begin{figure*}
  \includegraphics[scale=1.0]{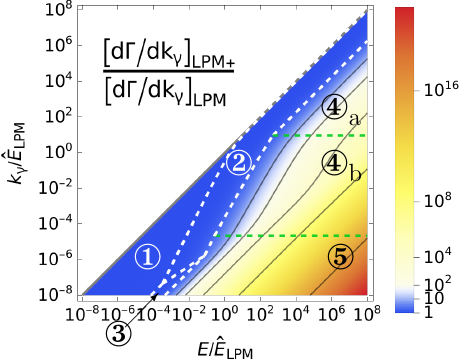}
  \caption{
     \label{fig:LPM+overLPM}
     The enhancement factor $\Sxtra \equiv {\protect\LPMplus}/\LPM$
     of the $\protect\LPMplus$ rate compared to the LPM rate
     for the infinite-volume case analyzed in this paper.
  }
\end{figure*}

Following ref.\ \cite{softqed2},
we define the enhancement factor $\Sxtra$ as the ratio
$\LPMplus$/LPM of the $\LPMplus$ rate over the LPM rate.
In the infinite-volume case, this is the ratio of fig.\ \ref{fig:overBH}b
to fig.\ \ref{fig:overBH}a and is shown here for reference in
fig.\ \ref{fig:LPM+overLPM}.  
We will focus on the transition
between the deep-LPM region 2 and the deep-$\LPMplus$ regions 4.

Ref.\ \cite{softqed1} reviews qualitative arguments%
\footnote{
  See section 2 of ref.\ \cite{softqed1}.
}
that the LPM suppression
of bremsstrahlung can be roughly understood (parametrically) as
the blurring together of all small-angle elastic collisions of the
electron with the medium into a single opportunity for bremsstrahlung,
in contrast to Bethe-Heitler where each collision is treated as
an independent opportunity for bremsstrahlung; that this causes
the LPM rate
to be suppressed compared to Bethe-Heitler by roughly the number of
collisions $\Delta N_\coll$ during the quantum duration $t_{\rm form}$
of the bremsstrahlung process; and so that
LPM/BH $\sim 1/{\Delta N}_\coll \propto 1/\tform^\LPM$.
We should clarify that, parametrically, the relevant
``collisions'' in this discussion are
(neglecting effects contributing only to logarithms) individual
collisions with the medium that have momentum transfers
$\gtrsim m$.%
\footnote{
  See section 2.1.2 of ref.\ \cite{softqed1}.
}
The mean free path for such collisions is
$\tau_{\el,m} \sim m^2/\qhat$.

Ref.\ \cite{softqed1} then argued that the effect of pair production
is to limit the duration of bremsstrahlung to at most
$1/\Gamma_\pair$, so that the number of collisions blurred together and
the duration of bremsstrahlung (outside of the
Bethe-Heitler regime) is
\begin {equation}
  \Delta N_\coll^\LPMplus
  \sim
  \frac{\tform^\LPMplus}{\tau_{\el,m}} \sim
  \frac{ \min( \tform^\LPM, \Gamma_\pair^{-1} ) }{ \tau_{\el,m} }.
\label {eq:tformLPMplus}
\end {equation}
Then the $\Sxtra \equiv \LPMplus$/LPM enhancement factor of
$\LPMplus$ over LPM is
\begin {equation}
   \Sxtra \sim \frac{(\Delta N^\LPMplus)^{-1}}{(\Delta N^\LPM)^{-1}}
   \sim \frac{ (\tform^\LPMplus)^{-1} }{ (\tform^\LPM)^{-1} }
   \sim \max( 1 , \Gamma_\pair \tform^\LPM ).
\label {eq:Sxtra}
\end {equation}
That qualitative expectation has been verified by the explicit
calculations in ref.\ \cite{softqed1} and this paper.

What then happens to the qualitative picture
if the medium has finite width $L$ with vacuum outside?
Then the number of collisions that can
be blurred together by the LPM effect cannot exceed the total number
of collisions over distance $L$, and so
(\ref{eq:tformLPMplus}) should be replaced by roughly
\begin {equation}
  \Delta N_\coll^\LPMplus
  \sim
  \frac{ \min( \tform^\LPM , \Gamma_\pair^{-1}, L ) }{ \tau_{\el,m} }
\label {eq:blur+}
\end {equation}
where $\tform^\LPM$ is still the $L{=}\infty$ LPM formation
time (\ref{eq:tformLPM}), and where we implicitly assume that
$L \gg \tau_{\el,m}$ so that there is at least one collision
with the medium.  If we ignore the possibility of pair production,
then (\ref{eq:blur+}) would instead be%
\footnote{%
  From (\ref{eq:blur}),
  the suppression of the LPM rate compared to the Bethe-Heitler rate
  is
  LPM/BH $\sim 1/(\Delta N_\coll^\LPM) \sim \tau_{\el,m} / \min( \tform^\LPM, L)$,
  which is LPM/BH $\sim \tau_{\el,m} / L$ in the case $L \ll \tform^\LPM$.
  Up to factors of logarithms, this is parametrically consistent with
  the more-detailed finite-$L$ LPM analysis of ref.\
  \cite{ShulgaFomin}.
  [Compare the $\gamma\vartheta{\gg}1$ case of their (3.4) for LPM
  with their (2.8) for BH.]
  We also found useful the summary of this result by
  eq.\ (13) of ref.\ \cite{CERNlpm2}, where
  their $\kappa$ means $(\LPM/\BH)^{-1}$ and their symbol ``$k_\gamma$''
  means $4\pi L/\alpha X_0$,
  which is $\sim L/\tau_{\el,m}$ (up to logarithms)
  in our notation.
}
\begin {equation}
  \Delta N_\coll^\LPM
  \sim
  \frac{ \min( \tform^\LPM, L ) }{ \tau_{\el,m} } \,,
\label {eq:blur}
\end {equation}
and so (\ref{eq:Sxtra}) becomes
\begin {equation}
  \Sxtra \sim 
  \frac{
    \min( \tform^\LPM,L )
  }{
    \min( \tform^\LPM , \Gamma_\pair^{-1}, L )
  }
  = \max\!\left( 1 , \, \Gamma_\pair \, \min(\tform^\LPM, L) \right).
\label {eq:SxtraL}
\end {equation}

\pagebreak[4] 


\subsubsection{Case: $L \ll 1/\Gamma_\pair$}

In the case that $L \ll 1/\Gamma_\pair$, pair production of the bremsstrahlung
photon is unlikely, and so there will be no significant $\LPMplus$ effect,
meaning
that $\Sxtra$ is always very close to 1.


\subsubsection{Case: $L \gg 1/\Gamma_\pair$}

In this case, (\ref{eq:SxtraL}) gives that
the condition $\Sxtra \gg 1$ for being deep in the $\LPMplus$
regime will simply be $\tform^\LPM \gg 1/\Gamma_\pair$,
which (up to logarithms) is
the \textit{same} condition as in the $L{=}\infty$ case analyzed in this
paper.  So, the LPM$|\LPMplus$ boundary between
region 2 and region $4_{\rm a}{+}4_{\rm b}$ in
fig.\ \ref{fig:LPM+overLPM} will not change significantly.%
\footnote{
  We will not attempt here to sharpen the parametric estimates of this section
  to cover the finite-$L$ dependence of the collinear \textit{logarithms}
  appearing in the
  $2|4_{\rm a}$ and $2|4_{\rm b}$ entries of table \ref{tab:boundaries}.
}
However, there will be a ceiling
\begin {equation}
  \Sxtra^{\rm max} \sim \Gamma_\pair L
\end {equation}
on the growth of $\Sxtra \equiv \LPMplus$/LPM in fig.\ \ref{fig:LPM+overLPM}.

For $L \gg 1/\Gamma_\pair$, the $\LPMplus$ rate itself is unaffected
deep in region 4.  The ceiling on
the ratio $\Sxtra \equiv \LPMplus/\LPM$ arises only because
of the $L$-dependence of the denominator.  Consequently, the boundary
$4_{\rm b}|5$ between the deep-$\LPMplus$ and dielectric$\Plus$ regions in
fig.\ \ref{fig:LPM+overLPM} will not change significantly.


\subsubsection{Case: $L \sim 1/\Gamma_\pair$}

This is like the $L \gg 1/\Gamma_\pair$ case but $\Sxtra$ cannot grow
extremely large compared to 1.
In the $\LPMplus$ region of fig.\ \ref{fig:LPM+overLPM}, the
$\LPMplus$ correction to $\LPM$ would be neither very large nor
negligibly small.


\section{Radiation \lowercase{v}.\ ionization
  for small energy losses \boldmath$\ll E$}
\label {sec:collEloss}


At high energy $E$ (parametrically $E \gg m/Z\alpha$ up to logarithms),
the total rate of electron energy loss
is dominated by bremsstrahlung
as opposed to non-radiative (collisional) energy loss
via elastic scattering from atomic electrons.
But the comparison is subtler if one asks more specifically
about the \textit{differential}
rate $d\Gamma/d\omega$ to lose energy $\omega \ll E$.
Here, $\omega$ represents the $k_\gamma$ of the radiated photon
in the case of radiative energy loss
but represents the energy transfer to an atomic electron
in the non-radiative case.  In this notation,
the total energy loss rate $dE/dz$ (where $z$ is distance) is
\begin {equation}
  -\frac{dE}{dz}
  = \int d\omega \> \omega \, \frac{d\Gamma}{d\omega} \,.
\end {equation}

In discussing the $\LPMplus$ effect, we have only been concerned with
high-energy brem\-sstrah\-lung $\kgamma \gg m$, and so only collisional
energy loss for $\omega \gg m$ is relevant for comparison.
For such large energy transfer, the atomic electrons may be approximated
as free electrons at rest.
The corresponding collision rate is then given by the
ultra-relativistic limit of the M{\o}ller electron-electron
differential cross-section in the rest frame of one electron,
which is
\begin {equation}
  \left[ \frac{d\sigma}{d\omega} \right]_{ee\to ee}
  \simeq \frac{2\pi \alpha^2}{m\omega^2} \,,
\end {equation}
and so%
\footnote{
  $d\Gamma/d\omega$ may also be extracted from the appropriate
  high-energy limit
  of formulas from the literature on collisional energy loss derived from
  the M{\o}ller cross-section.  For instance, differentiating eq.\ 11 of
  Uehling \cite{Uehling} with respect to the lower cut-off $\eta$ on
  energy transfer in that formula gives, in Uehling's notation
  but here setting $\hbar{=}c{=}1$, that
  $-dE/dz\,d\eta = -2\pi n_e e^4/m\beta^2\eta$.
  Taking the high energy limit and realizing that
  Uehling works in units where $e^2 = \alpha$, that result in our notation is
  $\omega\,d\Gamma/d\omega \simeq 2\pi \alpha^2 n_e/m\omega$,
  equivalent to (\ref{eq:collEsigma0}).
}
\begin {equation}
  \left[ \frac{d\Gamma}{d\omega} \right]_\coll
  \simeq \frac{2\pi \alpha^2 n_e}{m\omega^2}
  = \frac{2\pi Z \alpha^2 n}{m\omega^2} \,.
\label {eq:collEsigma0}
\end {equation}
In our discussions of bremsstrahlung, we have characterized the medium
in terms of $\Elpm$ and $m_\gamma/m$.  Even though the $ee$ scattering
rate has nothing to do with the medium-induced mass $m_\gamma$ of the
photon, it depends on the same combination $Z\alpha^2 n$, and
we can simplify comparisons with bremsstrahlung formulas by
using (\ref{eq:mgamma}) to rewrite (\ref{eq:collEsigma0}) as
\begin {equation}
  \left[ \frac{d\Gamma}{d\omega} \right]_\coll
  \simeq \frac{\alpha m_\gamma^2}{2\omega^2} \,.
\label {eq:collErate}
\end {equation}

\begin{figure}
  \includegraphics[scale=0.9]{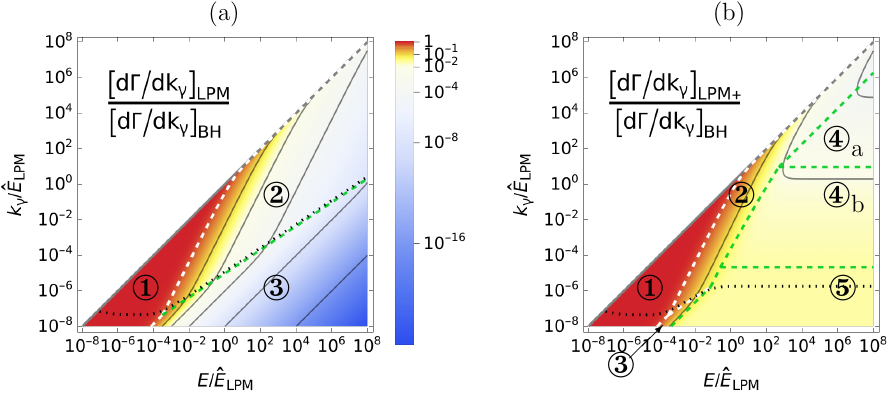}
  \caption{
     \label{fig:overBH-coll}
     The same as fig.\ \ref{fig:overBH} but here
     including black dotted curves below
     which the differential rate $[d\Gamma/d\omega]_\coll$ for collisional
     energy loss $\omega$ exceeds (a) the $\LPM$ rate
     $[d\Gamma/d\kgamma]_\LPM$, or (b) the $\protect\LPMplus$ rate
     $[d\Gamma/d\kgamma]_{\protect\LPMplus}$,
     for radiative energy loss $\kgamma{=}\omega$.
     Like previous figures,
     regions are labeled
     \cnum{1} Bethe-Heitler (BH),
     \cnum{2} deep LPM,
     \cnum{3} dielectric,
     \cnum{4} deep $\protect\LPMplus$, and
     \cnum{5} dielectric$\protect\Plus$.
  }
\end{figure}

So, for instance, comparing to the soft ($x_\gamma\ll 1$) limit of the
Bethe-Heitler rate (\ref{eq:BHrate}),
\begin {equation}
  \left[ \frac{d\Gamma}{d\omega} \right]_\BH \simeq 
  \frac{2\alpha \qhat}{3\pi m^2 \omega} \,
\label{eq:BHrateSoft}
\end {equation}
gives%
\footnote{
  Using (\ref{eq:mgamma}), the Bethe-Heitler limit $\Elpm \simeq 2\Elpma$
  from (\ref{eq:Elpmcases}), the standard definition
  $\Elpma = \alpha m^2 X_0/4\pi$ \cite{SpencerReview,RPP2024}, and the
  approximation $X_0 \simeq [4nZ^2\alpha^3 \ln(184 Z^{-1/3}) / m^2]^{-1}$,
  our (\ref{eq:COLLvBH}) above is equivalent to saying that, \textit{in
  the Bethe-Heitler region},
  bremsstrahlung energy loss dominates over collisional energy loss
  when $\omega \ge 3\pi m/[8 Z\alpha \ln(184 Z^{-1/3})]$.
  As a practical matter, this is satisfied for the case of
  high-energy bremsstrahlung $\kgamma \gg m$ (and so energy loss
  $\omega \gg m$) analyzed in this paper.
}
\begin {equation}
  \left[ \frac{d\Gamma}{d\omega} \right]_\coll \ge
  \left[ \frac{d\Gamma}{d\omega} \right]_\BH
  \quad \mbox{when} \quad
  \omega \le \frac{3\pi}{4} \left( \frac{m_\gamma}{m} \right)^2 \Elpm .
\label {eq:COLLvBH}
\end {equation}
Given that $m_\gamma/m \lesssim 10^{-4}$ for ordinary matter, this corresponds
to extremely small $\omega/\Elpm$, near the very bottom of the Bethe-Heitler
regions of fig.\ \ref{fig:overBH-coll}.

\pagebreak[4] 

A more interesting comparison arises if we compare the collisional differential
rate (\ref{eq:collErate}) to the limiting radiative rate (\ref{eq:deepDie+})
deep
inside the dielectric$\Plus$ region (region 5 of fig.\ \ref{fig:overBH-coll}b):
\begin {equation}
  \left[ \frac{d\Gamma}{d\omega} \right]_{\rm die+} \simeq
  \frac{\alpha}{2\pi E} \, P_{e\to\gamma}(x_\gamma) \, \Gamma_\pair
  \left[
    \ln \bigl( \tfrac{m^2}{m_\gamma^2} \bigr)
    + \tfrac{20}{21}
  \right]
  \simeq
  \frac{7\alpha^2\qhat}{18 \pi^2 m^2 \omega}
  \left[
    \ln \bigl( \tfrac{m^2}{m_\gamma^2} \bigr)
    + \tfrac{20}{21}
  \right] .
\end {equation}
This gives
\begin {equation}
  \left[ \frac{d\Gamma}{d\omega} \right]_\coll \ge
  \left[ \frac{d\Gamma}{d\omega} \right]_{\rm die+}
  \quad \mbox{when} \quad
  \omega \le
  \frac{ 9 \pi^2 }{ 7 \alpha }
  \left( \frac{m_\gamma}{m} \right)^2
  \,
  \frac{
    \Elpm
  }{
    \ln \bigl( \frac{m^2}{m_\gamma^2} \bigr)
    + \frac{20}{21}
  } \,.
\label {eq:COLLvDIE+}
\end {equation}
Compare this to the boundary between the $\LPMplus$ region
$4_{\rm b}$ and dielectric$\Plus$ region 5 of fig.\ \ref{fig:overBH}b.
That boundary can be estimated by equating the corresponding limiting
radiation rate formulas (\ref{eq:deepDie+}) and (\ref{eq:deep4b}),
giving%
\footnote{
  See table II of ref.\ \cite{softqed2}.
}
\begin {equation}
  \kgamma \sim \left(\frac{m_\gamma}{m}\right)^2 \frac{\Elpma}{\alpha}
  \qquad
  \mbox{($4_{\rm b}|5$ boundary).}
\end {equation}
Parametrically, this value of $\kgamma$ exceeds the upper limit on
$\omega$ in (\ref{eq:COLLvDIE+}) by a large logarithm, and so
the region (\ref{eq:COLLvDIE+}) of significant collisional differential
energy loss should be solidly in the dielectric$\Plus$ region.

To verify this conclusion, we have used the full LPM and $\LPMplus$ results
(rather than just limiting formulas deep inside various regions) to
add a black-dotted curve to fig.\ \ref{fig:overBH-coll} showing exactly
where $[d\Gamma/d\omega]_\coll$ equals radiative $d\Gamma/d\kgamma$.
In total, we see from the figure that the differential rate for
collisional energy loss dominates
over that for radiative energy loss (a) throughout the dielectric region of the
original LPM calculation and (b) in part of what we identify as the
dielectric$\Plus$ region of our $\LPMplus$ calculation.


\begin{acknowledgments}

We are grateful to Spencer Klein for alerting us to the role of
$\gamma A \to$ hadrons (section \ref{sec:gamAoverview})
at the very highest energies,
and to Francis Halzen for discussion of
his past work with Martin Block on analyzing the growth of the
total $\gamma$p cross section with energy.
The work of Arnold and Bautista was supported,
in part, by the National Science Foundation under Grant No.~2412362.
Elgedawy was supported at different times by the
National Natural Science Foundation of China under Grant No.\ 12447145
and by the European Union (ERC, QGPthroughEECs, grant agreement
No.\ 101164102).
Views and opinions expressed are however those of the authors
only and do not necessarily reflect those of the European Union or the European
Research Council. Neither the European Union nor the granting authority can be
held responsible for them.

\end{acknowledgments}

\medskip
\textit{Data availability}---Mathematica code that generates numbers
used in the plots is openly available \cite{data}.

\appendix

\section{Derivation of the effective photon mass from
  relativistic forward Compton scattering}
\label {app:mgamma}

\subsection{Background}

In the somewhat different context of an ionized plasma, the plasma
frequency is the lowest possible frequency for propagating electromagnetic
waves, given by
\begin {subequations}
\label {eq:plwave0}
\begin {equation}
  \omega_{\rm pl} \simeq \sqrt{ \frac{4\pi\alpha n_e}{m} } \,,
\label {eq:plasma0}
\end {equation}
where $n_e$ is the number
density of ionized electrons and $m$ is the electron mass.
The dispersion relation for
transversely-polarized electromagnetic waves is approximately
\begin {equation}
  \omega^2 \simeq \k^2 + \omega_{\rm pl}^2
  \qquad \mbox{($\perp$ polarization)},
\label {eq:dispersion0}
\end {equation}
\end {subequations}
and so $\omega_{\rm pl}$ is equivalent to
an effective mass for transversely-polarized photons.

For photon energies large compared to the binding energies of
the inner-most atomic electrons
[parametrically $\omega \gg (Z\alpha)^2m$ in the rest frame of the medium],
all atomic electrons are effectively free.
Then, keeping to the single-element case for simplicity,
(\ref{eq:plwave0})
applies with $n_e = Z n$, where $n$ is the number density of atoms.

In the context of this paper, we apply (\ref{eq:dispersion0}) at
extremely high photon energies.  One might wonder whether
(\ref{eq:dispersion0}) remains valid for
photon energies so high ($\omega \gg m$) that absorbing and re-emitting
the photon would make the electron temporarily relativistic.
Similarly, readers with a background in
ultra-relativistic plasmas may know that, in that context,
the effective
photon mass $m_{\gamma,\infty}$ of a
high-energy transverse photon ($|\k| \gg \omega_{\rm pl}$)
is $m_{\gamma,\infty} = \omega_{\rm pl} \sqrt{3/2}$ and not
$m_{\gamma,\infty} = \omega_{\rm pl}$.%
\footnote{
\label{foot:PiSigns}
  For textbook discussions, see, for example, section 6.7 of
  ref.\ \cite{KapustaGale} or sections 6.2.3 and 6.3.2 of
  ref.\ \cite{LeBellac}.  But beware that there are a wide variety of
  overall sign conventions for both $\Pi^{\mu\nu}$ and $\eta^{\mu\nu}$
  in zero- and finite-temperature gauge theory.
  We've attempted to be consistent in the convention we use here.
}
In this appendix, we discuss how to calculate the dispersion relation
in both the low-energy and
high-energy limits, for both non-relativistic and relativistic media,
by calculating the forward scattering amplitude for Compton scattering
$\gamma e \to \gamma e$.  We will see explicitly how a difference between
$m_{\gamma,\infty}$ and $\omega_{\rm pl}$ arises for relativistic media
but not for non-relativistic media.


\subsection{Derivation}

\begin {figure}[t]
\begin {center}
  \includegraphics[scale=0.7]{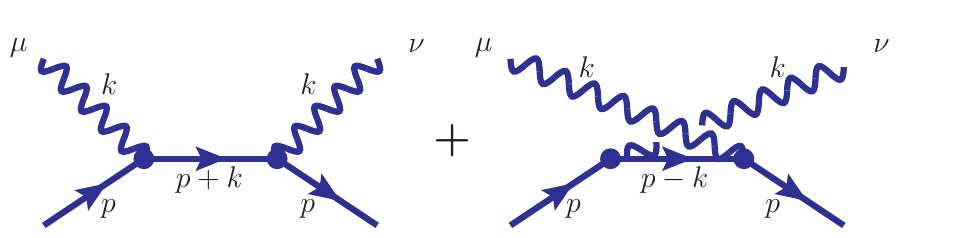}
  \caption{
     \label{fig:compton}
     Forward Compton scattering amplitude.
  }
\end {center}
\end {figure}

The dispersion relation for photons is%
\begin {equation}
  \bigl( k^2 \delta^\mu_\nu - k^\mu k_\nu - \Pi^\mu_\nu(k) \bigr) \vareps^\nu
  = 0,
\label {eq:disp}
\end {equation}
where $k^\mu=(\omega,\k)$, $\vareps$ is photon polarization,
and $\Pi^{\mu\nu}$ is the photon self-energy.
We use $({+}{-}{-}{-})$ sign convention
for the flat-space metric $\eta_{\mu\nu}$, in which case
$k^2 = \omega^2 - \k^2$ and the sign convention
for $\Pi^\mu_\nu$ in (\ref{eq:disp})
is the same as the sign convention would be for a photon mass term
$m_\gamma^2$. 
The medium's contribution to the self-energy is given by the forward-scattering
amplitude for scattering from charges in the medium, as depicted in
fig.\ \ref{fig:compton}.  (In the case of a non-relativistic medium,
Compton scattering will be dominated by the lowest-mass scatterers, which are
the atomic electrons.)  More specifically, $i \Pi^{\mu\nu}$ is given by
(i) multiplying the amputated forward scattering diagrams
by the phase-space density $f(\p)$ of electrons (and/or positrons)
to scatter from, and (ii) integrating over $\p$ with relativistic
phase-space measure.  That gives
\begin {equation}
  i\Pi^{\mu\nu}(k) =
  (-i e)^2 \int \frac{d^3p}{(2\pi)^3 2E_\p} \> f(\p) \,
  \bar u_\p
  \left[
    \gamma^\nu \, \frac{i}{\slashed{p}{+}\slashed{k}{-}m} \, \gamma^\mu
    +
    \gamma^\mu \, \frac{i}{\slashed{p}{-}\slashed{k}{-}m} \, \gamma^\nu
  \right]
  u_\p .
\end {equation}

The initial and final electron must have the same spin for a
forward-scattering process, and one may average over that spin to obtain
\begin {equation}
  \Pi^{\mu\nu}(k) =
  - \frac{e^2}{2} \int \frac{d^3p}{(2\pi)^3 2E_\p} \> f(\p)
  \left[
    \frac{\tr[ (\slashed{p}{+}m)
               \gamma^\nu (\slashed{p}{+}\slashed{k}{+}m) \gamma^\mu ]}
         {2p\cdot k + k^2}
    +
    (k \to -k; \mu\leftrightarrow\nu)
  \right],
\label {eq:Pi1}
\end {equation}
where the fact that the initial electron is on shell ($p^2{=}m^2$)
has been used
to expand the denominator $(p{+}k)^2-m^2$.  After taking the trace in the
numerator, the two terms in (\ref{eq:Pi1}) may be combined over a common
denominator and the result manipulated into the form
\begin {equation}
  \Pi^{\mu\nu}(k) =
  2e^2 \int \frac{d^3p}{(2\pi)^3 2E_\p} \> f(\p)
  \left\{
     \eta^{\mu\nu}
     +
     \frac{
        4 [ k^2 p^\mu p^\nu - (p^\mu k^\nu {+} k^\mu p^\nu)p\cdot k ]
        + \eta^{\mu\nu} (k^2)^2
     }{(2p\cdot k)^2 - (k^2)^2}
  \right\} .
\label {eq:PiGeneral}
\end {equation}

We now specialize to an isotropic medium, $f(\p) = f(|\p|)$.
We also focus on $\Pi^\mu_\nu$ for transverse photon polarizations.
By rotational invariance, $\Pi^\mu_\nu$ is independent of the direction
of the polarization in the plane perpendicular to $\k$, and its value
$\PiT$ may be obtained by
averaging over the basis of transverse polarizations.
Then (\ref{eq:PiGeneral}) gives
\begin {subequations}
\label {eq:dispT}
\begin {equation}
  \PiT(k) \equiv \Pi^1_1(k) = \Pi^2_2(k) =
  2e^2 \int \frac{d^3p}{(2\pi)^3 2E_\p} \> f(|\p|)
  \left\{
     1
     -
     \frac{
        k^2 (2\bpT^2 - k^2 )
     }{(2p\cdot k)^2 - (k^2)^2}
  \right\} ,
\label {eq:PiT}
\end {equation}
where here we take 1 and 2 to be the two spatial directions transverse
to $\k$, and $\bpT$ is the corresponding transverse momentum of the
electron.%
\footnote{
  In principle, one should carefully distinguish between
  (i) transverse to the direction of $\k$ (subscript ``T'' here) vs.\
  (ii) transverse to the $z$ axis
  (``$\perp$'' as in (\ref{eq:Heff}) here or the use of
  light-cone gauge $A^{+}{=}0$ in appendix D of ref.\ \cite{softqed1}).
  However, since our calculations in this paper
  ultimately
  choose the $z$ axis to be the direction $\k$ of the soft photon,
  the distinction will not be critical for this work.
}
The transverse-polarization photon dispersion relation is
\begin {equation}
  \omega^2 = \k^2 + \PiT(\omega,\k) .
\end {equation}
\end {subequations}

Remembering (i) that we are interested in solving the dispersion relation for
$k^2 \sim \omega_{\rm pl}^2 \ll m^2$ and (ii) that $\omega \ge \omega_{\rm pl}$,
the denominator in
(\ref{eq:PiT}) is \textit{generically}%
\footnote{
  We use the word ``generically'' because, if both
  the medium is ultra-relativistic
  \textit{and} $|\k| \gg \omega_{\rm pl}$ so that $\omega \simeq |\k|$,
  then $p\cdot k$ can be much smaller than (\ref{eq:estdenom}) in the limit
  that $\p$ and $\k$ are collinear.  But this is also a very small
  portion of the integration region for $\p$ for which $\p_{\rm T}$ in
  the numerator is also small, and it
  gives a parametrically
  suppressed contribution to the integral.
}
of order
\begin {equation}
  (2p\cdot k)^2 - (k^2)^2 \simeq (2p\cdot k)^2 \sim (E_\p \omega)^2
\label {eq:estdenom}
\end {equation}
and the numerator has
\begin {equation}
  |k^2(2\bpT^2-k^2)| \lesssim
  \max\bigl((E_\p \omega_{\rm pl})^2,\omega_{\rm pl}^4\bigr) .
\end {equation}
In the large-$\k$ limit ($\omega \simeq |\k| \gg \omega_{\rm pl}$),
eqs.\ (\ref{eq:dispT}) will then give
$\omega^2 \simeq \k^2 + m^2_{\gamma,\infty}$ with simply
\begin {equation}
  m^2_{\gamma,\infty} \simeq
  2e^2 \int \frac{d^3p}{(2\pi)^3 2E_\p} \> f(|\p|) .
\label {eq:mplasma}
\end {equation}
In contrast, in the low-energy limit $|\k| \ll \omega_{\rm pl}$,
the $\bpT^2$ term in (\ref{eq:PiT}) is possibly important, giving
$\omega^2 \simeq \k^2 + \omega^2_{\rm pl}$ with
\begin {equation}
  \omega^2_{\rm pl} \simeq
  2e^2 \int \frac{d^3p}{(2\pi)^3 2E_\p} \> f(|\p|) \,
  \left( 1 - \frac{\bpT^2}{2E_\p^2} \right)
  =
  2e^2 \int \frac{d^3p}{(2\pi)^3 2E_\p} \> f(|\p|) \,
  \left( 1 - \tfrac12 (\v_\p)_{\rm T}^2 \right) ,
\end {equation}
where $\v_\p \equiv \p/E_\p$ is the velocity of the
initial electron.  The angular integration allows
us to replace $1{-}\tfrac12 (\v_\p)_{\rm T}^2$ by its angular
average, and so
\begin {equation}
  \omega^2_{\rm pl} \simeq
  2e^2 \int \frac{d^3p}{(2\pi)^3 2E_\p} \> f(|\p|) \,
  \left( 1 - \tfrac13 |\v_\p|^2 \right) ,
\label {eq:wplasma}
\end {equation}

For a non-relativistic medium, the velocity $\v_\p$ of the initial electron
(in natural units) is negligible, and so
(\ref{eq:mplasma}) and (\ref{eq:wplasma}) give simply
$m_{\gamma,\infty} \simeq \omega_{\rm pl}$.
Also, $E_\p \simeq m$ and $f(\p) \simeq n_e \, (2\pi)^3 \delta^{(3)}(\p)$ for
the purpose of (\ref{eq:mplasma}), reproducing the usual
non-relativistic result (\ref{eq:plwave0}) even for very large photon energy.

In contrast, an ultra-relativistic medium
has $|\v_\p| \simeq 1$, and
so (\ref{eq:mplasma}) and (\ref{eq:wplasma}) give
$\omega_{\rm pl}^2 \simeq \tfrac23 m_{\gamma,\infty}^2$.
In that case,
the effective transverse photon mass is different in the low-energy
vs.\ high-energy limit.%
\footnote{
  For an ultra-relativistic gas of photons, electrons and positrons, our
  $f(\p)$ in the above discussion would be
  $f(\p) = 4 n_{\rm F}(|\p|)$,
  where $n_{\rm F}(E) = 1/(e^{\beta E}{+}1)$
  is the Fermi distribution, one factor of 2 is for spin, and the other
  includes positrons.
  This reproduces the standard results $\omega_{\rm pl} = e T/3$ and
  $m_{\gamma,\infty} = e T/\sqrt{6}$.
}


\subsection{Alternate Forms}

Finally, we make contact between our formula (\ref{eq:PiGeneral})
and related formulas for $\Pi^{\mu\nu}(k)$ we are familiar with from
the literature on ultra-relativistic QCD plasmas (without assuming
isotropy). Note that in the limits that we took, relevant
to solving the dispersion relation, the
$(k^2)^2$ terms in the numerator and denominator in 
(\ref{eq:PiGeneral}) could be ignored.  So replace (\ref{eq:PiGeneral})
by the approximation
\begin {equation}
  \Pi^{\mu\nu}(k) \simeq
  2e^2 \int \frac{d^3p}{(2\pi)^3 2E_\p} \> f(\p)
  \left\{
     \eta^{\mu\nu}
     +
     \frac{
        4 [ k^2 p^\mu p^\nu - (p^\mu k^\nu {+} k^\mu p^\nu)p\cdot k ]
     }{(2p\cdot k)^2}
  \right\} .
\label {eq:PiGeneralApprox}
\end {equation}
This can be rewritten more compactly as
\begin {equation}
  \Pi^{\mu\nu}(k) \simeq
  2e^2 \int \frac{d^3p}{(2\pi)^3 2E_\p} \> f(\p)
  \left\{
     \eta^{\mu\nu}
     -
     k \cdot\partial_p
     \frac{ p^\mu p^\nu }{p\cdot k}
  \right\} .
\end {equation}
It is also possible to integrate once ``by parts'' to
put this answer into the form%
\footnote{
  See the discussion of the relation between eqs.\ (4.1) and (4.2) of
  ref.\ \cite{AMYekt}, including the accompanying footnote there.
}
\begin {align}
  \Pi^{\mu\nu}(k) &\simeq
  -2e^2 \int \frac{d^3p}{(2\pi)^3 2E_\p} \>
  \frac{\partial f(\p)}{\partial p^i}
  \left\{
     p^\mu \eta^{i\nu}
     -
     \frac{k^i p^\mu p^\nu}{p\cdot k}
  \right\}
\nonumber\\
  &=
  -e^2 \int \frac{d^3p}{(2\pi)^3} \>
  \frac{\partial f(\p)}{\partial p^i}
  \left\{
     v_\p^\mu \eta^{i\nu}
     -
     \frac{k^i v_\p^\mu v_\p^\nu}{v\cdot k}
  \right\} ,
\label {eq:PiApprox3}
\end {align}
where $v^\mu \equiv (1,\v_\p)$ is a convenient short-hand notation even
though it is not a Lorentz 4-vector.
Eq.\ (\ref{eq:PiApprox3})
is a form that arises naturally from solving (linearized) Vlasov equations
for collisionless plasmas, which treat electromagnetic fields as
classical and the charges as (otherwise non-interacting)
classical particles.%
\footnote{
  See, for example, the review in ref.\ \cite{ALM} leading to
  that reference's eq.\ (2.8), or the earlier QCD discussion
  (requiring care in comparison of normalizations) of
  eq.\ (13) of ref.\ \cite{MrowThoma}.
  Note the warning about sign conventions in our footnote \ref{foot:PiSigns}.
}


\pagebreak[4] 

\section{Including helicity-flip amplitude in vertex factors}

\subsection{bremsstrahlung}
\label{app:BremWithFlip}

The vertex factors in (\ref{eq:Zrate0})
come from the usual relativistic matrix element
for high-energy nearly-collinear splitting $e \to e\gamma$:\,%
\footnote{
  For a textbook derivation, see eq.\ (17.92) of
  ref.\ \cite{Peskin}.  Remember that we are implicitly summing over
  final-state polarizations, and whether or not one then also averages over
  initial-state polarization makes no difference here.
}
\begin {equation}
  \Bigl|
  \langle -\P_\perp,(1{-}x_\gamma)E;\P_\perp,x_\gamma E
     | \delta H | {\bm 0},E \rangle_{\rm rel}
  \Bigr|^2
  = \frac{2e^2 P_\perp^2}{x_\gamma(1{-}x_\gamma)} \, P_{e\to\gamma}(x_\gamma) ,
\label {eq:calMrel}
\end {equation}
written here with the conventional choice that the $z$ axis is chosen
to be in the direction of the initial electron.
Following appendix C of ref.\ \cite{softqed1}, we've chosen to
write the state of a particle with momentum $\p$
as $|\p_\perp,p_z\rangle$, and the subscript ``rel'' above means that
relativistic normalization has been used for the states
(which differs from the non-relativistic normalization of states
used everywhere else in this paper by a factor of
$\sqrt{2\omega_\p} \simeq \sqrt{2p_z}$ for each particle).
We have labeled the transverse momentum of the final-state photon
and electron as $\pm\P_\perp$.  However, since
we find it convenient to choose the $z$ axis
as the direction of the photon instead of the initial electron,
we must perform a (small) rotation of (\ref{eq:calMrel}) to get
$\langle\p_\perp,(1{-}x_\gamma)E;{\bm 0},x_\gamma E|
    \delta H | \p_\perp,E \rangle_{\rm rel}$.
This replaces
\begin {equation}
  \P_\perp \longrightarrow x_\gamma \p_\perp
\label{eq:pi_to_p}
\end {equation}
on the right-hand side of (\ref{eq:calMrel}).
See ref.\ \cite{softqed1} for details.

Including the helicity-flip amplitude in the calculation
of (\ref{eq:calMrel}), one finds
\begin {equation}
  \Bigl|
  \langle -\P_\perp,(1{-}x_\gamma)E;\P_\perp,x_\gamma E
     | \delta H | {\bm 0},E \rangle_{\rm rel}
  \Bigr|^2
  = \frac{2e^2}{x_\gamma(1{-}x_\gamma)} \,
    \left[
       P_\perp^2 \, P_{e\to\gamma}(x_\gamma) + x_\gamma^3 m^2
    \right] ,
\end {equation}
corresponding to the replacement
\begin {equation}
  P_\perp^2 \, P_{e\to\gamma}(x_\gamma) \longrightarrow
  P_\perp^2 \, P_{e\to\gamma}(x_\gamma) + x_\gamma^3 m^2
\end {equation}
in (\ref{eq:calMrel}).
Now making the change (\ref{eq:pi_to_p}), so that $\k_{\gamma\perp}=0$,
gives the replacement
\begin {equation}
  p_\perp^2 \, P_{e\to\gamma}(x_\gamma) \longrightarrow
  p_\perp^2 \, P_{e\to\gamma}(x_\gamma) + x_\gamma m^2 .
\end {equation}
As discussed in ref.\ \cite{softqed1}, the $p_\perp^2$ above is really
a dot product $\p_\perp\cdot(\p_\perp')^*$ of the $\p_\perp$'s associated
with the vertex in the amplitude and the vertex in the conjugate amplitude;
these are different because we are in medium and $\p_\perp$ can
change over time; and the $\b$-space version corresponds to replacing
$p_\perp^2$ above by $\grad_\b\cdot\grad_{\b'}$, which gives
the replacement (\ref{eq:PwithFlip}) quoted in the main text.


\subsection{pair production}
\label{app:PairWithFlip}

For the case of pair production $\gamma\to e\bar e$, the relativistic matrix
element analogous to (\ref{eq:calMrel}) is%
\footnote{
  For a textbook derivation, see eq.\ (17.117) of
  ref.\ \cite{Peskin}.
}
\begin {equation}
  \Bigl|
  \langle \p_\perp,\yfrak\kgamma;-\p_\perp,(1{-}\yfrak)\kgamma
     | \delta H | {\bm 0},\kgamma \rangle_{\rm rel}
  \Bigr|^2
  = \frac{2e^2 p_\perp^2}{\yfrak(1{-}\yfrak)} \, P_{\gamma\to e}(\yfrak)
\end {equation}
when masses are ignored.
Since the usual convention takes the initial particle (in this case
the photon) to have transverse momentum zero, the above already represents
our desired choice $\k_{\gamma\perp}{=}0$ for the $z$ axis, and so no extra
rotation (and so no modification to the transverse momenta) is necessary.
Including the helicity-change amplitude in the calculation, one finds
\begin {equation}
  \Bigl|
  \langle \p_\perp,\yfrak\kgamma;-\p_\perp,(1{-}\yfrak)\kgamma
     | \delta H | {\bm 0},\kgamma \rangle_{\rm rel}
  \Bigr|^2
  = \frac{2e^2}{\yfrak(1{-}\yfrak)} \,
    \left[
       p_\perp^2 \, P_{\gamma\to e}(\yfrak) + m^2
    \right] .
\end {equation}
This is the source of (\ref{eq:pairPwithFlip}).


\section{Integrals}
\label {app:integrals}

This appendix derives some of the integrals used in the main text.


\subsection{Building blocks}

As we will see, the difficult integrals needed in the main text can
be related to the following two building blocks:
\begin {subequations}
\label {eq:fgints_nosub}
\begin {equation}
  \int_0^\infty d\tau \> \tau^\beta e^{-\lambda\tau} \coth\tau
  = 2^{-\beta} \, \Gamma(1{+}\beta) \,
       \zeta\bigl(1{+}\beta,\tfrac{\lambda}{2}\bigr)
    - \lambda^{-1-\beta} \, \Gamma(1{+}\beta) ,
\label {eq:f0int_nosub}
\end {equation}
\begin {equation}
    \int_0^\infty d\tau \> \tau^\beta e^{-\lambda\tau} \csch\tau
  = 2^{-\beta} \, \Gamma(1{+}\beta) \,
      \zeta\bigl(1{+}\beta,\tfrac12{+}\tfrac{\lambda}{2}\bigr) ,
\end {equation}
\end {subequations}
where $\zeta(s,q)$ is the Hurwitz $\zeta$ function
$\sum_{k=0}^\infty (k+q)^{-s}$.
One can find these integrals in tables, such as Gradshteyn and Ryzhik
\cite{GR} (3.551.3) and (3.552.1), but they are also easy to derive
by rewriting
\begin {equation}
  \coth\tau = \frac{e^{\tau}+e^{-\tau}}{e^\tau-e^{-\tau}}
   = \frac{1+e^{-2\tau}}{1-e^{-2\tau}}
   = 1 + 2 \sum_{n=1}^\infty e^{-2n\tau}
\end {equation}
or
\begin {equation}
  \csch\tau = \frac{2}{e^\tau-e^{-\tau}}
   = \frac{2 e^{-\tau}}{1-e^{-2\tau}}
   = 2 \sum_{n=1}^\infty e^{-(2n-1)\tau}
\end {equation}
in (\ref{eq:fgints_nosub}) and then integrating term by term.

Using (\ref{eq:fgints_nosub}), we also find the subtracted integrals
\begin {subequations}
\label {eq:fgints}
\begin {multline}
  f_0(\beta;\lambda) \equiv
  \int_0^\infty d\tau \> \tau^\beta \left[ \coth\tau - \tau^{-1} \right]
      e^{-\lambda\tau}
\\
  = 2^{-\beta} \, \Gamma(1{+}\beta) \,
       \zeta\bigl(1{+}\beta,\tfrac{\lambda}{2}\bigr)
    - \lambda^{-1-\beta} \, \Gamma(1{+}\beta)
    - \lambda^{-\beta} \, \Gamma(\beta) ,
\label {eq:f0int}
\end {multline}
\begin {equation}
  g(\beta;\lambda) \equiv
    \int_0^\infty d\tau \> \tau^\beta e^{-\lambda\tau}
      \left[ \csch\tau - \tau^{-1} \right]
  = 2^{-\beta} \, \Gamma(1{+}\beta) \,
      \zeta\bigl(1{+}\beta,\tfrac12{+}\tfrac{\lambda}{2}\bigr)
    - \lambda^{-\beta} \, \Gamma(\beta) .
\label {eq:gint}
\end {equation}
\end {subequations}

Using (\ref{eq:f0int}), we can also evaluate the integral
\begin {subequations}
\begin {equation}
  f(\beta;\lambda) \equiv
  \int_0^\infty d\tau \> \tau^\beta \left[ \csch^2\tau - \tau^{-2} \right]
      e^{-\lambda\tau}
\label {eq:fdef}
\end {equation}
by rewriting it as
${-}\int u\>dv$ with $u=\tau^\beta e^{-\lambda\tau}$
and $v = \coth\tau - \tau^{-1}$, and then integrate by parts to get
\begin {equation}
  f(\beta;\lambda) =
  \int_0^\infty d\tau \> (\beta \tau^{\beta-1} - \lambda \tau^\beta)
     \left[ \coth\tau - \tau^{-1} \right] e^{-\lambda\tau}
  =
  \beta f_0(\beta{-}1,\lambda) - \lambda \, f_0(\beta,\lambda)
\end {equation}
\end {subequations}

In what follows, we will need to know the
behavior of $\zeta(s,q)$ for two different limits of $s$.
The $s{\to}1$ behavior was given by (\ref{eq:Hurwitz}) as
\begin {equation}
  \zeta(1{+}\eps,q) =
  \frac{1}{\eps} - \psi(q) - \gamma_1(q) \, \eps + O(\eps^2) .
\end {equation}
The $s{\to}0$ behavior is%
\footnote{
  Eq.\ (\ref{eq:zetaseps}) follows, for example, from
  eqs.\ (9.531) and (9.533.3) of Gradshteyn and Ryzhik \cite{GR}
  (with $n{=}0$ in the first).
}
\begin {equation}
  \zeta(\eps,q) =
  \bigl(\tfrac12 - q\bigr)
    + \bigl[ \lnGamma(q) - \tfrac12 \ln(2\pi) \bigr] \eps + O(\eps^2) .
\label {eq:zetaseps}
\end {equation}

We then find the following useful limits:
\begin {subequations}
\label {eq:fglimits}
\begin {align}
  f_0(0,\lambda) &=
    - \lambda^{-1}
    -\psi\bigl(\tfrac{\lambda}{2}\bigr)
    + \ln\bigl(\tfrac{\lambda}{2}\bigr)
  = \lambda^{-1} - \PSI\bigl(1;\tfrac{\lambda}{2}\bigr) ,
\label {eq:f0zero}
\\
  g(0,\lambda) &=
    -\psi\bigl(\tfrac12{+}\tfrac{\lambda}{2}\bigr)
    + \ln\bigl(\tfrac{\lambda}{2}\bigr)
  \equiv -\PSI\bigl(\tfrac12;\tfrac{\lambda}{2}\bigr) ,
\label {eq:gzero}
\\
  f(0,\lambda) &= - \lambda \, f_0(0,\lambda)
  = -1 + \lambda\,\PSI\bigl(1;\tfrac{\lambda}{2}) ,
\label {eq:fzero}
\\[5pt]
  f_0(-1,\lambda) &=
  2 \LNGAMMA\kern-0.5pt\bigl( 0; \tfrac{\lambda}{2} \bigr) ,
\label {eq:f0minus1}
\\
  g(-1,\lambda) &=
  2 \LNGAMMA\kern-0.5pt\bigl( \tfrac12; \tfrac{\lambda}{2} \bigr) ,
\\[5pt]
  f_0(\eps,\lambda) &=
    \left[ 1 - \eps(\ln2+\gammaE) \right] f_0(0,\lambda)
    - \eps \, \bar{\bar\gamma}_1\bigl(0;\tfrac{\lambda}{2}\bigr)
    + O(\eps^2) ,
\\
  g(\eps,\lambda) &=
    \left[ 1 - \eps(\ln2+\gammaE) \right] g(0,\lambda)
     - \eps \bar\gamma_1\bigl(\tfrac12;\tfrac{\lambda}{2}\bigr)
    + O(\eps^2) ,
\label {eq:geps}
\\
  f(\eps,\lambda) &=
    \eps\,f_0(-1,\lambda) - \lambda\,f_0(\eps,\lambda)
    + O(\eps^2) .
\label {eq:feps}
\end {align}
\end {subequations}
Above,
\begin {subequations}
\label{eq:overlines}
\begin {equation}
  \LNGAMMA(r;z)
  \equiv
  \lnGamma(r{+}z)
  - 
  \left[
      \bigl(r{+}z{-}\tfrac12\bigr) \ln z
      - z + \tfrac12\ln(2\pi)
  \right]
\label {eq:LNGAMMA}
\end {equation}
can be thought of as $\lnGamma(r{+}z)$ minus
its large-$z$ behavior.  It behaves like $\LNGAMMA(z)\to 0$ as $|z| \to \infty$.
Similarly, the generalized Stieltjes coefficient $\gamma_1(z)$ appears
in our results in the combination
\begin {equation}
  \bar\gamma_1(r;z)
  \equiv
  \gamma_1(r{+}z)
  + \tfrac12 \ln^2 z ,
\label {eq:bargamma1}
\end {equation}
which has $\bar\gamma_1(z) \to 0$ as $|z| \to \infty$.
[See appendix \ref{app:asymptotic} for the large-$z$ expansion of
$\gamma_1(z)$.]
Finally, some results above involve the further-subtracted
version
\begin {equation}
  \bar{\bar\gamma}_1(r;z)
  \equiv
  \bar\gamma_1(r;z)
  + (r{-}\tfrac12) \, \frac{\ln z}{z} ,
\label {eq:barbargamma1}
\end {equation}
\end {subequations}
for which $z\,\bar{\bar\gamma}_1(z) \to 0$ as
$|z|\to\infty$.
Note that $\bar{\bar\gamma}_1(\tfrac12; z) = \bar\gamma_1(\tfrac12; z)$.

\pagebreak[3] 

We've introduced the subtracted notation (\ref{eq:overlines})
in order to make it clearer how results behave for large argument
$\lambda$, which will be relevant when we discuss numerics
in appendix \ref{app:numerics}.
Additionally, we simply think that it is interesting that
the integrals we need are given by various special functions with
the large-argument behavior subtracted away.%
\footnote{
  Some recursion relations for our subtracted special functions are
  \[
    \PSI(r{+}1;z) = \PSI(r;z) + \frac{1}{r{+}z} \,,
    \qquad
    \LNGAMMA(r{+}1;z) = \LNGAMMA(r;z) + \ln(r{+}z) - \ln z ,
  \]
  \[
    \bar{\bar\gamma}_1(r{+}1;z) =
    \bar{\bar\gamma}_1(r;z) - \frac{\ln(r{+}z)}{r{+}z} + \frac{\ln z}{z} \,.
  \]
  We originally chose to write the LPM bremsstrahlung rate (\ref{eq:LPMrate})
  in terms of $\PSI(1;z)$ and $\PSI(\tfrac12;z)$ --- rather than, for example,
  $\PSI(0,z)$ and $\PSI(\tfrac12;z)$ --- because then
  the $m{=}0$ result is cleanly separated from the $m{\not=}0$ corrections
  in (\ref{eq:LPMrate}).  Having made that choice, we have then also written our
  $\LPMplus$ corrections in terms of $\PSI(1;z)$ and $\PSI(\tfrac12;z)$
  as well, as in (\ref{eq:I12}).
}


\subsection{eqs.\ (\ref{eq:intLObrem})}
\label{app:intLObrem}

The integral on the left-hand side of (\ref{eq:bremint1}) is
\begin {equation}
  \int_0^\infty dt \>
    \left[ \Omega^2 \csc^2(\Omega t) - \frac{1}{t^2} \right] e^{-i h t} .
\end {equation}
Changing integration variable to $\tau \equiv i\Omega t$ then gives
\begin {equation}
  i\Omega \int_0^\infty d\tau \>
    \left[ \csch^2\tau - \frac{1}{\tau^2} \right] e^{-h\tau/\Omega}
  = i\Omega \, f\bigl(0,\tfrac{h}{\Omega})
  = -i\Omega + i h \, \PSI(1;\tfrac{h}{2\Omega}) ,
\end {equation}
which is the final result of (\ref{eq:bremint1}).

Similarly, making the change of integration variable $\tau = i\Omega t$ turns
the integral on the left-hand side of (\ref{eq:bremint2}) into
\begin {equation}
  \int_0^\infty d\tau \>
    \left[ \csch\tau - \frac{1}{\tau} \right] e^{-h\tau/\Omega}
    = g(0,\tfrac{h}{\Omega}) = -\PSI\bigl( \tfrac12;\tfrac{h}{2\Omega} \bigr) ,
\end {equation}
which is the final result quoted in (\ref{eq:bremint2}).


\subsection{\boldmath$I_1(c,a)$, $I_2(c,a)$ and $I_3(a)$}
\label{app:I1}

The integral
\begin {equation}
  I_1(c,a) \equiv
  \int_0^\infty d\tau\>
  \Bigl( \frac{1}{\sinh^2\tau} - \frac{1}{\tau^2} \Bigr)
  \ln(c\tau) \, e^{-a\tau}
\end {equation}
of (\ref{eq:I1def}) may be written in terms of the integral (\ref{eq:fdef})
as
\begin {equation}
  I_1(c,a) =
  f(0,a) \, \ln c
    + \frac{ \partial f(\eps,a) }{\partial \eps} \bigg|_{\eps=0} .
\end {equation}
Using (\ref{eq:fzero}) and (\ref{eq:feps}), this gives
\begin {equation}
  I_1(c,a) =
  \left[
     a\, \PSI\bigl(1;\tfrac{a}{2}\bigr)
     - 1
  \right]
  \left[ \ln\bigl(\tfrac{c}{2}\bigr) - \gammaE \right]
  + a \bar{\bar\gamma}_1 \bigl(0;\tfrac{a}{2}\bigr)
  + 2\LNGAMMA\bigl(0;\tfrac{a}{2}\bigr) .
\end {equation}
This is equivalent to (\ref{eq:I1}) except that in the main text
we have expanded the definitions (\ref{eq:overlines})
to avoid using too much obscure notation.

Similarly, the integral (\ref{eq:I2def}) can be written in terms of
the integral (\ref{eq:gint}) as
\begin {equation}
  I_2(c,a) =
  g(0,a) \, \ln c
    + \frac{ \partial g(\eps,a) }{\partial \eps} \bigg|_{\eps=0} .
\end {equation}
Using (\ref{eq:gzero}) and (\ref{eq:geps}),
\begin {equation}
  I_2(c,a) =
  -\PSI\bigl(\tfrac12;\tfrac{a}{2}\bigr)
  \left[ \ln\bigl(\tfrac{c}{2}\bigr) - \gammaE \right]
  - \bar\gamma_1 \bigl(\tfrac12;\tfrac{a}{2}\bigr) ,
\end {equation}
which is equivalent to (\ref{eq:I2}).

Finally, the integral (\ref{eq:I3def}) is
\begin {equation}
  I_3(a) \equiv
  \int_0^\infty \frac{d\tau}{\tau} \>
  \left( \frac{1}{\sinh\tau} - \frac{1}{\tau} \right) e^{-a\tau}
  = g(-1,a)
  = 2 \LNGAMMA\bigl(\tfrac12;\tfrac{a}{2}) ,
\end {equation}
which is equivalent to (\ref{eq:I3}).


\section{Numerical Integration}
\label {app:numerics}

Numerical evaluation of our results, as plotted in fig.\ \ref{fig:overBH}b,
involves (i) evaluation of the generalized Stieltjes function
$\gamma_1(q)$ and (ii) numerical integration over $\eta$ in
(\ref{eq:dGammaneq1}) for $[d\Gamma/dx_\gamma]_{n{=}1}$,
(\ref{eq:calGrate0}) for ${\cal G}_\pair$, and
(\ref{eq:Lrate}) for $[d\Gamma/dx_\gamma]_{\rm L}$.
We used Mathematica \cite{Mathematica}, which
provides a numerical implementation of $\gamma_1(q)$.

We do not pretend to be experts on numerical methods, but it may be useful
to others for us to outline difficulties that we faced with integration
and how we worked around them.

We used Mathematica's adaptive numerical integrator.
But problems can arise when integrating over $\eta$ because our formulas
for the integrands involve very large cancellations between terms as
$\eta \to 0$ (or symmetrically $1{-}\eta \to 0$),
and the true value of the integrand after those cancellations
gets swamped by machine-precision round-off error.
The resulting numerical noise in the integrand as $\eta \to 0$,
combined with singular (but integrable) behavior of the integrand
in that same limit, caused
the adaptive integrator to focus huge (and often unsuccessful) effort on
the $\eta\to 0$ region of the integration.

The small-$\eta$ behavior of our integrands turn out
to be expansions in powers of $\sqrt{\eta}$ (up to logarithms $\ln\eta$)
rather than $\eta$.
One way to smooth the integrands for more efficient integration is to
change integration variable from $\eta$ to $\sqrt{\eta}$\,;
so, our first step was to rewrite our integrals
(using the $\eta \leftrightarrow 1{-}\eta$ symmetry of our integrands) as
\begin {equation}
   \int_0^1 d\eta \> \cdots
   = 2 \int_0^{1/2} d\eta \> \cdots
   = 4 \int_0^{1/\sqrt2} \sqrt{\eta} \, d(\sqrt{\eta}) \> \cdots .
\end {equation}
However, the integrand still has singular $\ln(\sqrt{\eta})$ behavior after this
change.  Our next accommodations were to (i) introduce a small $\sqrt{\eta}$
cut-off on the integration region
(we chose $\sqrt{\eta_{\rm min}}=10^{-4}$)%
\footnote{
  Increasing $\sqrt{\eta_{\rm min}}$\, to $10^{-3}$ changes our numerical results
  by $\le$ a few $\times~0.01\%$.
}
and (ii)
use high-precision arithmetic,
as opposed to machine precision, in order
to reduce the round-off errors in our cancellations for small $\eta$.
(We used 60-digit precision arithmetic.)

That still was not good enough to efficiently
cover the entire range of fig.\ \ref{fig:overBH}b.
The large cancellations we have mentioned in the evaluations of
$I_1(c,a)$, $I_2(c,a)$, and $I_3(a)$ [(\ref{eq:I12}) and (\ref{eq:I3})]
occur when the argument
\begin {equation}
  a = \frac{\hpr_m}{\Omega_\pr}
    = \sqrt{ \frac{i \Elpm}{2\eta(1{-}\eta)\kgamma} }
\end {equation}
has large magnitude $|a| \gg 1$.
This happens as $\eta\to 0$ for fixed $\kgamma$,
but it also happens for all $\eta$ when $\kgamma \ll \Elpma$.
Using the large-argument asymptotic expansion of $\gamma_1(z)$
derived in appendix \ref{app:asymptotic}, and the known large-argument
expansions of $\psi(z)$ and $\lnGamma(z)$, one may find large-$a$ expansions
of the $I_n$.  For the purposes of making our plots, we find good numerical
control if we use asymptotic expansions of our integrands
through $O(\kgamma^2/\Elpm^2)$
when $\kgamma \le 0.04\,\Elpm$.%
\footnote{
  Where we switch to the asymptotic expansion at $\kgamma = 0.04$,
  the relative difference in the $\LPMplus$
  rate between using numerical integration vs.\ asymptotic expansions
  is less than 1\%.
  The accuracy of the asymptotic expansions will improve as $\kgamma$
  gets smaller.
}
Those asymptotic expansions are
\begin {align}
  {\cal G}_\pair &=
  \frac{\alpha\qhat}{18\pi m^2}
  \left[
    \tfrac{7}{2}
    + i \, \tfrac{59}{100} \bigl(\tfrac{\kgamma}{\Elpm}\bigr)
    -
    \tfrac{23}{49} \bigl(\tfrac{\kgamma}{\Elpm}\bigr)^2
    + O\!\left( \bigl(\tfrac{\kgamma}{\Elpm}\bigr)^3 \right)
  \right] ,
\\[4pt]
  \left[ \frac{d\Gamma}{dx_\gamma} \right]_{(n=1)}^{m_\gamma=0} &=
  \frac{\alpha^2\qhat}{18\pi^2 m^2} P_{e\to\gamma}(x_\gamma)
  \biggl\{
    \tfrac{7}{2} \left[
       \tfrac{5}{3}
       + \tfrac12 \ln\bigl( \tfrac{\Elpm}{8 x_\gamma^2\kgamma} \bigr)
       + \gammaE
    \right]
    -\tfrac{59\pi}{400} \bigl( \tfrac{\kgamma}{\Elpm} \bigr)
\nonumber\\[-7pt] & \hspace{3em}
    +\tfrac{1}{49} \left[
       \tfrac{30391}{2100}
       - 23 \left(
           \tfrac12 \ln\bigl( \tfrac{\Elpm}{8 x_\gamma^2\kgamma} \bigr)
           + \gammaE
         \right)
     \right]
     \bigl( \tfrac{\kgamma}{\Elpm} \bigr)^2
    + O\!\left( \bigl(\tfrac{\kgamma}{\Elpm}\bigr)^3 \right)
  \biggr\} ,
\\
  \left[ \frac{d\Gamma}{dx_\gamma} \right]_{\rm L} &=
  \frac{\alpha^2\qhat}{9\pi^2 m^2 x_\gamma}
  \left[
    1
    - \tfrac{62}{1225} \bigl(\tfrac{\kgamma}{\Elpm}\bigr)^2
    + O\!\left( \bigl(\tfrac{\kgamma}{\Elpm}\bigr)^4 \right)
  \right] .
\end {align}


\section {\boldmath Large and small $q$ expansions of $\gamma_1(q)$}
\label {app:limits}

\subsection{\boldmath Large-$q$ asymptotic expansion}
\label {app:asymptotic}

\subsubsection{The expansion}

To get the large-$q$ expansion of the generalized Stieltjes
constant $\gamma_1(q)$, start from the large-$q$ expansion of
the Hurwitz zeta function $\zeta(s,q)$:%
\footnote{
  See, for example, eq.\ (25.11.43) of the Digital Library of Mathematical
  Functions \cite{DLMF}, which cites eqs.\ (1.3) and (1.4) of
  \cite{Paris}, which in turn cites ref.\ \cite{Magnus}.
}
\begin {equation}
   \zeta(s,q) =
   q^{1-s} \left[
     \frac{1}{s{-}1} + \frac{1}{2q}
     + \sum_{k=1}^\infty
       \frac{\Gamma(2k{+}s{-}1)\, B_{2k}}{(2k)!\,\Gamma(s)\,q^{2k}}
     \right]
   =
   q^{1-s} \left[
     \frac{1}{s{-}1} + \frac{1}{2q}
     + \sum_{k=1}^\infty
       \frac{(s)_{2k-1}\, B_{2k}}{(2k)!\,q^{2k}}
     \right] ,
\label{eq:Hzeta}
\end {equation}
where $B_n$ are Bernoulli numbers.
Above, $(s)_n$ is the Pochhammer symbol
\begin {equation}
  (s)_n \equiv s(s{+}1)(s{+}2)\cdots(s{+}n{-}1) .
\end {equation}
\pagebreak[4] 
Now expand (\ref{eq:Hzeta}) in $s{-}1$ and compare, order by order,
to (\ref{eq:Hurwitz}).  The order $(s{-}1)^0$ terms give the known
asymptotic expansion of the digamma function,%
\footnote{
  Though we write them with ``equals'' signs, it should be remembered that the
  large-$q$
  expansions (\ref{eq:asymp}) are asymptotic expansions and do not converge.
}
\begin {subequations}
\label {eq:asymp}
\begin{equation}
  \psi(q) =
  \ln q - \frac{1}{2q} - \sum_{k=1}^\infty \frac{B_{2k}}{2k\,q^{2k}} \,.
\end {equation}
The order $(s{-}1)^1$ terms give
\begin{equation}
  \gamma_1(q) =
  -\tfrac12\ln^2 q + \frac{\ln q}{2q}
    + \sum_{k=1}^\infty \frac{B_{2k}}{2k\,q^{2k}} \, (\ln q - H_{2k-1}) ,
\end {equation}
\end {subequations}
where
\begin {equation}
  H_n \equiv \sum_{j=1}^n \frac{1}{j}
\end {equation}
is the $n$th harmonic number.

If one makes the subtractions of (\ref{eq:bargamma1}) and
(\ref{eq:barbargamma1}), then the expansion of
$\bar{\bar\gamma}_1(0;q)$ is
\begin{equation}
  \bar{\bar\gamma}_1(0;q) =
    \sum_{k=1}^\infty \frac{B_{2k}}{2k\,q^{2k}} \, (\ln q - H_{2k-1}) .
\end {equation}


\subsubsection{Derivation of (\ref{eq:Hzeta}) from (\ref{eq:f0int_nosub})}

We mention in passing that (\ref{eq:Hzeta}) can be derived from
(\ref{eq:f0int_nosub}), which can be rearranged into the form
\begin {equation}
  \zeta(s,q) =
  \frac{q^{-s}}{2} +
  \frac{2^{s-1}}{\Gamma(s)} \int_0^\infty d\tau \> \tau^{s-1} e^{-2 q\tau} \coth\tau
  .
\label {eq:funint}
\end {equation}
Using the Taylor series expansion
\begin {equation}
  \coth\tau
  = \sum_{k=0}^\infty \frac{2^{2k} B_{2k}}{(2k)!} \, \tau^{2k-1}
\end {equation}
and then integrating term by term in (\ref{eq:funint}) yields (\ref{eq:Hzeta}).


\subsection{\boldmath Small-$q$ expansion}

For completeness, we also give the expansion of
$\gamma_1(q)$ about $q{=}0$.
It will be convenient to first find the expansion about
$q{=}1$ and then relate the two using
\begin {equation}
  \zeta(s,q) = q^{-s} + \zeta(s,1{+}q) ,
\label {eq:HurwitzPlus1}
\end {equation}
which follows from the definition
$\zeta(q,s) \equiv \sum_{k=0}^\infty (k+q)^{-s}$
of the Hurwitz $\zeta$ function.


\subsubsection{Expansion about $q{=}1$}

Differentiating the definition of the Hurwitz $\zeta$ function
with respect to $q$ gives
\begin {equation}
  \frac{\partial \zeta(s,q)}{\partial q} = -s\,\zeta(s{+}1,q).
\end {equation}
Note also that $\zeta(s,1) = \zeta(s)$, where $\zeta(s)$ is the
Riemann $\zeta$ function.  Then the Taylor expansion of
$\zeta(s,q)$ about $q{=}1$ is \cite{DLMF}%
\footnote{
  The expansion (\ref{eq:taylor1}) and explanation of
  its derivation have been taken
  from eq.\ (25.11.10) of ref.\ \cite{DLMF} and its annotations.
}
\begin {equation}
   \zeta(s,1{+}\delta) =
   \sum_{n=0}^\infty \frac{(-)^n(s)_n}{n!} \, \zeta(n{+}s) \, \delta^n .
\label {eq:taylor1}
\end {equation}
Now Taylor expand (\ref{eq:taylor1}) about $s{=}1$ to first order in
$s{-}1$ and compare to (\ref{eq:Hurwitz}).
At order $(s{-}1)^0$, this gives the known Taylor expansion%
\footnote{
  See, for example, eq.\ (6.3.14) of ref.\ \cite{AS}.
}
\begin {equation}
  \psi(1{+}\delta) = -\gammaE - \sum_{n=1}^\infty (-)^n \zeta(1{+}n)\,\delta^n .
\label {eq:psi1taylor}
\end {equation}
At order $(s{-}1)^1$, we get
\begin {equation}
  \gamma_1(1{+}\delta) =
  \gamma_1
  - \sum_{n=1}^\infty (-)^n
    \left[ H_n\,\zeta(1{+}n) + \zeta'(1{+}n) \right] \delta^n ,
\label {eq:1taylor}
\end {equation}
where here $\gamma_1 = \gamma_1(1) = -0.0728158\cdots$ is the first
(ungeneralized) Stieltjes constant defined by the Laurent expansion
of the Riemann zeta function $\zeta(s) = \zeta(s,1)$ about its
$s{=}1$ pole.


\subsubsection{Expansion about $q{=}0$}

Expanding both sides of (\ref{eq:HurwitzPlus1}) as in (\ref{eq:Hurwitz})
gives the known recursion relation
\begin {equation}
  \gamma_n(q) = \frac{\ln^n q}{q} + \gamma_n(1{+}q) .
\label{eq:gammanplus1}
\end {equation}
Then, using (\ref{eq:1taylor}),
\begin {equation}
  \gamma_1(q) =
  \frac{\ln q}{q} + \gamma_1
  - \sum_{n=1}^\infty (-)^n \left[ H_n\,\zeta(1{+}n) + \zeta'(1{+}n) \right] q^n .
\label {eq:smallq}
\end {equation}


\section{Comparison of (\ref{eq:ourBKlog}) with Baier and Katkov}
\label {app:BK}

\subsection{The overall comparison}

In this appendix, we extract the piece of
Baier and Katkov \cite{BaierKatkov} (BK) that should match
our (\ref{eq:ourBKlog}), and we discuss how the
argument of their logarithm differs parametrically from ours.
For the reasons discussed in section \ref{sec:logs}, we restrict
attention to region 2 of fig.\ \ref{fig:overBH}b and to $\kgamma{\ll}\Elpma$
for this comparison.

As explained in section \ref{sec:logs}, our logarithms arise
from pair production $\gamma\to e\bar e$ initiated by the
vacuum-like photon content $\gamma$ of the initial electron,
as in (\ref{eq:LLOrate}).  This is what BK call direct pair production,
as opposed to a ``cascade'' process of medium-induced bremsstrahlung
followed later by independent (non-overlapping) medium-induced
pair production.  They were focused on the case of a medium whose
length $l$ is small compared to the radiation length, and so the
probability of their direct and cascade processes
were proportional to $l$ and $l^2$ respectively.  And because their $l$ was
relatively small (in contrast to our assumption that the medium
is large), they also considered processes where the photon arises
from boundary radiation.
But one can isolate the part of their calculation that should, in
principle, correspond to our calculation at leading-log order.
It is captured by what they call
$d\mbox{\sl w}_2^m/dz\>dy$ in BK eq.\ (5).
Their notation $\mbox{\sl w}_2$ refers to vacuum-like splitting $e \to e\gamma$
followed by medium-induced pair production via
a single interaction with the medium (corresponding to
the Bethe-Heitler limit for pair production and so the $\kgamma \ll \Elpma$
limit of our analysis).
The superscript $m$ on $\mbox{\sl w}_2^m$ is their notation for
also including the influence of multiple scattering on the initial electron,
which they only claim to handle at leading-log order.
Their $y$ and $z$ are our $x_\gamma$ and $(1{-}\eta) x_\gamma$, and
table \ref{tab:BK} summarizes several such translations of notation between
their work and ours.
Keeping only logarithmic-enhanced terms, BK (5) and (6) give
\begin {equation}
  \frac{\mbox{\sl w}_2^m}{dz\,dy} \approx
  \frac{\alpha l}{\pi L} \frac{(1{-}y)}{y^2} \, P(y,z)
  \, \ln\Bigl( \frac{1}{\xi(1{+}v_1)} \Bigr)
  \simeq
  \frac{\alpha l}{\pi L y^2} \, P(y,z)
  \, \ln\Bigl( \frac{1}{\xi v_1} \Bigr) .
\label {eq:dw2}
\end {equation}
In the last equality, and throughout this appendix,
we take the soft-photon ($y{\ll}1$)
and deep-LPM ($v_1{\gg}1$) limits since both apply to our
limiting formula (\ref{eq:ourBKlog}).


\begin {table}[t]
\begin {center}
\begin{tabular}{ccl}
\toprule
  Baier+Katkov
  & our notation
  & source or notes
\\ \hline
  $\varepsilon$ & $E$ & third paragraph of BK \\
  $\omega$ & $\kgamma$ & BK (2) \\
  $\varepsilon_+$, $\varepsilon_-$ & $(1{-}\yfrak)\kgamma$, $\yfrak\kgamma$
     & second paragraph of BK \\
  $\varepsilon_e$ & $\simeq \Elpma \simeq \tfrac12\Elpm$
     & $\frac{m^2}{\varepsilon_e} = \cdots$ in BK (6); $L_0$ in BK (2) \\
  $y$, $z$ & $x_\gamma$, $(1{-}\yfrak)x_\gamma$
     & BK (2) \\
  $\beta$ & $\yfrak(1{-}\yfrak)$ & BK (2) \\
  $l$ & & length of the medium \\
  $\frac{1}{L}$ & $\simeq \frac{\alpha\,\qhat(m^{-1})}{2\pi m^2}$
     & compare BK (2) to our (\ref{eq:qhat}) with (\ref{eq:BHlog}) \\
  $n(y,z)$ &
    $~\approx \frac{\alpha}{2\pi} \, P_{e\to\gamma}(x_\gamma)
     \, \ln\bigl( \frac{q^2_\max}{q^2_\min} \bigr)$
     & $\gamma$ content of initial $e$ (BK (7); leading-log for us) \\
  $\mbox{\sl w}_c$ & & probability of cascade process \\
  $\mbox{\sl w}_2$ &
     & prob.\ of direct process (single scattering) \\
  $\mbox{\sl w}_2^m/l$
     & $\Delta\Gamma_{e\to ee\bar e}$
     & rate for direct process (multiple scattering)$^{(*)}$ \\
  $\mbox{\sl w}_b$ &
     & prob.\ of cascade process from boundary radiation\\[-4pt]
  $n_a$ & $n$ & number density of atoms \\
  $\omega_0$ & $m_\gamma$ & ignored by BK except for text before BK (13)
  \\ \hline
  $\xi = \frac{z(y{-}z)^{\strut}}{1{-}y}$
     & $\frac{\eta(1{-}\eta)x_\gamma^2}{1{-}x_\gamma}$ & BK (2)\\[3pt]
  $v_1 = \frac{\varepsilon(1{-}y)}{\varepsilon_e y}$ 
    & $ \simeq \frac{E(E{-}k)}{\Elpma\kgamma} $
    & BK (6); 
              deep LPM is $v_1{\gg}1$ \\[8pt]
  $\omega_c = \frac{\varepsilon^2}{\varepsilon+\varepsilon_e}$
    & $\frac{E^2}{\Elpma+E}$
    & $\omega \ll \omega_c$ is deep LPM \\[5pt]
  $\frac{d\mbox{\scriptsize\sl w}_{2_{\vphantom{.}}}^m}{l\,dz\,dy}$
    & $\approx \frac{d\Gamma}{x_\gamma\,dx_\gamma\,d\yfrak}$ \\
  $P(y,z) \approx 1-\frac43\beta$ & $\frac13 [ 2P_{\gamma\to e}(\yfrak) + 1 ]$
    & BK (6); $\propto$ DGLAP+spin-flip for BH $\gamma \to e\bar e$
      (\ref{eq:BHpair}) \\[-4pt]
\\ \botrule
\end{tabular}
\caption{
  \label{tab:BK}
  Select translations of Baier and Katkov's notation
  \cite{BaierKatkov} to ours.
  As elsewhere in our paper, $\approx$ indicates leading-log order.
  (*) The $\Delta$ in our notation $\Delta \Gamma_{e\to ee\bar e}$ above
  is borrowed from refs.\ \cite{seq,qedNf} to mean
  the correction to the $e\to \gamma e \to ee\bar e$ rate
  from overlapping formation times.
}
\end {center}
\end {table}


Dividing (\ref{eq:dw2})
by length $l$ to convert probability $\mbox{\sl w}_2^m$ into a rate,
and then converting to our notation,
their result is
\begin {multline}
  \left[ \frac{d\Gamma}{dx_\gamma\,d\eta} \right]_\BK \approx
  \frac{\alpha^2 \, \qhat(m^{-1})}{6\pi^2 m^2 x_\gamma}
     \, [ 2P_{\gamma\to e}(\yfrak) + 1 ]
     \, \ln\Bigl( \frac{m^4}{\eta(1{-}\eta)\qhat\kgamma} \Bigr)
\\
  \simeq
  \frac{\alpha}{2\pi} \,
  P_{e\to\gamma}(x_\gamma) \left[ \frac{d\Gamma}{d\eta} \right]_\pair
     \, \ln\Bigl( \frac{m^4}{\eta(1{-}\eta)\qhat\kgamma} \Bigr)
\label {eq:BKrate1}
\end {multline}
In the last equality, we
have used the soft limit $P_{e\to\gamma}(x_\gamma) \simeq 2/x_\gamma$
and the low-energy $(\kgamma{\ll}\Elpma)$ limit (\ref{eq:BHpair}) of
the pair production rate.  To compare (\ref{eq:BKrate1}) to
(\ref{eq:ourBKlog}), we just need to integrate over $\eta$.
That integration is dominated by democratic pair production
[$\eta(1{-}\eta) \sim 1$], and so we
can drop the $\eta$ dependence of the logarithm
at leading-log order, giving
\begin {equation}
  \left[ \frac{d\Gamma}{dx_\gamma} \right]_\BK \approx
  \frac{\alpha}{2\pi} \,
  P_{e\to\gamma}(x_\gamma) \, \Gamma_\pair
     \, \ln\Bigl( \frac{m^4}{\qhat\kgamma} \Bigr) .
\label {eq:BKrate2}
\end {equation}
For comparison,
taking the soft-photon limit of (\ref{eq:MOmega0}) for $\Omega_0$,
our leading-log contribution (\ref{eq:ourBKlog}) arising from
$e \to e\gamma^* \to ee\bar e$ is instead
\begin {equation}
  \delta\left[ \frac{d\Gamma}{dx_\gamma} \right] \approx
  \frac{\alpha}{2\pi} \,
  P_{e\to\gamma}(x_\gamma) \, \Gamma_\pair
     \, \ln\Bigl( \frac{m^2}{x_\gamma\sqrt{\qhat\kgamma}} \Bigr) .
\label {eq:ourBKrate2}
\end {equation}


\subsection{Difference: lower bound on photon virtuality}

Baier and Katkov express their logarithm in terms of the range of
the photon virtuality $q^2$ (which in our notation would be $-k^\mu k_\mu$).
Their logarithm in (\ref{eq:BKrate2}) is
\begin {equation}
   \ln\left( \frac{\qmax^2}{\qmin^2} \right) .
\label {eq:logq}
\end {equation}
In our analysis, which uses time-ordered perturbation theory, we discuss
the off-shellness $\Delta E$ in energy of
intermediate states instead of the virtuality of individual particles.
For high-energy
bremsstrahlung, $\Delta E$ is of order the kinetic energy difference
$E_{e^-} + E_\gamma - E_{e^+}$ given by everything excluding
the potential term in our effective 3-particle Hamiltonians (\ref{eq:Heff})
and (\ref{eq:HeffHO}).  (See the brief review of the origin of those terms
in section 3.1 of ref.\ \cite{softqed1}.)  At leading-log order, however,
we need only rough parametric estimates of $\qmax^2$ and $\qmin^2$ in
(\ref{eq:logq}), and in that context one may roughly translate
between $\Delta E$ and photon virtuality by identifying
\begin {equation}
   q^2 \equiv -k^\mu k_\mu \simeq 2 \kgamma \, \Delta E .
\end {equation}


\subsubsection {$\qmax^2$}

In section \ref{sec:logs}, we identified $(\Delta t)_{\rm min}$ as
$\tform^\pair$, and so
\begin {equation}
   \qmax^2 \sim \kgamma (\Delta E)_{\rm max}
   \sim \frac{\kgamma}{\tform^\pair} \,.
\end {equation}
As also discussed in section \ref{sec:logs},
$\tform^\pair \sim 1/\hpr_m \sim \kgamma/m^2$ for the case $\kgamma{\ll}\Elpma$
here, and so
\begin {equation}
  \qmax^2 \sim m^2 ,
\label {eq:qmax2}
\end {equation}
which is the threshold at which the medium-induced pair production rate
$\Gamma_\pair$ from an \textit{on-shell} photon ceases to be a
good approximation to pair production from a photon with virtuality $q^2$.
In Baier and Katkov's notation, their $\qmax^2$ is
$Q^2 = m^2\omega^2/\varepsilon_+\varepsilon_-$, which in our notation
is
\begin {equation}
   \qmax^2 = \frac{m^2}{\eta(1{-}\eta)} \,.
\label {eq:BKqmax2}
\end {equation}
Since total pair production is dominated by $\eta(1{-}\eta) \sim 1$,
this matches our (\ref{eq:qmax2}).


\subsubsection {$\qmin^2$}

The difference between our logarithm in (\ref{eq:ourBKrate2}) and
BK's logarithm in (\ref{eq:BKrate2}) comes from the scale of $\qmin^2$.
Our estimate in (\ref{eq:ourBKlog}) was that
$(\Delta t)_{\rm max} \sim \tform^\brem$, which is
$~1/|\Omega_0|$ in the deep-LPM (and small overlap correction) region being
discussed here, where $|\Omega_0| \ll \Gamma_\pair$.
That translates to photon virtuality
\begin {equation}
  \qmin^2 \sim \frac{\kgamma}{\tform^\brem} \sim \kgamma |\Omega_0|
  \sim x_\gamma \sqrt{\qhat\kgamma} .
\label {eq:qmin2}
\end {equation}

In this context, BK refer to the minimum photon virtuality as $q_c^2$.
The easiest way to identify their
value is to identify their logarithm in (\ref{eq:BKrate2}) above
as (in our notation) $\ln(\qmax^2/\qmin^2)$ and then use
(\ref{eq:BKqmax2}) to get%
\footnote{
  Beware BK eqs.\ (8) for $q_c^2 = \qmin^2$, which are specialized to the case
  where multiple scattering from the medium is \textit{not} included.
}
\begin {equation}
  \bigl[\qmin^2\bigr]_\BK
  \sim \frac{\qhat\kgamma}{m^4} \, \qmax^2
  \sim \frac{\qhat\kgamma}{\eta(1{-}\eta) m^2} \,.
\end {equation}
Again specializing to the case of democratic pair production (dominating
our total pair-production rate), this is
\begin {equation}
  \bigl[\qmin^2\bigr]_\BK
  \sim \frac{\qhat\kgamma}{m^2} \,,
\label {eq:BKqmin2}
\end {equation}
which is very different from our (\ref{eq:qmin2}).


\subsection {Discussion}

We have not fully absorbed BK's argument for their value of $\qmin^2$ and
so cannot definitely pinpoint why their analysis does not reproduce ours
at leading-log order.  But we note that their $\qmin^2$ in
(\ref{eq:BKqmin2}) is parametrically equal to the squared kick to $\p_\perp$
that an electron would experience over the duration of democratic
pair production $\gamma \to e\bar e$.  That's because
$(\Delta p_\perp)^2 \simeq \qhat\,\Delta t$ by definition of $\qhat$,
and the duration of pair production for $\kgamma{\ll}\Elpma$ is
$(\Delta t)_\pair \sim 1/\hpr_m \sim \kgamma/m^2$.
However, in our calculation, the duration of the entire process
$e \to e\gamma^* \to ee\bar e$ ranged from $(\Delta t)_\pair$ to
$(\Delta t)_\brem^\LPM$ and so can be much longer than $(\Delta t)_\pair$
[corresponding in our analysis to a smaller value of $\Delta E$
for vacuum-like bremsstrahlung and so a smaller value of $\qmin^2$
(in the deep LPM regime)].

Finally, we emphasize again that the full
calculation in this paper was not restricted to leading-log approximations
and did not require the independent estimates of $\qmin^2$ and $\qmax^2$
presented in this appendix nor the related qualitative arguments of
section \ref{sec:logs}.
In particular, our logarithm (\ref{eq:ourBKrate2}) originates from
the leading-log piece of (\ref{eq:neq1low}), which arose as the
$k_\gamma{\ll}\Elpma$ limit of the more general formula
(\ref{eq:dGammaneq1}).%
\footnote{
  Readers might find confusing that (\ref{eq:neq1low}) is the
  result for the $n{=}1$ case of fig.\ \ref{fig:LPM+diags}a, which is
  seemingly a diagram contributing to overlap of bremsstrahlung
  $e \to e\gamma$ with \textit{virtual} pair production and not
  a diagram for real pair production.
  As briefly summarized at the start of section
  \ref{sec:diags} and discussed in more detail in section 4.1.1 of
  ref.\ \cite{softqed1}, fig.\ \ref{fig:LPM+diags} nonetheless is equal
  to the \textit{sum} of both real and virtual pair production processes.
  Physically, it is the \textit{real}
  pair production process that generates the logarithms in our results,
  as discussed in section 7 of ref.\ \cite{softqed1}.
  See in particular footnote 55 of ref.\ \cite{softqed1}, which
  explains how to explicitly verify this claim
  (albeit in the $k{\gg}\Elpma$ limit rather than the $k{\ll}\Elpma$ limit)
  from the work of ref.\ \cite{qedNfenergy}.
}
(Other contributions to overlapping
bremsstrahlung and pair production are either sub-leading
deep in region 2 of fig.\ \ref{fig:overBH}b
or do not generate large logarithms in the limits
taken.)


\end {document}